\documentclass[amsmath,amssymb,showpacs,preprint]{revtex4-1}
\usepackage{hyperref}
\usepackage[utf8]{inputenc}
\numberwithin{equation}{section}

\usepackage{multirow}
\usepackage{float}
\usepackage{soul}
\usepackage{graphicx}
\usepackage{xcolor}
\graphicspath{ {./images/} }

\begin{document}

\title{Using physics-informed neural networks to compute\\quasinormal modes}

\author{Alan S Cornell}
\email{acornell@uj.ac.za}
\author{Anele Ncube}
\email{ncubeanele4@gmail.com}
\author{Gerhard Harmsen}
\email{gerhard.harmsen5@gmail.com}
\affiliation{Department of Physics, University of Johannesburg, PO Box 524, Auckland Park 2006, South Africa}

\date{\today}

\begin{abstract}
In recent years there has been an increased interest in neural networks, particularly with regard to their ability to approximate partial differential equations. In this regard, research has begun on so-called physics-informed neural networks (PINNs) which incorporate into their loss function the boundary conditions of the functions they are attempting to approximate. In this paper, we investigate the viability of obtaining the quasi-normal modes (QNMs) of non-rotating black holes in 4-dimensional space-time using PINNs, and we find that it is achievable using a standard approach that is capable of solving eigenvalue problems (dubbed the eigenvalue solver here). In comparison to the QNMs obtained via more established methods (namely, the continued fraction method and the 6th-order Wentzel, Kramer, Brillouin method) the PINN computations share the same degree of accuracy as these counterparts. In other words, our PINN approximations had percentage deviations as low as $(\delta\omega_{_{Re}}, \delta\omega_{_{Im}}) = (<0.01\%, <0.01\%)$. In terms of the time taken to compute QNMs to this accuracy, however, the PINN approach falls short, leading to our conclusion that the method is currently not to be recommended when considering overall performance.
\end{abstract}


\maketitle

\section{Introduction}

\par In recent years there has been an increased interest in the use of neural networks (NNs) as functional approximators \cite{hornik1989multilayer, wu2000dynamic, nielsen2015neural}. The interest lies in the fact that NNs are versatile as demonstrated in their success in various applications such as natural language processing, image recognition and, more recently, scientific computing \cite{Raissietal2019, Luetal2020, Nordstrom2021}. In scientific computing, they have been shown to be robust and data-efficient solvers of partial differential equations that govern diverse systems studied in mathematics, science and engineering \cite{Raissietal2019}. In general, NNs can be trained once and used in a variety of situations that are within the scope of the problem it was trained on. The advantage of applying a trained NN is that it expedites the computation of later solutions whereas, by contrast, more traditional numerical approximating methods would require an inefficient process beginning from scratch each time a solution is derived. Furthermore, NNs are also natively parallelisable, which adds to their higher computational efficiency compared to other numerical approximations.

\par A new technique has recently been developed to assist in creating NNs that can act as functional approximators, which takes inspiration from boundary type problems where the boundary conditions of the function are used to solve for the underlying function \textit{as-is}; namely, physics-informed neural networks (PINNs) \cite{Raissietal2019, Jinetal2020}. In this regard, we are interested in determining if these types of NNs could be used to compute the quasi-normal modes (QNMs) of black holes. The QNMs of black holes have been studied for many years and it is well-known that they are correlated to the parameters of the black holes that generate them, and as such, they act as a \textit{telltale} sign to probe the properties of black holes \cite{KokkotasSchmidt1999, KonoplyaZhidenko2011, Bertietal2009}.

\par Over the years numerous techniques have been used to determine the numerical values of black holes using the radial equations that govern the perturbations of black holes \cite{KonoplyaZhidenko2011}. Some examples are the Wentzel, Kramer, Brillouin (WKB) method, asymptotic iteration method (AIM), and the continued fraction method (CFM) \cite{IyerWill1987, Leaver1985, Choetal2012}. Although all of these approaches have been successful in solving the radial equations of black holes to determine the numerical values of the QNMs; however, they do have computational limitations \cite{KokkotasSchmidt1999}. The WKB, in particular, becomes progressively difficult to apply when more accurate results are needed since achieving this requires painstaking derivations of higher-order approximations. In this work, we intend to show that PINNs can potentially supplement the extant techniques as a new alternative method for obtaining the black hole QNMs, with its unique advantages and limitations. Furthermore, we will compare the accuracy of PINNs to the already established methods and test their generalisability when applied to black hole perturbations equations.

\par Our motivation for using the equations of QNMs to test the usefulness of PINNs is that the equations that govern the QNMs are based on only a few parameters, namely a black hole's physical properties, and their boundary conditions are well-defined for the system. As such, the boundary conditions act as a regularisation mechanism that sufficiently limits the space of admissible solutions and contributes to the NN's stability \cite{Raissietal2019}. Furthermore, in astrophysical circles, there has been an increased interest in black hole QMNs given the recent landmark detections of gravitational waves at the VIRGO and LIGO detectors \cite{Isietal2019, Ghoshetal2021}.

\par The paper is set out in the following manner. In the next section, we describe the equations that govern the QNMs for various space-times. In section \ref{SEC: Numerical methods} we present the currently accepted methods for determining the QNMs and proceed to touch on the new PINN approach in section \ref{SEC: PINNS}. Finally, in sections \ref{SEC: Results} and \ref{SEC: Discussion}, we discuss the results obtained from applying PINNs and compare them to the QNMs obtained from the canonical methods.

\section{The radial perturbation equations of black holes}

\par In this section, we will derive the equations required to determine the numerical value of QNMs beginning with the simplest space-times and then building up to more complex ones, which will eventually be encoded into the numerical methods. We begin with the Schwarzschild metric:

\subsection{The asymptotically flat Schwarzschild solution}\label{SEC: Schwarzschild}

\par We consider scalar-type perturbations (and later, electromagnetic, Dirac and gravitational perturbations) then in order to derive the radial equations required to determine the QNMs we begin by considering the equation of motion, which is given by the Klein-Gordon equation \cite{CarrollTraschen2005, Mcmahon2008}:
\begin{equation}
\label{EQ: Klein-Gordon}
\partial_{\mu}\partial^{\mu}\Phi + m^2\Phi = \frac{1}{\sqrt{|g|}}\partial_{\mu}(\sqrt{|g|} g^{\mu\nu} \partial_{\nu}\Phi) +m^2\Phi = 0,
\end{equation}

\noindent where $\Phi$ is a scalar field with mass $m$ perturbing the black hole's space-time as given by the metric $g$. In the case of the Schwarzschild black hole, the metric is given as:
\begin{eqnarray}
\label{EQ: Schwarzchild_metric}
ds^2 = - fdt^2 + \frac{1}{f}dr^2 + r^2(d\theta^2 + \mathrm{sin}^2\theta d\phi),
\end{eqnarray}

\noindent where $f = 1 - 2M/r$ is the metric function, with $M$ and $r$ representing the mass of the black hole and the radial distance from the centre of the black hole, respectively. The last two terms on the right-hand side of this equation represent the metric of a 2-sphere \cite{CarrollTraschen2005}. As $r \rightarrow \infty$, we expect to recover a weak-field approximation of the metric wherein the components of the metric tensor can be decomposed into the flat Minkowski metric tensor $\eta_{\mu\nu}$ plus a small perturbation $|h_{\mu\nu}| \ll 1$; that is: $g_{\mu\nu} \approx \eta_{\mu\nu} + h_{\mu\nu}$ \cite{CarrollTraschen2005}.

\par Considering the massless form of the Klein-Gordon equation, where $m=0$ in equation (\ref{EQ: Klein-Gordon}), and plugging in it the metric given in equation (\ref{EQ: Schwarzchild_metric}) we obtain:
\begin{alignat}{2}
\label{EQ:Schwarz_Klein-Gordon}
&& \frac{1}{\sqrt{|g|}} \partial_{\mu}(\sqrt{|g|}g^{\mu\nu}\partial_{\nu}\Phi) &= -\left(1 -\dfrac{2M}{r}\right)^{-1}\dfrac{\partial^2\Phi}{\partial t^2} + \dfrac{1}{r^2}\dfrac{\partial}{\partial r}\left[ r^2 \left(1 -\dfrac{2M}{r}\right)\dfrac{\partial\Phi}{\partial r} \right]\notag\\
&&&\qquad +\ \dfrac{1}{r^2\sin{\theta}}\dfrac{\partial}{\partial \theta}\left[\sin{\theta}\dfrac{\partial\Phi}{\partial \theta} \right] + \dfrac{1}{r^2\sin^2{\theta}}\dfrac{\partial^2\Phi}{\partial\phi^2} = 0.
\end{alignat}

\noindent In this explicit form, we can derive the equation of massless scalar fields in the Schwarzschild background in terms of the radial coordinate $r$ via a separation of variables \cite{Iyer1987}. By mapping the resulting one-dimensional differential co-ordinate into an infinite domain given by a tortoise co-ordinate, $x$, we find \cite{ReggeWheeler1957, IyerWill1987}:
\begin{equation}
\label{2.4}
\dfrac{d^2 \psi}{dx^2} + \left\lbrace \omega^2 - V(r)\right\rbrace\psi =\ 0,
\end{equation}

\noindent where: 
\begin{equation}
\label{2.5}
V(r) = \left(1 - \dfrac{2M}{r}\right)\left[ \frac{\ell(\ell + 1)}{r^2} + \dfrac{2M}{r^3} \right],
\end{equation}
\begin{equation}
\label{2.6}
f(r) = \frac{dr}{dx} = \left(1 - \frac{2M}{r}\right)^{-1}.
\end{equation}

\noindent Here\ $n$,\ $\ell$ and\ $m$\ are the principal, multipole, azimuthal and numbers, respectively \cite{KonoplyaZhidenko2011}. The tortoise coordinate maps the location of the event horizon of the Schwarzschild black hole from $r = 2M$ (in geometric units) to $x = -\infty$. As such, it maps the space from a semi-infinite domain to an infinite one. Note that equation (\ref{2.4}) is quite similar in form to the one-dimensional time-independent Schr\"{o}dinger equation, but in this case, $V(r)$ is the effective potential for a which scalar field perturbs an asymptotically flat Schwarzschild metric \cite{Choetal2012, Iyer1987}. The QNM frequencies, $\omega$, are complex-valued solutions to equation (\ref{2.4}), which is a non-Hermitian problem, unlike the Schr\"{o}dinger equation \cite{Bertietal2009}. For asymptotically flat astrophysical black holes, the eigenfunctions, $\psi$, that solve this equation have asymptotic behaviour governed by \cite{Choetal2012}:
\begin{equation}
\label{2.7}
\psi(x) = \begin{cases}
e^{-i\omega x},\quad &x \rightarrow -\infty\\
e^{+i\omega x},\quad  &x \rightarrow +\infty\\
\end{cases}.
\end{equation}

Transforming from the infinite domain of the tortoise co-ordinate to the finite domain of a new variable, $\xi = 1 - 2M/r$, it can be shown that equation (\ref{2.4}) takes the form \cite{Choetal2012}:
\begin{equation}
\label{2.8}
\chi^{\prime\prime} =  \lambda_{0}(\xi)\chi^{\prime}\ +\ s_{0}(\xi)\chi,
\end{equation}
\noindent where

\begin{alignat}{2}
\label{2.9}
&& \lambda_{0}(\xi)\quad &=\quad \frac{4Mi\omega(2\xi^2 - 4\xi + 1) - (1 - 3\xi)(1 - \xi)}{\xi(1 - \xi)^2},\\
\label{2.10}
&& s_{0}(\xi)\quad &=\quad \frac{16 M^2\omega^2(\xi - 2) - 8 Mi\omega(1 - \xi) - \ell(\ell + 1) + (1 - s^2)(1 - \xi)}{\xi(1 - \xi)^2},
\end{alignat}

\noindent and $\chi(\xi)$ is a complex-valued scale factor. The boundary conditions have been incorporated into equation (\ref{2.8}). The importance of this transformation is that it maps the domain from one that is infinite, i.e. $-\infty < x < +\infty$ to one that is finite, i.e. $0 \leq \xi < 1$. As such, it is now possible to numerical solve the perturbation equation since the domain is now finite and the QNM boundary conditions are implicitly accounted for.

For electromagnetic field perturbations of Schwarzschild black holes, the same Schr\"{o}dinger-like radial equations are obtained by following the same procedure for deriving the massless scalar fields. However, in this case, the equation of motion considered is the source-free Gauss-Amp\`{e}re law of Maxwell's equations \cite{CarrollTraschen2005, DeyChakrabarti2019}:
\begin{equation}\label{EQ: Maxwell}
F^{\mu\nu}_{\ \ ;\nu} = \frac{1}{\sqrt{|g|}} \partial_{\nu}\left(\sqrt{|g|} F^{\mu\nu} \right) = 0,
\end{equation} 

\noindent where $F^{\mu\nu}$ is the electromagnetic field tensor. Applying the components of the electromagnetic field tensor $F^{\mu\nu}$, we can determine the radial perturbation equation from Maxwell's equations:
\begin{equation}
\label{2.12}
\dfrac{\partial}{\partial t} F^{\mu t} + \frac{1}{r^2}\dfrac{\partial}{\partial r}(r^2 F^{\mu r}) + \frac{1}{\mathrm{sin}\theta}\dfrac{\partial}{\partial \theta}(\mathrm{sin}\theta F^{\mu\theta}) + \dfrac{\partial}{\partial \phi} F^{\mu\theta} = 0.    
\end{equation}

\noindent We can simplify the equations with indices $\mu=\theta$ and $\mu=\phi$ to obtain the Schr\"{o}dinger-like perturbation equations. In short, we arrive at:
\begin{equation}
\label{2.13}
- \frac{\partial^2 a_0(t,r)}{\partial t^2} + f^2 \frac{\partial^2 a_0(t,r)}{\partial r^2} - \frac{f \ell(\ell + 1)}{r^2} a_0(t,r) = 0,
\end{equation}

\noindent where $a_0(t,r)$ represents the electromagnetic field perturbations. Thus, if we have \newline $a_0(t,r) = a_0(r)e^{i\omega t}$, converting to tortoise co-ordinates we retrieve equation (\ref{2.4}), where $\psi(r) = a_0(r)$ and $V(r) = \ell(\ell + 1) f(r)/r^2$ is the effective potential of an asymptotically Schwarzschild black hole perturbed by an electromagnetic field. For gravitational perturbations, the equations have the same form except for the effective potential $V(r)$. Refs. \cite{CardosoLemos2001, DeyChakrabarti2019} outline concisely the steps for arriving at the wave equations for these direct metric perturbations on a Schwarzschild black hole. 

\subsection{The Schwarzschild (anti)-de Sitter solution}
\label{SEC: SdS}

\par We shall also consider asymptotically curved space-times that are solutions to Einstein's equations with a non-zero cosmological constant. The cosmological constant, denoted by $\Lambda$, encodes the curvature of space-time via the relation $\Lambda = \pm 3/a^2$, where $a$ is the cosmological radius \cite{Molina2003, CardosoLemos2003}. The metric in this case is:
\begin{equation}
\label{2.14}
ds^2 = -\left(1 - \frac{r_s}{r} - \frac{\Lambda r^2}{3}\right) dt^2 + \left(1 - \frac{r_s}{r} - \frac{\Lambda r^2}{3}\right)^{-1} dr^2 + r^2d\Omega^2.
\end{equation}

\noindent With this metric as a starting point, the radial perturbation equation derived for a 4-dimensional (anti)-de Sitter Schwarzschild black hole is the same form as equation (\ref{2.4}) but, with a more general effective potential given as \cite{Choetal2012}:
\begin{equation}
\label{2.15}
V(r) = f(r)\left[\frac{\ell(\ell + 1)}{r^2} + (1 - s^2)\left( \frac{2M}{r^3} - \frac{(4 - s^2) \Lambda}{6}\right)\right],
\end{equation}

\noindent where $f(r) = 1 - 2M/r - (\Lambda r^2)/3$ is the metric function for (anti)-de Sitter Schwarzschild space-times and $s = 0,\ 1/2,\ 1$ and $2$ denote the spins of scalar, Dirac, electromagnetic and gravitational fields, respectively.

\subsection{Near extremal Schwarzschild and Reissner-Nordstr\"{o}m-de Sitter solutions}\label{SEC: Near Extremal}

\par A final case we shall consider are Reissner-Nordstr\"{o}m-de Sitter black holes, albeit in the near extremal case. The metric of a Reissner-Nordstr\"{o}m-de Sitter black hole is \cite{Molina2003}:
\begin{equation}
\label{2.16}
ds^2 = -\left(1 - \frac{r_s}{r} + \frac{r^2_Q}{r^2}  - \frac{\Lambda r^2}{3}\right) dt^2 + \left(1 - \frac{r_s}{r} + \frac{r^2_Q}{r^2}  - \frac{\Lambda r^2}{3}\right)^{-1} dr^2 + r^2d\Omega^2,
\end{equation}

\noindent where $r_{s} = 2M$ (the Schwarzschild radius) and $r_Q^2 = Q^2 /4\pi\epsilon_0$. Generally, when solving radial perturbation equations, the nature of the effective potentials preclude applying a direct, analytical approach to deriving exact QNMs \cite{IyerWill1987}. However, in special cases, such as this one involving near extremal Schwarzschild and Reissner-Nordstr\"{o}m-de Sitter black holes, the effective potentials can be transformed to yield differential equations with known analytic solutions \cite{KonoplyaZhidenko2011}. 

\par To obtain the effective potentials of non-rotating black holes in the near extremal (anti)-de Sitter case, we consider the relevant metric function, $f(r) = 1 - 2M/r - \Lambda r^2/3$. The solutions to $f(r) = 0$ are $r_b$ and $r_c$,  which are the black hole's event horizon and the space-time's cosmological radius, respectively (where $r_c > r_b$). For $r_0 = -(r_b + r_c)$, the metric function can be given as \cite{CardosoLemos2003}:
\begin{equation}
\label{2.17}
f(r) = \frac{1}{a^2 r}(r - r_b)(r_c - r)(r - r_0),
\end{equation}

\noindent where $a^2 = r_b^2 + r_br_c+ r_c^2$ and $2Ma^2 = r_b r_c (r_b + r_c)$. The surface gravity, $\kappa$, associated with the black hole event horizon $r = r_b$ is defined as \cite{CardosoLemos2003}: 
\begin{equation}
\label{2.18}
\kappa = \left. \frac{1}{2}\frac{df}{dr}\right\vert_{r=r_b} = \frac{(r_c - r_b)(r_b - r_0)}{2a^2r_b}.
\end{equation}

\noindent In the near extremal de Sitter case, the cosmological horizon $r_c$ of the space-time is very close (in the co-ordinate $r$) to the black hole horizon $r_b$ so that $(r_c - r_b)/ r_b \ll 1$, and the following approximations apply \cite{CardosoLemos2003}:
\begin{equation}
\label{2.19}
r_0 \sim -2r_b^2;\ a^2 \sim 3r_b^2;\ M \sim \frac{r_b}{3};\ \kappa \sim \frac{r_c - r_b}{2r_b^2}.
\end{equation}

\noindent Also since the domain of $r$ is within ($r_b$, $r_c$) and $r_b \sim r_c$, we find that $r - r_0 \sim r_b - r_0 \sim 3r_0$. In turn the metric function equation (\ref{2.17}) becomes:
\begin{equation}
\label{2.20}
f \sim \frac{(r - r_b)(r_c - r)}{r_b^2}.    
\end{equation}

\noindent With this new form of the metric, the relation between the tortoise co-ordinate and the radial co-ordinate (\ref{2.6}) reduces to:
\begin{equation}
\label{2.21}
r = \frac{r_c e^{2\kappa x} + r_b}{1 + e^{2\kappa x}}.
\end{equation}

\noindent Substituting this expression for $r$ into the $f(r)$ equation (\ref{2.17}), we find the expression for $f(x)$ as \cite{CardosoLemos2003}:
\begin{equation}
\label{2.22}
f(x) = \frac{(r_c - r_b)^2}{4r_b^2 \mathrm{cosh}^2(\kappa x)}.    
\end{equation}

\noindent With this metric function, the effective potential of a near extremal Schwarzschild-de Sitter black hole is an inverted P\"{o}schl-Teller potential \cite{CardosoLemos2003}:
\begin{equation}
\label{2.23}
V(x) = \frac{V_0}{\mathrm{cosh}^2(\kappa x)},
\end{equation}

\noindent where $V_0 = \kappa^2 \ell(\ell + 1)$ for massless scalar and electromagnetic perturbations and \newline $V_0 = \kappa^2 (\ell + 2)(\ell - 1)$ for gravitational perturbations. With the effective potential in this form, the perturbation equation (\ref{2.4}) can now be solved analytically to derive the QNMs of near-extremal Schwarzschild-de Sitter black holes.

\par For astrophysical near-extremal de Sitter black holes, the asymptotic behaviour of the solution is similar to that of an asymptotically flat Schwarzschild black hole equation (\ref{2.7}); considering that they force the solution near the event horizon (cosmological horizon) not to generate outgoing (incoming) waves. 

\par Considering the boundary conditions for astrophysical black holes, equation (\ref{2.7}), Ref.~\cite{FerrariMashhoon1984} determined the analytic expressions of the QNM eigenfunctions and eigenfrequencies \cite{CardosoLemos2003, Molina2003, FerrariMashhoon1984} as:
\begin{equation}
\label{2.24}
\psi(x) =[\xi(\xi -1)]^{i\omega/2\kappa} \cdot  {}_2F_1\left(1 + \beta + i\frac{\omega}{\kappa}, - \beta + i\frac{\omega}{\kappa}; 1 + i\frac{\omega}{\kappa}; \xi\right),
\end{equation}
\begin{equation}
\label{2.25}
\frac{\omega}{\kappa} = \sqrt{\left(\ell(\ell + 1) - \frac{1}{4}\right)} - i\left(n + \frac{1}{2}\right),\quad n = \mathbb{Z}^+_0,
\end{equation}

\noindent where $\xi^{-1} = 1 + \exp{(-2\kappa x)}$ and $\beta = -1/2 + (1/4 - V_0/\kappa^2)^{1/2}$. 

Extending from near extremal Schwarzschild-de Sitter black holes, Ref.~\cite{Molina2003} showed that an inverted P\"{o}schl-Teller potential can also be used to represent the effective potential of Reissner-Nordstr\"{o}m black holes perturbed by scalar fields. This is due to the fact that for any de Sitter black hole in the near extremal limit, the metric function $f(r)$ is given as \cite{Molina2003}:
\begin{equation}
\label{2.26}
f(r(x)) =  \frac{(r_2 - r_1)\kappa_1}{2 \mathrm{cosh}^2{\kappa_1 x}} + \mathcal{O}(\delta^3),  
\end{equation}

\noindent where $\delta = (r_2 - r_1)/r_1$, $\kappa_1$ is the surface gravity at the horizon, $r_1$ and $r_{2}$ are two consecutive positive roots of $f(r)$,  and $x$ is the tortoise coordinate whose domain lies within ($r_1$, $r_2$). 

\par For both Schwarzschild and Reissner-Nordstr\"{o}m-de Sitter cases, the terms $r_1$ and $r_2$ are the event and cosmological horizons, respectively, with $r_2 > r_1$. In the near extremal limit where $r_2 \sim r_1$, the metric function for a near-extremal Reissner-Nordstr\"{o}m-de Sitter black hole would take the same form as equation (\ref{2.22}). Therefore, when considering the near extremal limit, non-rotating black holes share the same mathematical expression for the metric function, which in turn results in the same expression for the effective potential, equation (\ref{2.23}). From that, we can infer that the corresponding analytic expressions for QNMs of a near extremal Reissner-Nordstr\"{o}m-de Sitter black holes are the same as for Schwarzschild-de Sitter black holes as given by equations (\ref{2.24} - \ref{2.25}). 

\section{Established numerical methods for determining QNMs}\label{SEC: Numerical methods}

\par  The perturbation equations of near extremal non-rotating black holes are among a few known cases with exact QNMs as solutions. As we shall use these equations to measure the accuracy of the PINN approach applied in the context of QNMs. More generally, though, the radial perturbations equations of Schwarzschild and Reissner-Nordstr\"{o}m black hole perturbations are difficult to solve analytically, though not as challenging as cases involving rotating black holes, which require a more arduous investigation. Therefore, approximation techniques have been employed in the past to determine QNMs. We outline here a few prominent techniques used in the literature on black hole QNMs. 

\subsection{Ferrari and Mashhoon approach}

Ref.~\cite{FerrariMashhoon1984} showed the connection between the QNMs of black holes and the bound states of inverted black hole effective potentials. The effective potential, denoted by $U$ in Ref.~\cite{FerrariMashhoon1984} is parametrised by some constant $p$ and is invariant under the transformations $p \rightarrow p^{\prime} = \Pi(p)$ and $x \rightarrow -ix$, as in:
\begin{equation}
\label{3.1}
U(-ix; p^{\prime}) = U(x; p).
\end{equation}

\noindent By considering $x \rightarrow -ix$, the Schr\"{o}dinger-like perturbation equation (\ref{2.4}) transforms to:
\begin{equation}
\label{3.2}
\frac{d^2 \phi}{dx^2} + (-\Omega^2 + U)\phi = 0,
\end{equation}

\noindent where $\phi(x; p) = \psi(-ix; p^{\prime})$ and $\Omega(p) = \omega(p^{\prime})$. The QNM boundary conditions then become:
\begin{equation}
\label{3.3}
\phi \rightarrow \mathrm{exp}(\mp \Omega x),\quad \mathrm{as}\quad  x \rightarrow \pm \infty.
\end{equation}

\noindent In this new form, the problem has become a bound state problem with the original black hole effective potential inverted to $-U$. The transformed boundary conditions, equation (\ref{3.3}), now correspond to vanishing states at both infinities as expected for bound state problems. After solving this problem to find $\Omega$ and $\phi$, the corresponding QNMs can then be found using inverse transformations: 
\begin{equation}
\label{3.4}
\omega(p) = \Omega(\Pi^{-1}(p)), \psi(x;p) = \phi(ix;\Pi^{-1}(p)).
\end{equation}

\noindent The values of $\omega$, that are determined from the bound states $\Omega$, are known as proper QNMs. Ref.~\cite{FerrariMashhoon1984} demonstrated this approach using an inverted P\"{o}schl-Teller potential to approximate the effective potential of a Schwarzschild black hole. The former was used because the bound states of a P\"{o}schl-Teller potential are well-known and could then provide approximate analytic formulas for the QNMs of the Schwarzschild black hole \cite{FerrariMashhoon1984}.

\subsection{WKB Method}

The WKB method is a semi-analytic technique that has been used to approximately solve the radial equation of black hole perturbations since 1985, as first proposed by Schutz and Will \cite{SchutzWill1985}, where they computed the QNMs of an asymptotically flat Schwarzschild black hole. It had already been established as an approximating technique for solving the time-independent Schr\"{o}dinger equation.

\subsection{Continued Fraction Method}

In a 1985 paper \cite{Leaver1985}, Leaver put forward the method of continued fractions (previously used to compute the electronic spectra of the hydrogen molecule ion) to compute the QNM spectra of both stationary and rotating black holes. Overall, this approach was found to be very accurate for higher-order $n$ modes, especially after the improvement made by Nollert \cite{Nollert1993}. It has been used in the context of Schwarzschild, Kerr and Reissner-Nordstr\"{o}m black holes \cite{KokkotasSchmidt1999, Leaver1985, Leaver1990}. 

\subsection{Asymptotic Iteration Method}

The AIM is another semi-analytic technique for solving black hole perturbations. In the context of black hole QNMs, this approach was developed by Ref.~\cite{Choetal2012} who made improvements to a more traditional algorithm to make it markedly more efficient. In Ref.~\cite{Choetal2012} the improved AIM was used to compute of QNMs for cases involving (A)dS, Reissner-Nordstr\"{o}m and Kerr black holes. In later research, it was used to calculate QNMs of general dimensional and non-singular Schwarzschild black holes \cite{Konoplyaetal2019, Daghighetal2020}. Compared to other extant approximation techniques, the improved AIM was shown to be as accurate as Leaver's CFM \cite{Choetal2012}.  

\section{Physics-informed neural networks}\label{SEC: PINNS}
As briefly recapped above, there are several techniques that already exist for solving radial equations in order to obtain the QNMs of black holes. To supplement them, we now introduce PINNs as an alternative to these methods. Firstly, we introduce the idea of deep neural networks and how they can act as universal function approximators. We then introduce PINNs and how they can be used to solve ordinary differential equations (ODEs) and partial differential equations (PDEs). 

\subsection{Deep Neural Networks}\label{SEC: DNN}

\par Deep neural networks are a system of interconnected computational nodes loosely based on biological neural networks and, mathematically, can be formulated as compositional functions \cite{Nordstrom2021, Luetal2021}. In contrast to shallow neural networks, which are networks with just a single hidden layer, these NNs are composed of two or more hidden layers \cite{nielsen2015neural}. In many applications, the latter are favoured because they are capable of replicating the complexity of functions and, at the same time, generalise well to unseen data better than shallow models \cite{Montufar2014}.

\par Of several available types of structures (or architectures) of deep neural networks, the simplest and most common one is the feed-forward neural network (FNN).\newline
\newline
\noindent\textbf{Definition 4.1.} The FNN is comprised of neurons that hold single numerical values (called activations) combine to form a NN $\mathcal{N}^L(\mathbf{x})$ that is a series of $L$ layers with $N_{\ell}$ neurons in the $\ell$-th layer. There are $L - 1$ hidden layers, $N_0$ number of neurons in the input layer ($\ell = 0$) and $N_L$ number of neurons in the output layer ($\ell = L$). The transformations combining the neurons in the ($\ell - 1$)-th layer to those in the $\ell$-th layer are weight matrices and bias vectors $\mathbf{W}^{\ell} \in \mathbb{R}^{N_{\ell}\times N_{\ell - 1}}$ and  $\mathbf{b}^{\ell} \in \mathbb{R}^{N_{\ell}}$, respectively. With these transformations, a FNN is generally structured as follows \cite{Luetal2021}:
\begin{eqnarray}
\text{input layer:}&&\quad \mathcal{N}^0(\mathbf{x}) = \mathbf{x} \in \mathbb{R}^{N_0},\notag\\
\text{hidden layers:}&&\quad \mathcal{N}^{\ell}(\mathbf{x}) = \sigma(\mathbf{W}^{\ell} \mathcal{N}^{\ell - 1}(\mathbf{x}) + \mathbf{b}^{\ell}) \in  \mathbb{R}^{N_{\ell}},\quad \text{for}\quad 1 \leq \ell \leq L-1, \notag\\
\text{output layers:}&&\quad \mathcal{N}^L(\mathbf{x}) =  \sigma(\mathbf{W}^{\ell} \mathcal{N}^{L - 1}(\mathbf{x})  + \mathbf{b}^{L}) \in  \mathbb{R}^{N_{L}},\notag
\end{eqnarray}

\noindent where $\sigma$ denotes nonlinear activation functions that operate on $\mathbf{W}^{\ell} \mathcal{N}^{\ell - 1}(\mathbf{x}) + \mathbf{b}^{\ell}$ element-wise. Examples of frequently used activation functions are the hyperbolic tangent ($\mathrm{tanh}$) and the logistic sigmoid $1/(1 + e^{-x})$. Given that these are nonlinear functions, this makes values at each of the output nodes nonlinear combinations of the values at the nodes in the hidden and input layers \cite{Alundetal2021}.

\par Key seminal research on NNs, such as Refs. \cite{Horniketal1990, Hornik1991, Pinkus1999}, has shown that deep neural networks are universal function approximators. That is to say, when NNs have a sufficient number of neurons they can approximate any function and its partial derivatives \cite{Luetal2021}, though in practice this is constrained by the limit in the size of NNs that can be set up before they lead to overfitting. In such cases, the NN model gives the illusion of a good model that captures the underlying pattern in data, while a true test of its accuracy by means of exposing it to an unseen test dataset reveals a fallible model that gives poor predictions and a high generalisation error \cite{nielsen2015neural, Luetal2021}. In general, training deep NNs entails minimising a loss function that measures the deviation of its approximations from the expected solutions. Analogous to linear least squares regression, the loss function is minimised via tuning of the many parameters in the deep neural network (which are the elements of its weight matrices and bias vectors) with the effect of steering their approximations closer to the target functions.   

\par Mathematically, the weights and biases are tuned according to the equations:
\begin{eqnarray}
\label{4.1}
w^{\ell}_{jk} &\rightarrow& w^{\ell}_{jk} - \frac{\eta}{m} \sum_x \frac{\partial C_x}{\partial w^{\ell}_{jk}},\\
\label{4.2}
b^{\ell}_j &\rightarrow& b^{\ell}_j  - \frac{\eta}{m} \sum_x \frac{\partial C_x}{\partial b^{\ell}_{j}},
\end{eqnarray}

\noindent where $C_x$ is the loss function of the FNN computed for a single training example $x$ that is taken from a minibatch of $m$ training examples, which in turn are taken from a training dataset with $n$ samples. These equations govern stochastic gradient descent optimisation, an algorithm that entails randomly selecting different minibatches from the training dataset of $n$ examples until all of them are exhausted (this constitutes one epoch of training). In equations (\ref{4.1}) and (\ref{4.2}), $\eta$ is a small, positive parameter known as the learning rate. Ultimately, the Adam optimiser is employed in our investigation of PINNs. It is a standard optimisation algorithm that extends from classical methods of stochastic gradient descent \cite{KingmaBa2017, nielsen2015neural}.

\subsection{Physics-informed neural networks}\label{SEC: PINNS2}

Inspired by deep neural networks, PINNs follow the same modus operandi as traditional NNs. Similar to traditional NNs, PINNs are trained through gradient-descent optimisation, whereby the partial derivatives of the loss function (with respect to the network's weight and biases) are minimised by tuning the weights and biases of the FNN. However, the difference is in the constraints that are embedded within the loss function of the PINNs which enable them to solve PDEs. These constraints are the PDEs themselves (or the governing equations) and their associated initial/boundary conditions \cite{Chenetal2021}. 

Autodiff is a technique that is used in PINNs to compute the partial derivatives of the NN approximations and thus embed the governing PDEs and associated boundary conditions in the loss function. Given that it facilitates ``mesh-less'' numerical computations of derivatives, it endows PINNs with several advantages over traditional numerical discretisation approaches for solving PDEs (such as the finite difference and finite element methods) that can be computationally expensive due to complex mesh-generation \cite{Karniadakisetal2021, Nordstrom2021, Luetal2021}. 

For example, Refs. \cite{Nordstrom2021, Luetal2021} demonstrated the advantage of applying NN-aided techniques over using traditional mesh-based techniques to approximate solutions with steep gradients. The latter give rise to unphysical oscillations when the meshes have low resolution, hence higher resolutions are required to remove these undesirable oscillations, which can be prohibitively expensive and lead to excessive execution times \cite{Nordstrom2021}. Remarkably, the same level of accuracy that is achieved by higher resolution meshes (in mesh-based schemes) can be achieved more efficiently in PINNs. In such cases, PINNs could be a viable alternative for solving PDEs.

\par It is worth noting that derivatives of Padé approximations can be utilised as an alternative to PINNs and autodiff, which is a part of the PINN algorithm. They have indeed been applied in extensions of the WKB method in computing black hole QNMs \cite{Konoplyaetal2019}; therefore, the focus has been to compare their performance (and that of other established approaches in black hole QNMs) with the novel PINNs in this physical context.

\begin{figure}[ht!]
\begin{center}
\includegraphics[scale=1.]{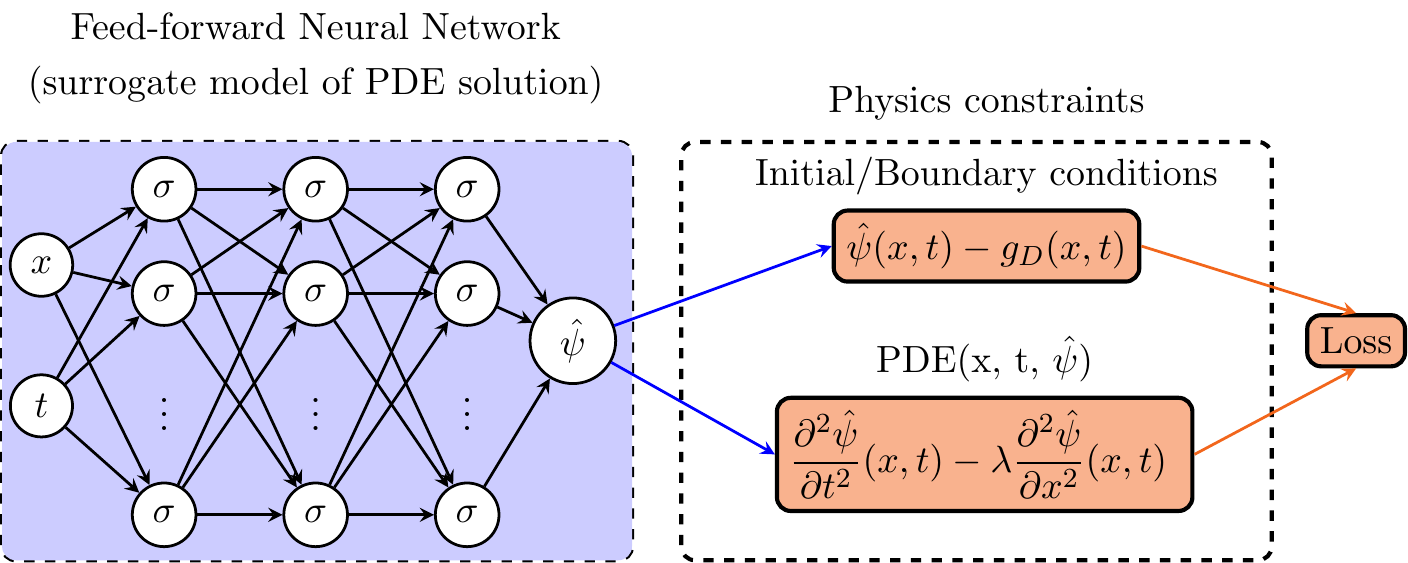}
\caption{\label{FIG: TYPPINN} A schematic of a typical PINN. For the sake of illustration, a well-known hyperbolic type PDE (one-dimensional wave equation), $\partial^2 \psi/\partial t^2 - \lambda \partial^2 \psi/\partial x^2 = 0 $, and initial/boundary conditions $\psi = g_D(x,t)$ are embedded in the loss function.}   
\end{center}
\end{figure}

The basic structure of PINNs can be divided into two components \cite{Luetal2021, Karniadakisetal2021}:

\begin{enumerate}
    \item A deep neural network with a particular architecture, such as a FNN. It represents the NN approximation of the PDE's solution (figure \ref{FIG: TYPPINN} (left)).
   \item A \textit{loss-function} that measures the deviation of the FNN solution from the physical constraints of the problem (figure \ref{FIG: TYPPINN} (right)). The NN learns the solution of the PDE through gradient-based optimisation, an algorithm that minimises the loss function through an iterative tuning of the weights and biases in the deep neural network.
\end{enumerate}

\par In general, to expand on the methodology, PINNs solve PDEs that are parameterised by $\mathbf{\lambda}$, satisfied by a dependent variable\footnote{``Dependent variable'' or ``eigenfunction'' is used in this paper rather than ``solution'', to avoid misunderstading PINNs whose goal may not always be the find $u(\mathbf{x})$ as a solution; but, in some cases, a labelled dataset of the predictors $\mathbf{x}$ and associated $u(\mathbf{x})$ may be used to find unknown parameters in a PDE for inverse problems (for example).} $u(\mathbf{x})$, and are expressed generally as \cite{Luetal2021}:
\begin{equation}
\label{4.3}
f(\mathbf{x}; \frac{\partial u}{\partial x_1}, \ldots, \frac{\partial u}{\partial x_d} ; \frac{\partial^2 u}{\partial x_1 \partial x_1}, \ldots, \frac{\partial^2 u}{\partial x_d \partial x_d} ; \mathbf{\lambda} ) = 0\ \text{on}\ \Omega, 
\end{equation}

\noindent where $\mathbf{x} = (x_1, \ldots, x_d)$ defined on a domain $\Omega \subset \mathbb{R}^d$. Along with a given PDE are its boundary conditions:
\begin{equation}
\label{4.4}
\mathcal{B}(u, \mathbf{x}) = 0\ \text{on}\ \partial\Omega,
\end{equation}

\noindent where $\mathcal{B}(u, \mathbf{x})$ stands for Dirichlet, Neumann Robin or periodic boundary conditions. Note that both steady-state and dynamic systems can be solved using PINNs; where, for the latter, time $t$ is considered to be special component of $\mathbf{x}$ and $\Omega$ contains the time domain. As such, initial conditions are treated as a type of Dirichlet boundary condition on the spatio-temporal domain \cite{Luetal2021}.\newline
\newline
\noindent\textbf{Remark 4.1.} It is worth noting the special nature of PINNs compared to traditional NNs, particularly in the case of classification and regression problems. While many FNNs are typically data-driven and highly dependent on labelled datasets, PINNs, however, are suitable within the scant data regime provided the physical laws governing a system are known \cite{Karniadakisetal2021}. In fact, PINNs are unsupervised and learn from based purely on the PDEs and boundary conditions in the case of forward problems and eigenvalue problems. The goal of forward problems is to find the dependent variable $u(\mathbf{x})$ for every $\mathbf{x}$ provided $\mathbf{\lambda}$ are known parameters. Eigenvalue problems are more challenging because both $u(\mathbf{x})$
and  $\mathbf{\lambda}$ are unknown. In the case when PINNs utilise a labelled dataset these are inverse problems, where the goal is to determine $\mathbf{\lambda}$ given a dataset (which can be small) of $u(\mathbf{x})$ at given points $\mathbf{x} \subset \Omega$.

\subsubsection{The PINN algorithm for solving PDEs}\label{SEC: PINNAlgorithm}

PINNs follow these steps when solving forward, inverse and eigenvalue problems \cite{Luetal2021}:

\begin{enumerate}
\item\textbf{Build a neural network $\hat{u}(\mathbf{x}; \mathbf{\theta})$ with parameters $\mathbf{\theta}$:} 
The neural network $\hat{u}(\mathbf{x}; \mathbf{\theta})$ takes in $\mathbf{x}$ as input and is a surrogate of the function $u(\mathbf{x})$ that satisfies the governing PDE and boundary/initial conditions. Whereas, $\mathbf{\theta}=\{\textbf{W}^{\ell}, \textbf{b}^{\ell}\}_{1 \leq \ell \leq L}$ is a set of all weight matrices and vectors in the neural network \cite{Luetal2021}.
\item\textbf{Specify the training dataset $\mathcal{T}$:}
In the case of forward and eigenvalue problems, we specify a dataset of ``unlabelled'' randomly distributed points in the domain (also known as residual points). The points within the $\mathcal{T}_f \subset \Omega$ are used to restrict the NN approximation $\hat{u}(\mathbf{x}; \mathbf{\theta})$ to satisfy the physics imposed by the PDE. Similarly, the boundary points of the spatio-temporal domain $\mathcal{T}_b \subset \partial\Omega$ are used to restrict the NN to satisfy the physics represented by the initial/boundary conditions. For inverse problems, since $\mathbf{\lambda}$ is missing from the PDE, a ``labelled'' dataset of $u(\mathbf{x})$, denoted by $\mathcal{T}_o$, is required in addition.
\item\textbf{Specify a loss function by adding the weighted Euclidean norm of the PDE, boundary conditions, and other regularisation functions:} In general, the loss function of PINNs will be given as \cite{Luetal2021}:
\begin{equation}
\label{4.5}
\mathcal{L}(\mathbf{\theta}; \mathcal{T}) =  w_f\mathcal{L}_f(\mathbf{\theta}; \mathcal{T}_f) + w_b\mathcal{L}_b(\mathbf{\theta}; \mathcal{T}_b) + w_r\mathcal{L}_r(\mathbf{\theta}; \mathcal{T}_r),  
\end{equation}
where $w_f, w_b, w_r$ are weights and still $\mathbf{\theta} = \{\textbf{W}^{\ell}, \textbf{b}^{\ell}\}_{1 \leq \ell \leq L}$, with $\ell$ specifying a hidden layer as defined in section \ref{SEC: DNN} \cite{Luetal2020}. Additionally $\mathcal{L}_f$ and $\mathcal{L}_b$ are loss terms due to the PDE and initial/boundary conditions, respectively:
\begin{eqnarray}
\label{4.6}
\mathcal{L}_f(\mathbf{\theta}; \mathcal{T}_f) &=& \frac{1}{|\mathcal{T}_f|} \sum_{\mathbf{x} \in \mathcal{T}_f} \left\lVert f(\mathbf{x}; \frac{\partial \hat{u}}{\partial x_1}, \ldots, \frac{\partial \hat{u}}{\partial x_d} ; \frac{\partial^2 \hat{u}}{\partial x_1 \partial x_1}, \ldots, \frac{\partial^2 \hat{u}}{\partial x_d \partial x_d} ; \hat{\mathbf{\lambda}} ) \right\rVert^2_2,\\
\label{4.7}
\mathcal{L}_b(\mathbf{\theta}; \mathcal{T}_b) &=& \frac{1}{|\mathcal{T}_b|} \sum_{\mathbf{x} \in \mathcal{T}_b} \left\lVert \mathcal{B}(\hat{u}, \mathbf{x})\right\lVert^2_2,
\end{eqnarray}
where the circumflex in $\hat{u}$ and $\hat{\mathbf{\lambda}}$ denotes that these are the NN's approximations of the dependent variable and any unknown PDE paramters of inverse problems. The loss term $\mathcal{L}_r$ represents regularisation functions in general. For example, for forward problems this term is left out while for inverse problems it is the error between the NN approximations and a ``labelled dataset'' of $u(\mathbf{x})$:
\begin{equation}
\label{4.8}
\mathcal{L}_f(\mathbf{\theta}; \mathcal{T}_f) = \frac{1}{|\mathcal{T}_r|} \sum_{\mathbf{x} \in \mathcal{T}_r} \left\lVert u(\mathbf{x}) - \hat{u}(\mathbf{x}) \right\rVert^2_2.    
\end{equation}
\item\textbf{Train the FNN towards the optimal weights and biases $\mathbf{\theta}^{*}$ by minimising the loss function $\mathcal{L}(\mathbf{\theta}; \mathcal{T})$}: The goal of training is to optimise $\mathbf{\theta}$, $\hat{u}$ and $\hat{\mathbf{\lambda}}$ such that we have:
\begin{equation}
\label{4.9}
\mathbf{\theta}^*, \hat{u}^*, \hat{\mathbf{\lambda}}^* = \mathrm{argmin}_{\theta, u, \mathbf{\lambda}} \mathcal{L} (\mathbf{\theta}, \hat{u}, \hat{\mathbf{\lambda}}; \mathcal{T})   
\end{equation}
Note that the loss function is highly nonlinear and nonconvex with respect to $\mathbf{\theta}$, thus gradient-descent optimisers such as Adam are often used during training. The disadvantage of a nonconvex optimisation problems is the difficulty to find unique solutions compared to traditional numerical methods of solving PDEs \cite{Luetal2021}.
\end{enumerate}

\noindent\textbf{Remark 4.2.} Two important differences between PINNs and typical NNs are worth noting. Firstly, the former has the approximate function $\hat{u}(\mathbf{x})$ bound by the domain $\Omega$ where the governing PDE is defined. Secondly, PINNs learn from their own predictions (which is to say the governing PDEs and initial/boundary conditions are sufficient to optimise $\hat{u}(\mathbf{x},\mathbf{})$ with respect to $\mathrm{\theta}$. The ``unlabelled'' dataset of points randomly selected from $\Omega$ are split into training and validation/test sets, which is not done in some variations of PINNs such as the eigenvalue solvers (see section \ref{SEC: Eigenvalue Solver}) since the FNN hyperparamters are fixed. In the case of inverse problems, when a dataset of the true values of $u(\mathbf{x})$ is available, part of that dataset can be used for validating the approximate $\hat{u}(\mathbf{x})$ by computing the $L_2$ relative error, one example of a test metric.

In the following, we discuss two examples of Python libraries which have been employed to construct PINNs; namely, DeepXDE \cite{Luetal2021} and Pytorch \cite{Jinetal2020}. 

\subsubsection{The \textsc{DeepXDE} package}\label{SEC: DeepXDE}

The DeepXDE package is customised primarily for constructing PINN models. To help elaborate on the DeepXDE package, we consider here a toy problem that was discussed in Ref.~\cite{Choetal2012}, which involves the same Schr\"{o}dinger-like differential equation in equation (\ref{2.4}) but with an inverted symmetric P\"{o}schl-Teller potential $V_{PT}(x)$ \cite{Choetal2012}:
\begin{equation}
\label{4.10}
V_{PT}(x) = \frac{1}{2\mathrm{cosh}^2(x)}.
\end{equation}

\noindent In the tortoise co-ordinate $x$, the domain of our problem is infinite, i.e. $x \in (-\infty, +\infty)$, where the QNM boundary conditions are given by equation (\ref{2.7}). Via quasi-exactly solvable theory, Ref.~\cite{ChoHo2007} found the exact solutions of equation (\ref{2.4}) with $V = V_{PT}$ to be given as \cite{Choetal2012}:
\begin{equation}
\label{4.11}
\psi_n(x) = (\mathrm{cosh}(x))^{(i + 1)/2}\chi_{\textsubscript{$n$}}(\mathrm{sinh}(x)), 
\end{equation}
\begin{equation}
\label{4.12}
\omega_{\textsubscript{$n$}}= \pm \frac{1}{2} - i(n + \frac{1}{2}),
\end{equation}

\noindent where $\chi_{\textsubscript{$n$}}$ is a polynomial of degree $n$ in $\mathrm{sinh}(x)$ and $n = \mathbb{Z}^+_0$ . 

As a first step to finding the approximate solutions using PINNs, we need to change to a new coordinate $y = \mathrm{tanh(x)}$, which maps the infinite domain $-\infty < x < +\infty$ to a finite domain of $-1 < y < +1$, so that equation (\ref{2.4}) becomes \cite{Choetal2012}: 
\begin{equation}
\label{4.13}
(1 - y^2)^2 \frac{d^2\psi(y)}{dy^2} - 2 y (1 - y^2) \frac{d\psi(y)}{dy} + \left[ \omega^2 - \frac{1}{2} (1 - y^2)\right]\psi(y) = 0.
\end{equation}

In this form, numerical implementation of this problem in PINNs becomes possible. We test the feasibility of solving equation (\ref{4.13}) given as an inverse problem using DeepXDE. We specify $\omega$ as an unknown to be tuned while the PINN undergoes training. The total loss function $\mathcal{L}(\theta;\mathcal{T})$ of the PINN, in this case, is a weighted sum of the squared Euclidean ($L^2$-) norm of the physical constraints, similar to equation (\ref{4.5}) \cite{Luetal2021}:
\begin{equation}
\label{4.14}
\mathcal{L}(\theta;\mathcal{T}) =  w_f\mathcal{L}_f(\theta;\mathcal{T}_f) + w_b\mathcal{L}_b(\theta; \mathcal{T}_b) + w_o\mathcal{L}_o(\theta;\mathcal{T}_o), 
\end{equation}
\noindent where:
\begin{eqnarray}
\label{4.15}
\mathcal{L}_f(\theta;\mathcal{T}_f) &=& \frac{1}{|\mathcal{T}_f|} \sum_{y \in \mathcal{T}_f} \left\lVert (1 - y^2)^2 \hat{\psi}^{\prime\prime} - 2 y (1 - y^2) \hat{\psi}^{\prime} + \left[\hat{\omega}^2 - \frac{(1 - y^2)}{2} \right]\hat{\psi} \right\rVert^2_2,\\
\label{4.16}
\mathcal{L}_b(\theta; \mathcal{T}_b) &=& \frac{1}{2} \sum_{y \in \mathcal{T}_b} \left\lVert \hat{\psi}(y) - \psi_{b}(y)   \right\lVert^2_2,\ 
\mathcal{L}_o(\theta; \mathcal{T}_o) =  \frac{1}{|\mathcal{T}_o|} \sum_{y \in \mathcal{T}_o} \left\lVert \hat{\psi}(y) - \psi(y)   \right\lVert^2_2.
\end{eqnarray}

\noindent Note that $w_f, w_b, w_o$ are weights that are typically set to one and $\mathbf{\theta}$ is as defined as in section \ref{SEC: PINNAlgorithm}. $\mathcal{T} = \{y_1, y_2, ..., y_{|\mathcal{T}|} \}$ is a set which consists of all training points randomly selected from our 1D spatial domain ($-1 < y < 1$). The subset $\mathcal{T}_f$ are points chosen from the domain to train the FNN based on the governing equation (\ref{4.13}). The subsets $\mathcal{T}_b (= \{-1,1\}), \mathcal{T}_o$ are the boundary points for training on the boundary conditions, and the dataset of the true values of the dependent variable (\ref{4.16}), respectively. Training of this PINN proceeds as outlined in section \ref{SEC: PINNAlgorithm}.

\begin{figure}[t!]
\begin{center}
\includegraphics[scale=1.]{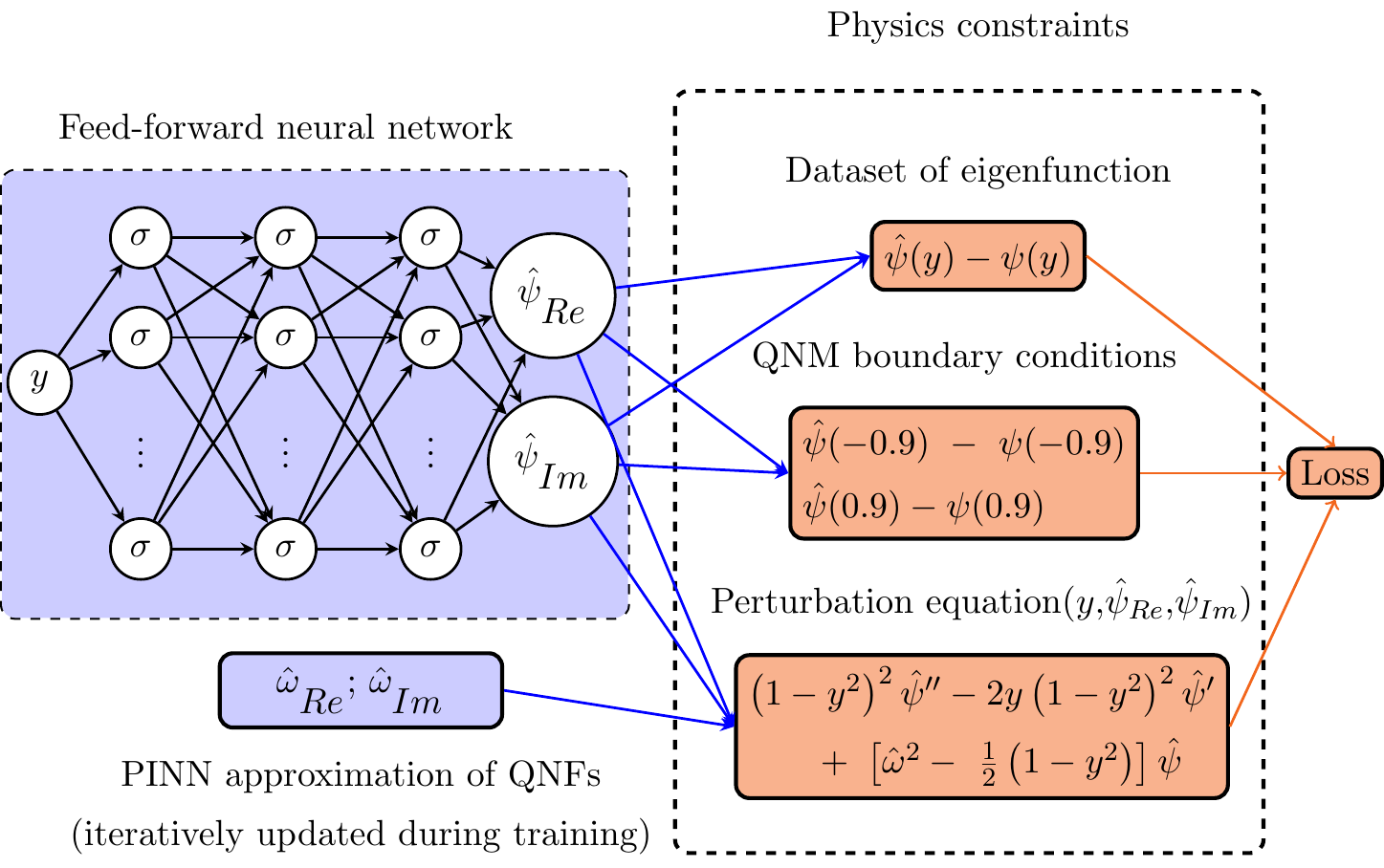}
\caption{\label{FIG: Poschl-TellerPINN} The structure of PINN for solving equation (\ref{4.6}).}  
\end{center}
\end{figure}

Figure \ref{FIG: Poschl-TellerPINN} illustrates the PINN for solving this problem. The input layer of the FNN consists of one input for co-ordinate $y$, while the output layer has two output nodes for real and imaginary parts of the approximate solution $\hat{\psi}$. 

In building PINNs, the code

we used mirrors the two-component structure of PINNs discussed in section \ref{SEC: PINNS2}. The code is fairly intuitive as it is a high-level representation that closely resembles the mathematical formulation \cite{Luetal2021}. Beginning with the physics constraints, our ODE is defined using the DeepXDE functions for executing the first and second-order derivatives via auto diff; that is, \texttt{dde.grad.jacobian} and \texttt{dde.grad.hessian}, respectively. We define $\omega$ with the function \texttt{tf.Variable} and have represented it with $\hat{\omega}_{Re}$ and  $\hat{\omega}_{Im}$ in figure \ref{FIG: Poschl-TellerPINN}.

To provide Dirichlet boundary conditions and a labelled dataset, as needed to solve our inverse problem, we define both the real and imaginary parts of the known eigenfunction $\psi(y)$ that satisfies equation (\ref{4.13}). Numerically, at the true boundary points, $y=-1$ and $y=1$, the solution $\psi(y)$ yields a complex-infinity. As such, a narrower domain $-0.9 < y <0.9$ is specified in the definition of the domain of our problem using the function \texttt{dde.geometry.Interval(-0.9, 0.9)}. The exact values of $\psi$ at these artificial boundary points are considered to be the Dirichlet boundary conditions. The DeepXDE function for defining these boundary conditions is \texttt{dde.DirichletBC}. To create a labelled dataset to train our PINN, we generate 50 equidistant points in the domain $(-0.9,0.9)$ and their associated exact eigenfunctions using equation (\ref{4.11}). This dataset is the set $\mathcal{T}_o$ in equation (\ref{4.16}).  

At this stage, we have defined the physics constraints of the PINN, but for completeness, we set up the deep neural network (our surrogate model). In the code we also define a FNN with one input node, two output nodes and three hidden layers with 20 nodes per layer. In each of the hidden layer nodes, we use the nonlinear activation function ``tanh'' considering that it is a smooth, infinitely differentiable function \cite{Salvatoretal2022}. Generally for PINNs, ``smooth'' activation functions are preferred over the ReLU-like non-smooth activation functions since the former have demonstrated significant empirical success \cite{Gnanasambandam2022}. For this reason, the tanh function is chosen here by default, however it is worth noting that (of late) adjustable, smooth function such as Swish have proven to outperform fixed functions such as tanh in terms of convergence rate and accuracy \cite{CChengetal2021, Salvatoretal2022}. Swish is defined by $x\cdot$ Sigmoid($\beta x$), where $\beta$ is a trainable parameter.   

\begin{figure}[t!]
\begin{minipage}{18pc}
\includegraphics[width=18pc]{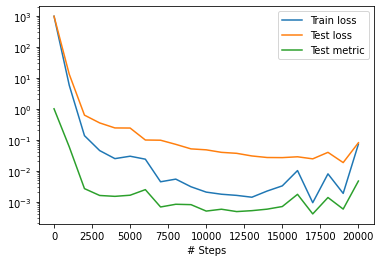}
\end{minipage}\hspace{0pc}%
\begin{minipage}{18pc}
\includegraphics[width=18pc]{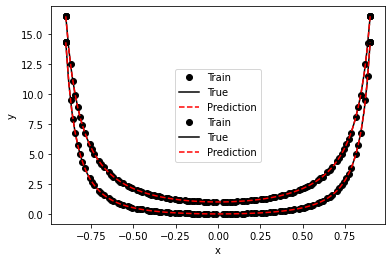}
\end{minipage}
\caption{\label{FIG: DeepXDE} Plot of the train loss, test loss and $L_2$ relativer error (test metric) over a period of 20~000 epochs (left panel); Plot of the NN prediction of $\psi(y)$ superimposed with the true function and training points selected from the domain, $-0.9 < y < 0.9$ (right panel).}
\end{figure}

\par The loss function \texttt{dde.Model} combines the FNN with the physical constraints to form a complete PINN. We also add the ``callback'' function \texttt{dde.callback} in the algorithm so as to keep track of the FNN approximations of $\omega$ during training. Finally, our PINN model is compiled and trained. Compilation defines the learning rate and algorithm for optimising our model. For training the model, we choose 20~000 training epochs wherein the model will be iteratively tuned based on the physics constraints. Figure \ref{FIG: DeepXDE} displays the evolution of the loss function and model accuracy over 20~000 epochs and compares the NN approximation of $\psi(y)$ with the exact function. The PINN algorithm in DeepXDE illustrated here works well for inverse problems where $\psi$ is known at some points in the domain. However, for more general scenarios of black hole QNMs, where both $\omega$ and $\psi$ are unknown, we require an algorithm capable of solving eigenvalue problems.

\subsubsection{The eigenvalue solver}\label{SEC: Eigenvalue Solver}

One such algorithm that we have investigated was initiated in Ref.~\cite{Jinetal2020} to solve quantum eigenvalue problems using unsupervised NNs (also called, data-free surrogate models). The authors experimented with their ``eigenvalue solvers'' on well-known equations in quantum mechanics; namely, the time-independent Schr\"{o}dinger equation with an infinite square well potential and, in another case, a quadratic potential function of a quantum harmonic oscillator. Although their approach is similar to the PINNs, in terms of embedding learning biases in the loss function, there is an additional feature which allows the eigenvalue solver to scan the eigenvalue space in a scheduled manner and progressively find several eigenvalues in a single training.

To help visualise this approach, we consider one well-known bound-state eigenvalue problem \cite{PoschlTeller1933}:
\begin{equation}
\label{4.17}
-\frac{1}{2}\psi^{\prime\prime}(x) + V(x)\psi(x) = E\psi(x),    
\end{equation}

\noindent where:
\begin{equation}
\label{4.18}
V(x) = -\frac{\lambda(\lambda + 1)}{2}\mathrm{sech}^2(x),    
\end{equation}

\noindent which is a P\"{o}schl-Teller potential and $\lambda = 1, 2...$\ . We can now change to a new co-ordinate $u = \mathrm{tanh}(x)$. As such, equation (\ref{4.17}) can be written in the form of a Legendre differential equation:
\begin{equation}
\label{4.19}
[(1 - u^2)\psi^{\prime}(u)]^{\prime} + \lambda(\lambda + 1)\psi(u) + \frac{2E}{1 - u^2}\psi(u) = 0,
\end{equation}

\noindent which is solved exactly by associated Legendre functions, i.e. $\psi(x) = P^{\mu}_{\lambda}(\mathrm{tanh}(x))$ with $E = -\mu^2/2$ and $\mu= 1, 2, 3..., \lambda$. These are bound states that vanish at the boundaries of the eigenvalue problem, i.e. $\psi(x=\pm \infty) = 0$ or $\psi(u=\pm 1) =0$.

The eigenvalue solvers in Ref.~\cite{Jinetal2020} are built using the PyTorch library. To solve equation (\ref{4.19}) using the eigenvalue solvers, we embed them in the loss function of the NN along with some regularisation terms, similar to equation (\ref{4.5}):
\begin{equation}
\label{4.20}
\mathcal{L}(\theta;\mathcal{T}) = \mathcal{L}_{ODE}(\theta;\mathcal{T}) + \mathcal{L}_{reg}(\theta;\mathcal{T}),  
\end{equation}

\noindent where: 
\begin{eqnarray}
\label{4.21}
\mathcal{L}_{ODE}(\theta;\mathcal{T}) &=& \frac{1}{|\mathcal{T}|} \sum_{u \in \mathcal{T}} \left[ ((1 - u^2)\hat{\psi}^{\prime}(u))^{\prime} + \lambda(\lambda + 1)\hat{\psi}(u) + \frac{2\hat{E}}{1 - u^2}\hat{\psi}(u)\right]^2,\\
\label{4.22}
\mathcal{L}_{reg}(\theta;\mathcal{T}) &=& w_f\mathcal{L}_f(\theta;\mathcal{T}) + w_E\mathcal{L}_E(\theta;\mathcal{T}) +
w_{drive}\mathcal{L}_{drive}(\theta;\mathcal{T}).
\end{eqnarray}

\begin{figure}[t!]
\begin{center}
\includegraphics[scale=1.]{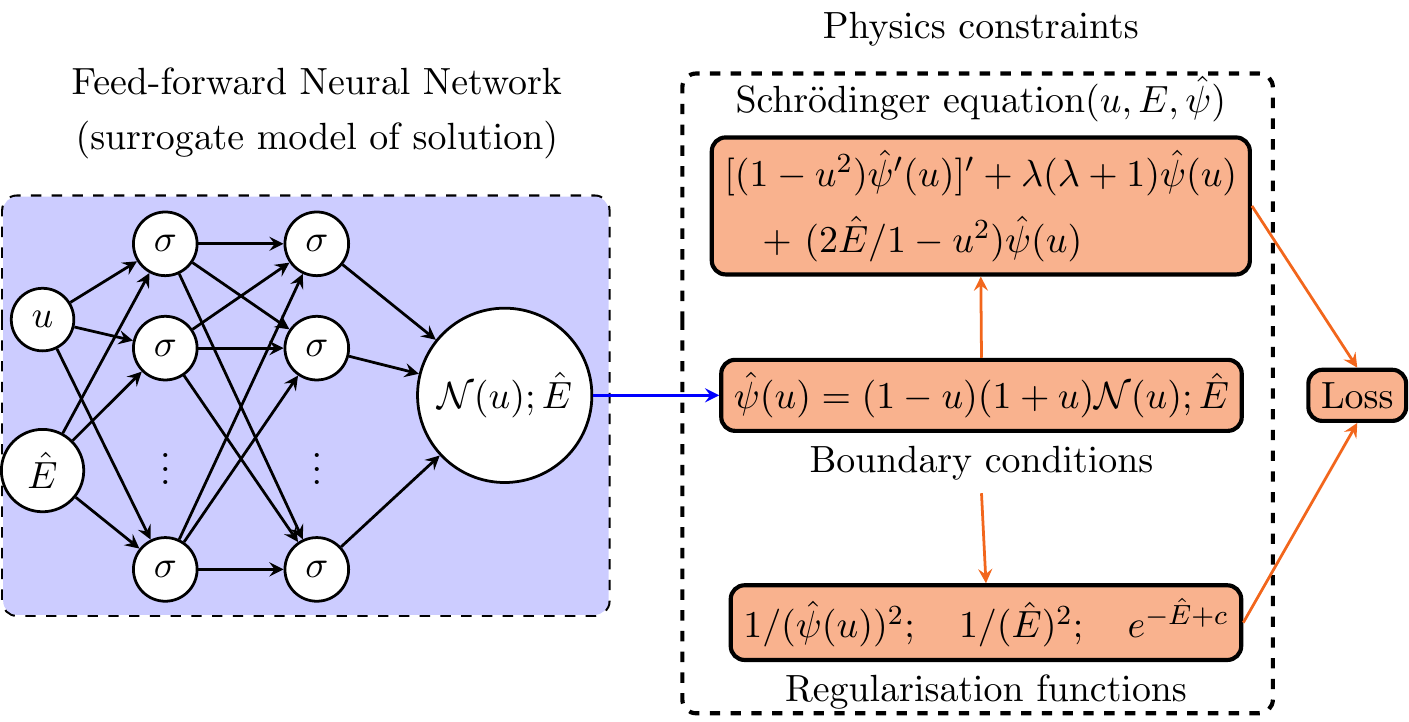}
\caption{\label{FIG: Schrodinger} The structure of the eigenvalue solver. Unlike the PINNs used for inverse problems} in DeepXDE, these FNNs are unsupervised. Instead, the unknown eigenpairs can be determined only from the governing equations and boundary conditions (that are enforced using a parametric function $(1 - u)(1 + u)$, a ``hard constraint'', to ensure they are satisfied exactly \cite{Luetal2021}). 
\end{center}
\end{figure}

\noindent As defined in section \ref{SEC: PINNAlgorithm}, $\mathcal{T}$ is a set of training points randomly selected from the domain $u \in (-1, 1)$. Figure \ref{FIG: Schrodinger} illustrates how the boundary conditions (i.e. a vanishing solution at the boundary points) are enforced using a parametric function $(1 - u)(1 + u)$. Note also the absence of the observational bias term (the reason our eigenvalue solver is called a data-free model). 

In equation (\ref{4.22}) $\mathcal{L}_{reg}$ is a weighted sum of regularisation functions, where the weights $w_f, w_E, w_{drive}$ are typically set to one \cite{Jinetal2020}. Individually, the regularisation functions are:
\begin{equation}
\label{4.23}
\mathcal{L}_f = \frac{1}{|\mathcal{T}|} \sum_{u \in \mathcal{T}} \frac{1}{\hat{\psi}^2},\quad  \mathcal{L}_{E} = \frac{1}{|\mathcal{T}|} \sum_{u \in \mathcal{T}} \frac{1}{\hat{E}^2},\quad  \mathcal{L}_{drive} = \frac{1}{|\mathcal{T}|} \sum_{u \in \mathcal{T}} \mathrm{exp}(-\hat{E} + c), 
\end{equation}

\noindent where $\mathcal{L}_f$ and $\mathcal{L}_{E}$ steer the learning algorithm away from zero as a possible value for the eigenfunction and eigenvalue, respectively. For this purpose, the mathematical form of these loss terms have the PINN approximations ($\hat{\psi}$ and $\hat{E}$) inversely proportional to the loss so that as they approach zero they are penalised by high loss values. The crucial term in these unsupervised NNs is $\mathcal{L}_{drive}$, which motivates the NN to scan through the space of eigenvalues. This is achieved by adding within the training algorithm a mechanism that increases the constant $c$ in $\mathcal{L}_{drive}$ at regular intervals, after an arbitrary number of training epochs. 

It is important to note that without the $\mathcal{L}_{drive}$ loss component the PINNs lack the necessary constraint to learn other eigenvalues than the first energy level it initially gravitates towards during training, which is often but not always the ground energy level. In this case, the algorithm has more limitations, similar to the original, baseline PINN loss function (equation \ref{4.5}) where forward and inverse approaches can learn, respectively, only one of the eigenfunctions and eigenvalues at a time, and only when at least one of the other eigenpairs is known. As a consequence, a classification approach (with, for example, output nodes of the PINN representing the dependent variable) may not be applicable because the loss function will only have the wherewithal to learn a single eigenstate in each training, regardless of the input data since it is unlabelled and is randomly selected from a domain where one eigenstate cannot be separated spatially from the other solutions.

The key Pytorch functions used in defining our physics constraints include\newline \texttt{torch.autograd}, which executes automatic differentiation to find the first and second derivatives in $\mathcal{L}_{ODE}$ given by equation (\ref{4.21}). With the physics constraints defined, we set the structure of our FNN: 2 input nodes, 1 output node and 2 hidden layers with 50 nodes each (see figure \ref{FIG: Schrodinger}), where our chosen activation function is the trigonometric function, $\mathrm{sine}$. This activation function has been found to greatly accelerate the NN's convergence to eigenstates compared to more common functions, e.g. sigmoid and ReLU \cite{Jinetal2020, Mattheakisetal2022}.

\begin{figure}[t!]
\begin{minipage}{17pc}
\includegraphics[width=17pc]{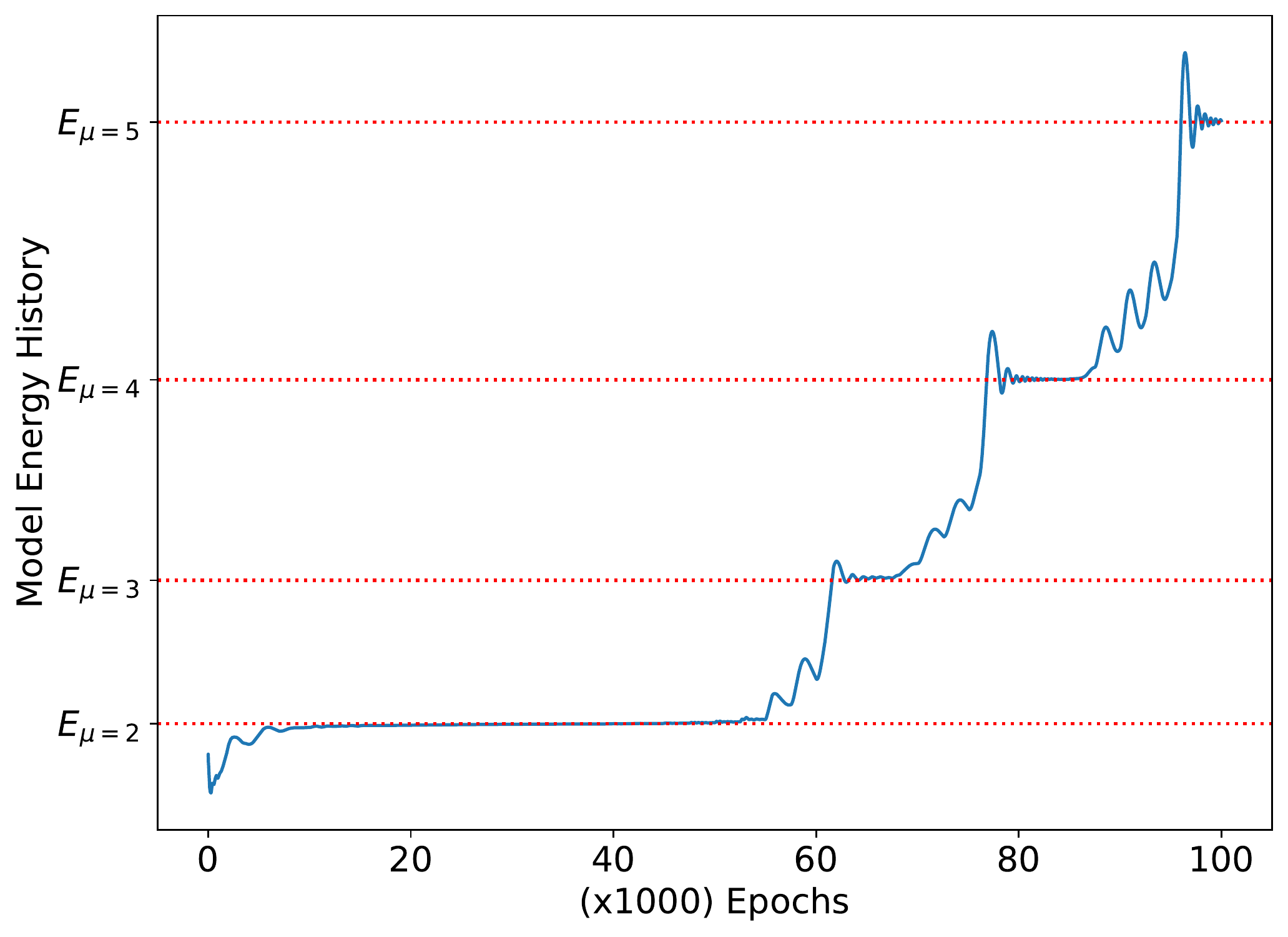}
\end{minipage}\hspace{0pc}%
\begin{minipage}{17pc}
\includegraphics[width=17pc]{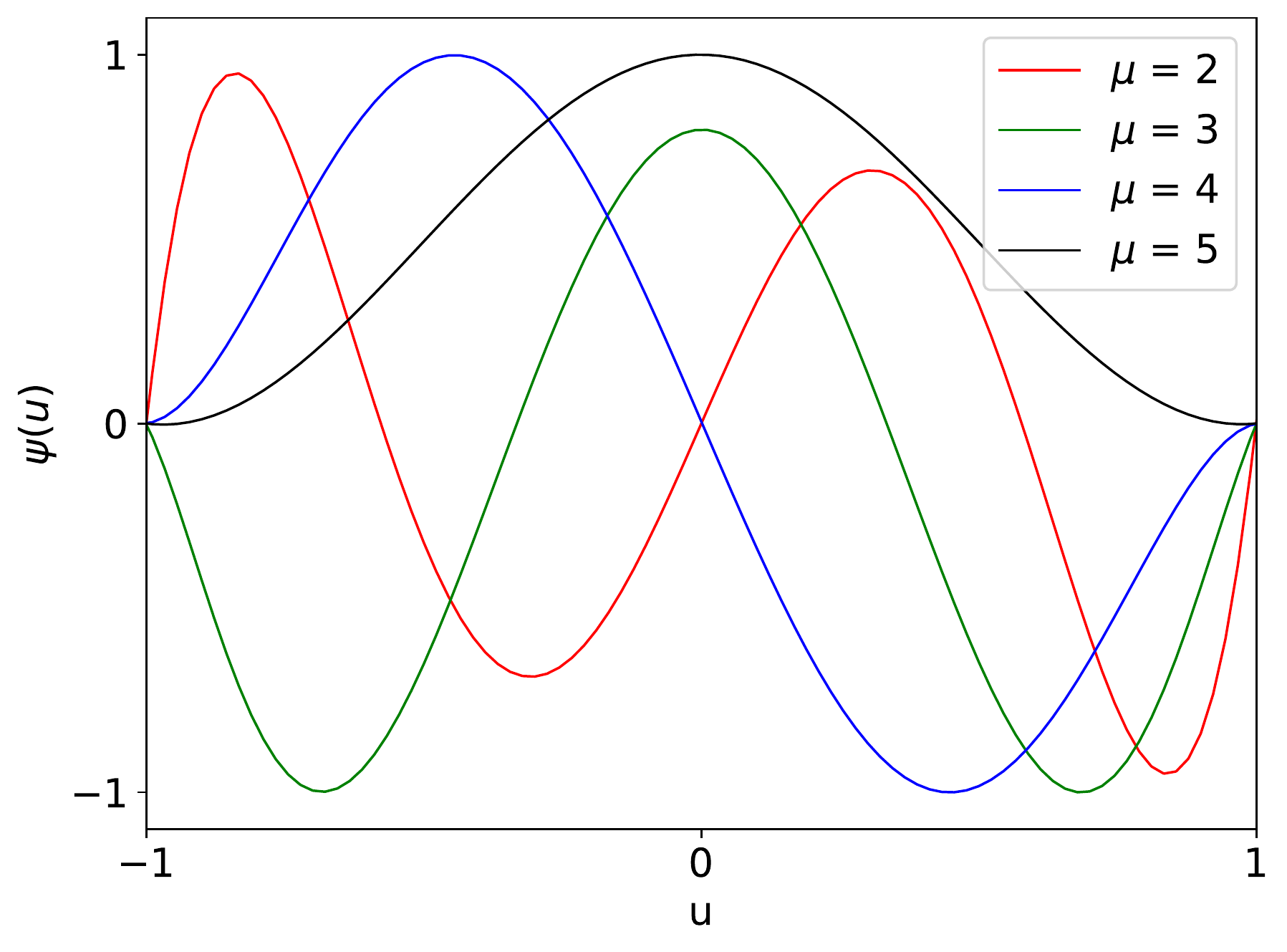}
\end{minipage}
\caption{\label{FIG: Approximations} Plot of the NN approximations of the $\psi$ and $-E$, the P\"{o}schl-Teller bound states. Notice that the plot of $-E$ plateaued whenever the NN found an eigenvalue given by $-E =\mu^2/2$ for $\mu = 1, 2, 3,...$ As expected, the bound states $\psi(u)$ vanish at the boundary points.}
\end{figure}

\par Compared to the code in DeepXDE, the eigenvalue solvers provided more flexibility when customising the training algorithm. The total loss function in our training algorithm was defined according to equations (\ref{4.20} - \ref{4.23}). To generate \texttt{n\_train} points from the domain of our example problem $u \in (-1 , 1)$,  we used the Pytorch function \texttt{torch.linspace}. In terms of optimisation, the standard Adam optimiser is applied \cite{KingmaBa2017}.

\par Ultimately, the training phase follows after all parameters for training the model (such as the number of training epochs) have been defined. In our case, we chose the following parameters: 100 training points, 100~000 training epochs and a learning rate of $8\times10^{-3}$. Figure \ref{FIG: Approximations} shows the resulting NN approximations of the eigenvalues and eigenfunctions. Note that the $\mathcal{L}_{drive}$ is only included in the loss function of this example, for complete demonstration of the method. However, it is not applied in the QNM computations resulting in the PINNs converging on one eigenvalue (as we will see), rather than several (as in the many plateaus of figure \ref{FIG: Approximations}). As seen in the example, the flips between eigenvalues occur arbitrarily, without any method of controlling when they occur. Therefore, this loss term requires further investigation, outside this present work, to make it less random. 
    	
\section{Results: QNM computations with the eigenvalue solver} \label{SEC: Results}

The results from our investigation of the performance of PINNs when applied to the computation of QNMs shall now be presented, where it is important to note that, generally for deep neural networks, there are no set rules for customising them since they are statistical tools with too many parameters to admit any meaningful physical interpretability. Taking this into account in this work, we have carried out grid-search-like experimentation of the NN hyperparameters to discover the most optimal choices with the best performance. Specifically, we considered a range of values for three hyperparameters, which are the number of training points, number of training epochs, and the number of nodes per layer, keeping the other hyperparameters (e.g. optimiser and activation function) fixed. In this work, we have focussed on computing the QNMs of a Schwarzschild black hole in the asymptotically flat and near extremal de Sitter cases. For the former, we considered massless scalar, Dirac, electromagnetic and gravitational field perturbations; while, for the latter, we only considered massless scalar fields where the equations look the same as near extremal Reissner-Nordstr\"{o}m-de Sitter black holes. Due to this, the QNMs of near extremal Schwarzschild-de Sitter black holes can be more generally treated as the QNMs of near extremal non-rotating de Sitter black holes. 

\subsection{Scanning hyperparameters}
\label{SEC: QNMs1}

\begin{figure}[t!]
\begin{minipage}{13pc}
\includegraphics[width=1\linewidth]{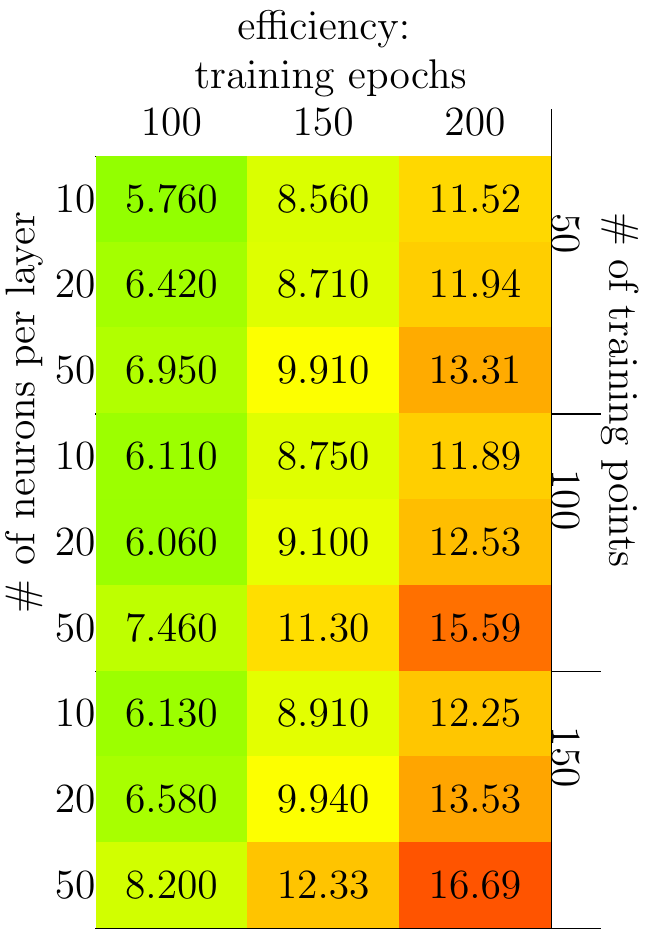}
\end{minipage}
\begin{minipage}{22pc}
\includegraphics[width=1\linewidth]{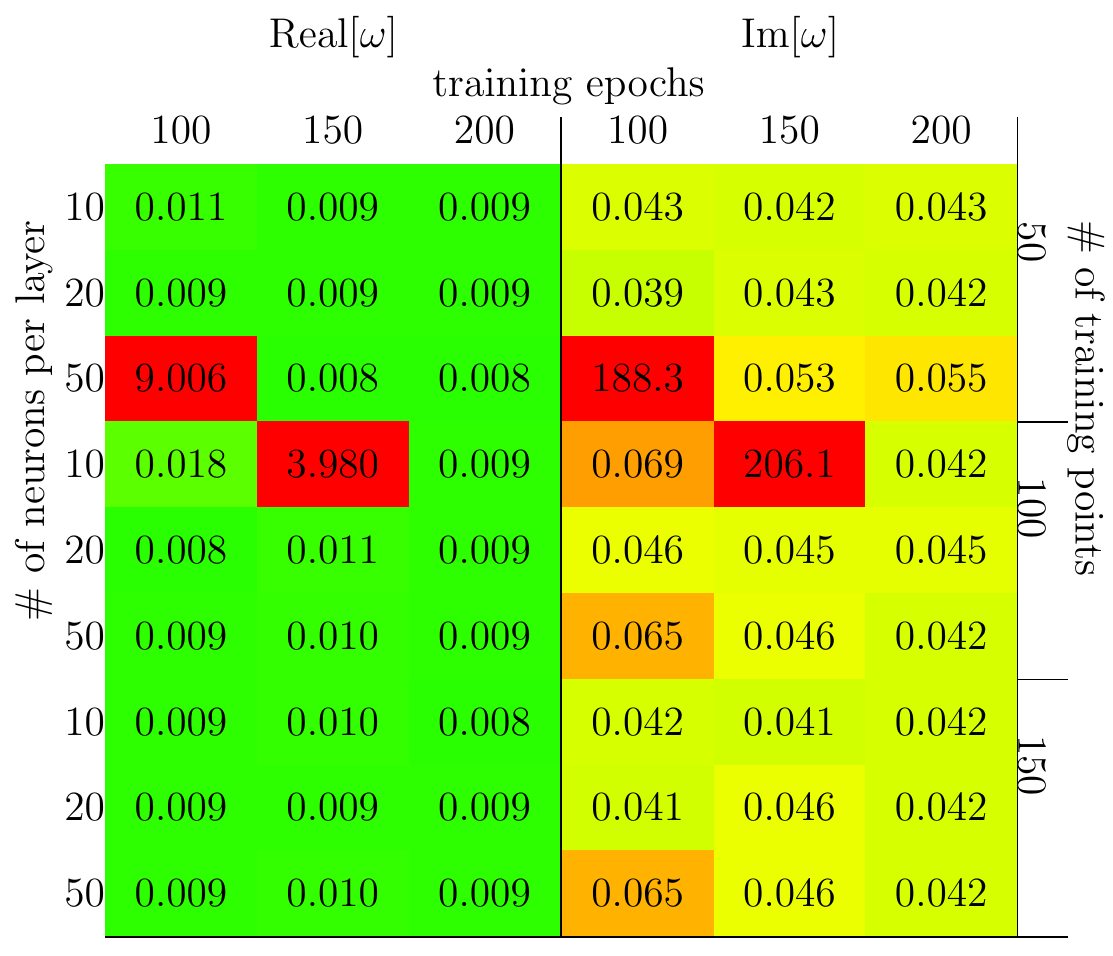}
\end{minipage}

\caption{\label{FIG: HYPERP} The training times, in minutes, (right panel) and percentage deviations (left panel) obtained for different hyperparameter choices. To compute the QNMs of asymptotically flat Schwarzschild BHs ($s=0,\ \ell=2,\ n=0$), we tested different permutations of the number of training points, number of neurons per layer and number of training epochs ($\times1000$).}
\end{figure}

Figure \ref{FIG: HYPERP} graphs the results we obtained from testing different hyperparameter configurations for computing the QNMs of an asymptotically flat Schwarzschild black hole ($s=0,\ \ell=2,\ n=0$). The accuracy of the NN approximations (measured in terms of percentage deviation) and the execution times for training our NNs have been measured as a function of the number of training points, number of training epochs, and number of nodes per layer. The fixed hyperparameters were: learning rate of $8\times10^{-3}$, $2$ hidden layers, and $\mathrm{sine}$ as the activation function.

Note that the accuracy values measure the deviation of the NN approximations from Leaver's QNMs, whose precision is up to 4 decimal places \cite{Leaver1985, Iyer1987}. As seen in figure \ref{FIG: HYPERP}, the percentage deviations of our computations remain the same across all hyperparameter configurations. But for a few cases, the percentage deviations for the real and imaginary parts of the QNMs hover around about $0.009\%$ and $-0.042\%$, respectively. Both these values correspond to a 4 decimal place precision, making the NN approximations as good as Leaver's CFM. Note that beyond 4 decimal places we cannot reliably determine the accuracy of our NN approximations based on the QNMs given in the literature \cite{Leaver1985, Iyer1987}.

\par The red cells given in the right panel of figure \ref{FIG: HYPERP} correspond to cases where the eigenvalue solvers veer from determining the QNMs with a minimum loss, which are the $n = 0$ modes. These are few in comparison to ``normal'' cases where the eigenvalue solvers converge to a loss minimising solution. The displayed training times and percentage deviations were obtained by iterating the eigenvalue solver algorithm automatically and scanning through the specified range of hyperparameter combinations. Note that the total loss was set as:
\begin{equation}
\label{5.1}
\mathcal{L}(\theta;\mathcal{T}) = \mathcal{L}_{ODE}(\theta;\mathcal{T}) + \mathcal{L}_{f}(\theta;\mathcal{T}),  
\end{equation}

\noindent where: 
\begin{eqnarray}
\label{5.2}
\mathcal{L}_{ODE}(\theta;\mathcal{T}) &=& \frac{1}{|\mathcal{T}|} \sum_{\xi \in \mathcal{T}} \left[ \chi^{\prime\prime} -  \lambda_{0}(\xi)\chi^{\prime} -  s_{0}(\xi)\chi\right]^2,\\
\label{5.3}
\mathcal{L}_{f}(\theta;\mathcal{T}) &=& \frac{1}{|\mathcal{T}|} \sum_{\xi \in \mathcal{T}} \frac{1}{\hat{\chi}^2}.
\end{eqnarray}

\noindent Here $\theta, \mathcal{T}, $$\hat{\chi}$ and $\hat{\omega}$ have their definitions from sections \ref{SEC: Schwarzschild} and \ref{SEC: PINNAlgorithm}. We have considered the ODE given by Ref.~\cite{Choetal2012}, which is a transformation of the radial perturbation equation (\ref{2.4}) to a finite domain of the coordinate $\xi \in (-1, 1)$. In this form of the loss function given by equation (\ref{5.1}), the NN is motivated to converge on the QNMs with the highest amplitude, $|\chi|$, because of the regularisation loss term $\mathcal{L}_f$. Incidentally, it turns out that the QNM with the highest $|\chi|$ (for any given multipole number) is the $n = 0$ mode. This is consistent with the fact that for black hole QNMs, the higher overtones are damped faster \cite{Isietal2019}.  

Some observations from figure \ref{FIG: HYPERP} are that varying the hyperparameters, as we did, has no significant effect on the accuracy. However, there is an increase in the training time with the number of epochs for a fixed number of training points and neurons per layer. Additionally, an increase in the number of neurons per layer also leads to a slight increase in the training time. Therefore, to obtain a favourable trade-off between accuracy and efficiency, one may train for 100~000 epochs instead of 200~000 to achieve the same level of accuracy in less time. This reduction in training time becomes significant when running a large batch of computations.

\subsection{QNMs of near extremal non-rotating black holes}
\label{SEC: QNMs2}

In our discussion of black hole perturbation equations in section \ref{SEC: Near Extremal}, we have seen a special case where the effective potential is given exactly by an inverted P\"{o}schl-Teller potential; namely, the near extremal Schwarzschild and Reissner-Nordstr\"{o}m-de Sitter black holes. In these cases, analytic expressions of the QNMs are known and we could reliably test the accuracy of our NN approximations compared to the exact QNMs given as:
\begin{equation}
\label{5.4}
\omega = \sqrt{\left(\ell(\ell + 1) - \frac{1}{4}\right)} - i\left(n + \frac{1}{2}\right),\quad n = 0, 1, 2, ... 
\end{equation}
\noindent where $\ell$ and $n$ are as defined in section \ref{SEC: Schwarzschild}. 

In table \ref{Table 1}, the exact QNMs for $n = 0$ and $\ell = 1, ..., 3$ are compared with the NN approximations ($\omega_{\textsubscript{$eigeNN$}}$). The latter were obtained by embedding the governing differential equation of near extremal non-rotating de Sitter black holes and extra regularisation terms in the loss function. In the last column of table \ref{Table 1} are values that were produced by adding to the loss function a seed value loss term given as:
\begin{equation}
\label{5.5}
\mathcal{L}_{seed}(\theta, \hat{\omega};\mathcal{T}) = \frac{1}{|\mathcal{T}|} \sum_{\xi \in \mathcal{T}} \left[ \hat{\omega} - \omega_{seed}\right]^2.
\end{equation}

\noindent The seed value loss term measures the deviation of the NN approximations from specific $n$ and $\ell$ dependent seed values close to the exact QNMs (i.e. accurate up to a certain number of decimal places, e.g. 2 decimal places, in this case). The goal of the seed loss term is to steer the NN towards specific QNMs of the several possible differential equation residual minimisers (or eigenstates) that exist for a chosen multipole number $\ell$.

\begin{figure}[H]
\begin{minipage}{18pc}
\includegraphics[width=18pc]{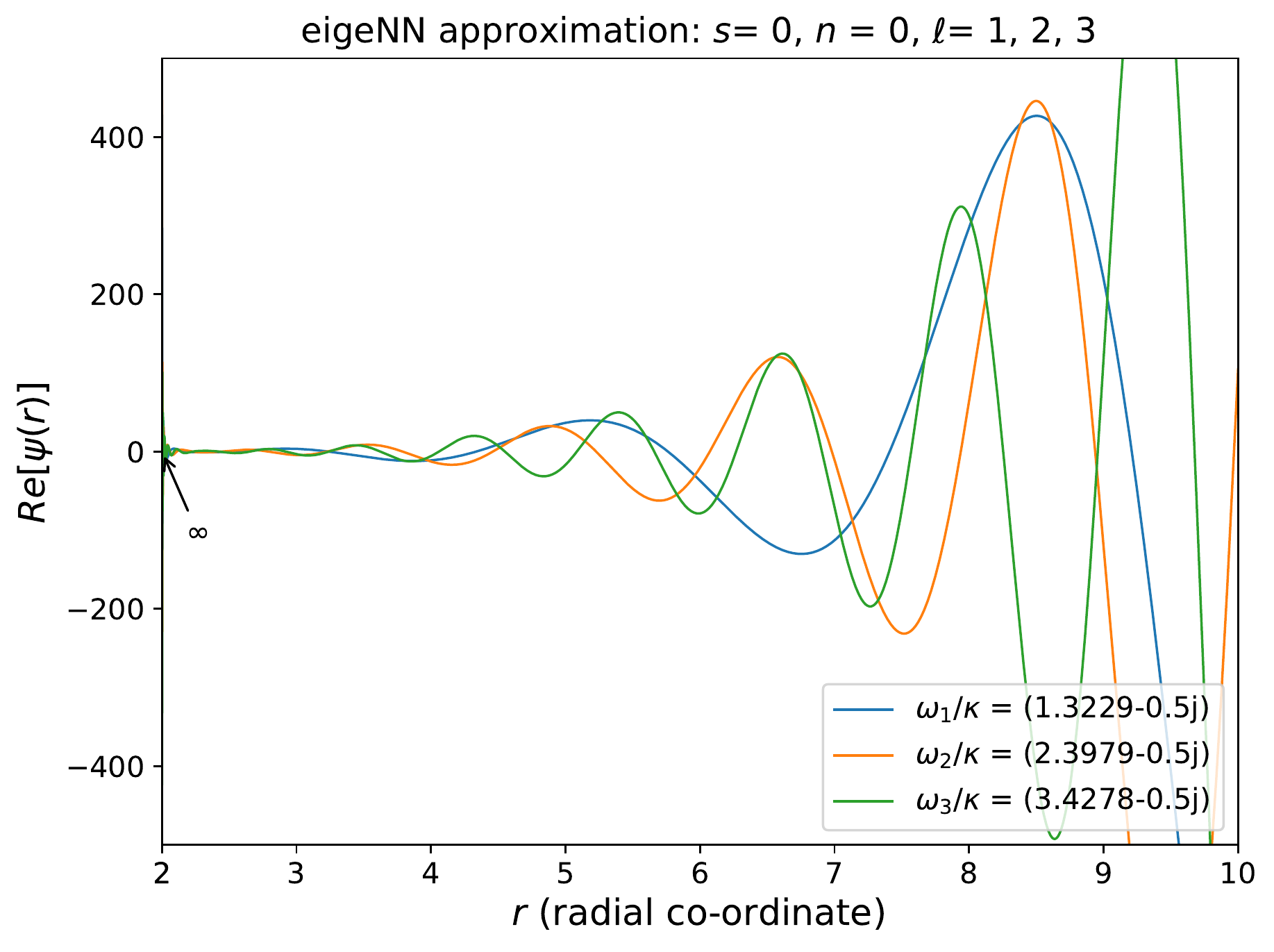}
\end{minipage}\hspace{0pc}%
\begin{minipage}{18pc}
\includegraphics[width=18pc]{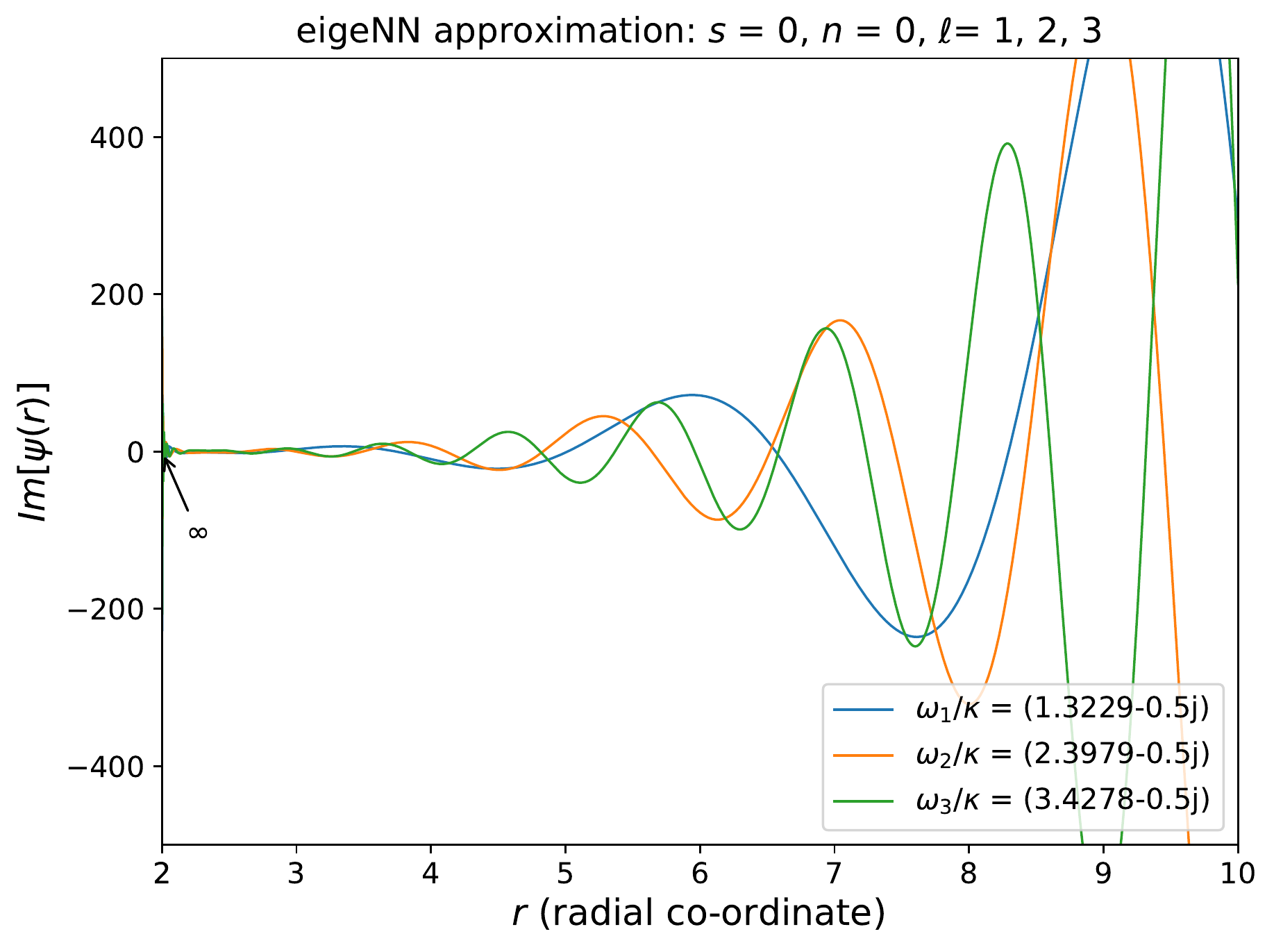}
\end{minipage}
~
\begin{minipage}{18pc}
\includegraphics[width=18pc]{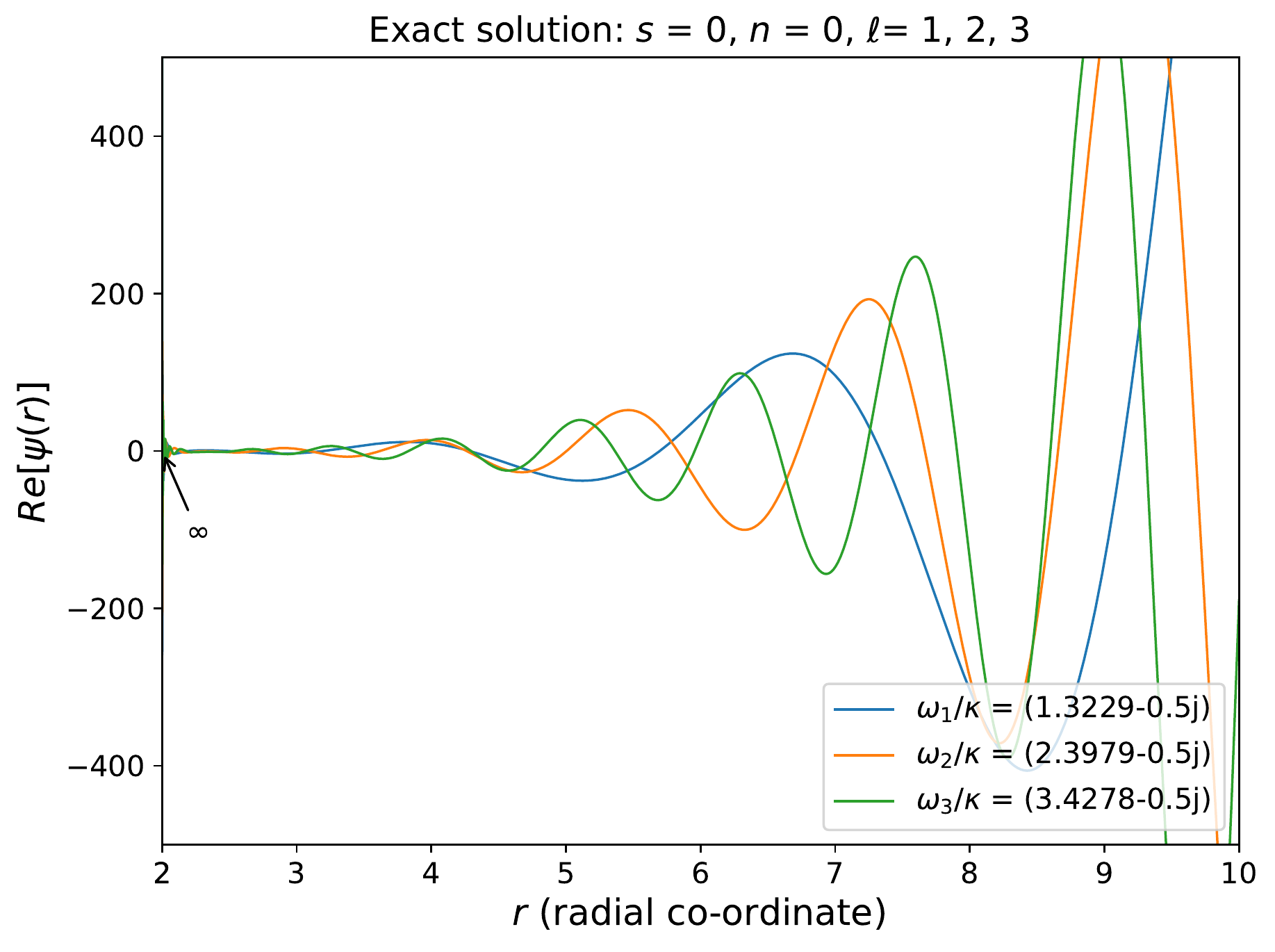}
\end{minipage}
\begin{minipage}{18pc}
\includegraphics[width=18pc]{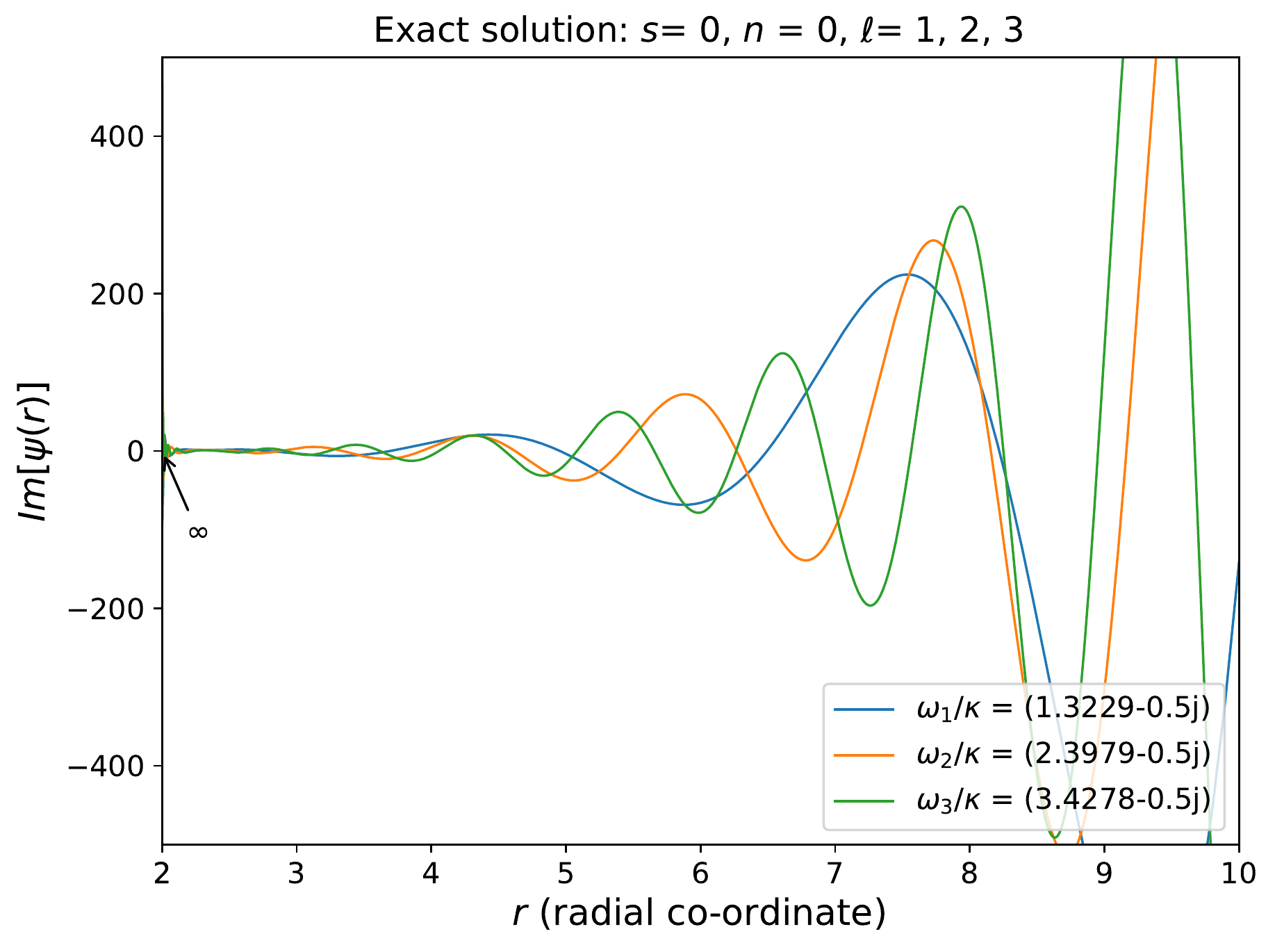}
\end{minipage}
~

\caption{\label{FIG: Output1} The exact QNM wave functions (cf. equation (\ref{2.21})) vs. the eigeNN approximations. 
}
\end{figure}

The plots in figure \ref{FIG: Output1} are the NN approximations of the eigenpairs $(\omega, \psi)$ associated with table \ref{Table 1}, where the first three multipole numbers for the $n = 0$ mode are superimposed. These are the QNM eigenfunctions that obey the asymptotic behaviour expected for astrophysical asymptotically de Sitter black holes \cite{KonoplyaZhidenko2011}. As was pointed out previously, this is:
\begin{equation}
\psi \sim \text{pure outgoing wave},\quad x\rightarrow +\infty.
\end{equation}

\begin{figure}[t!]
\begin{minipage}{18pc}
\includegraphics[width=18pc]{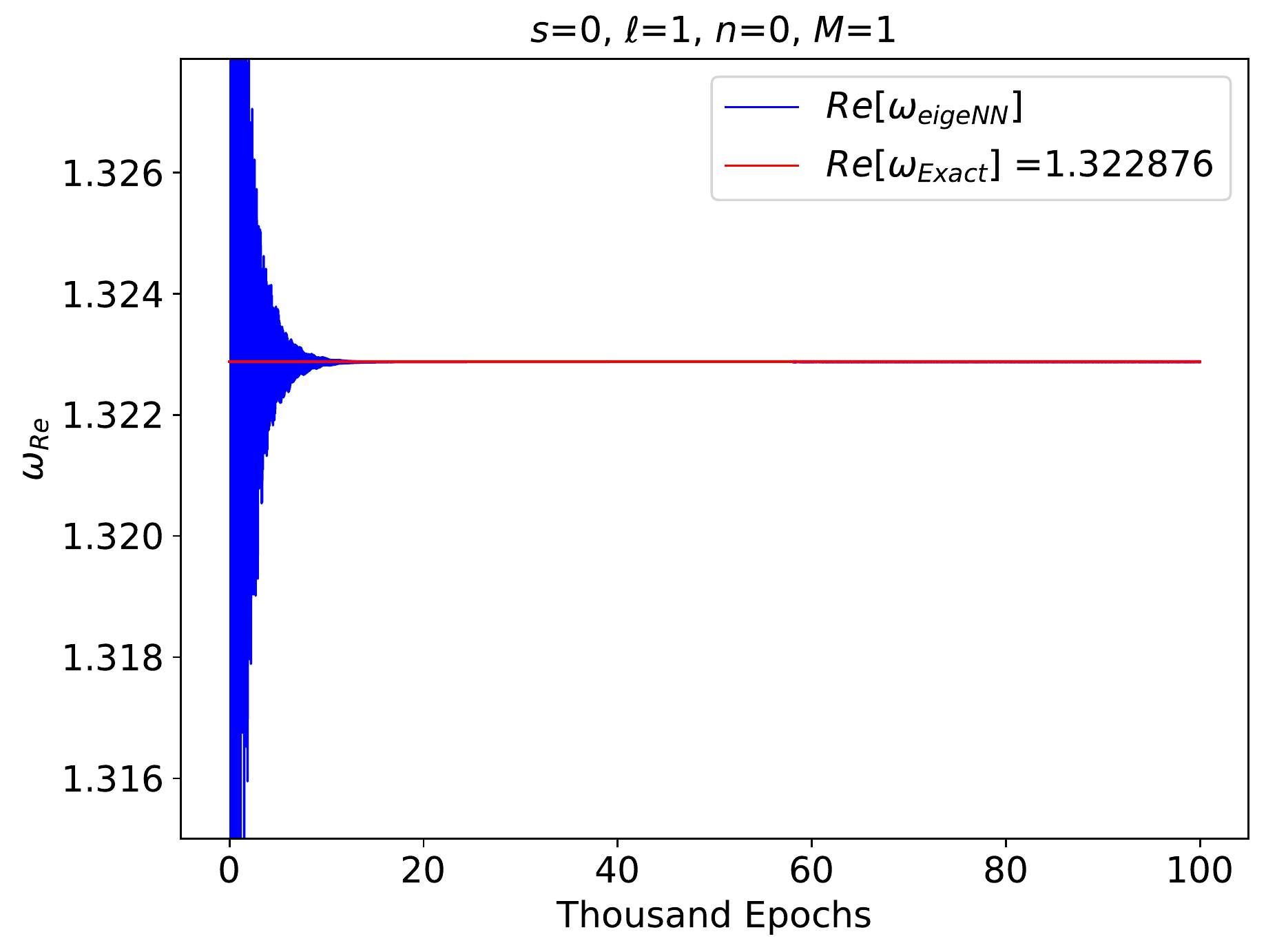}
\end{minipage}\hspace{0pc}%
\begin{minipage}{18pc}
\includegraphics[width=18pc]{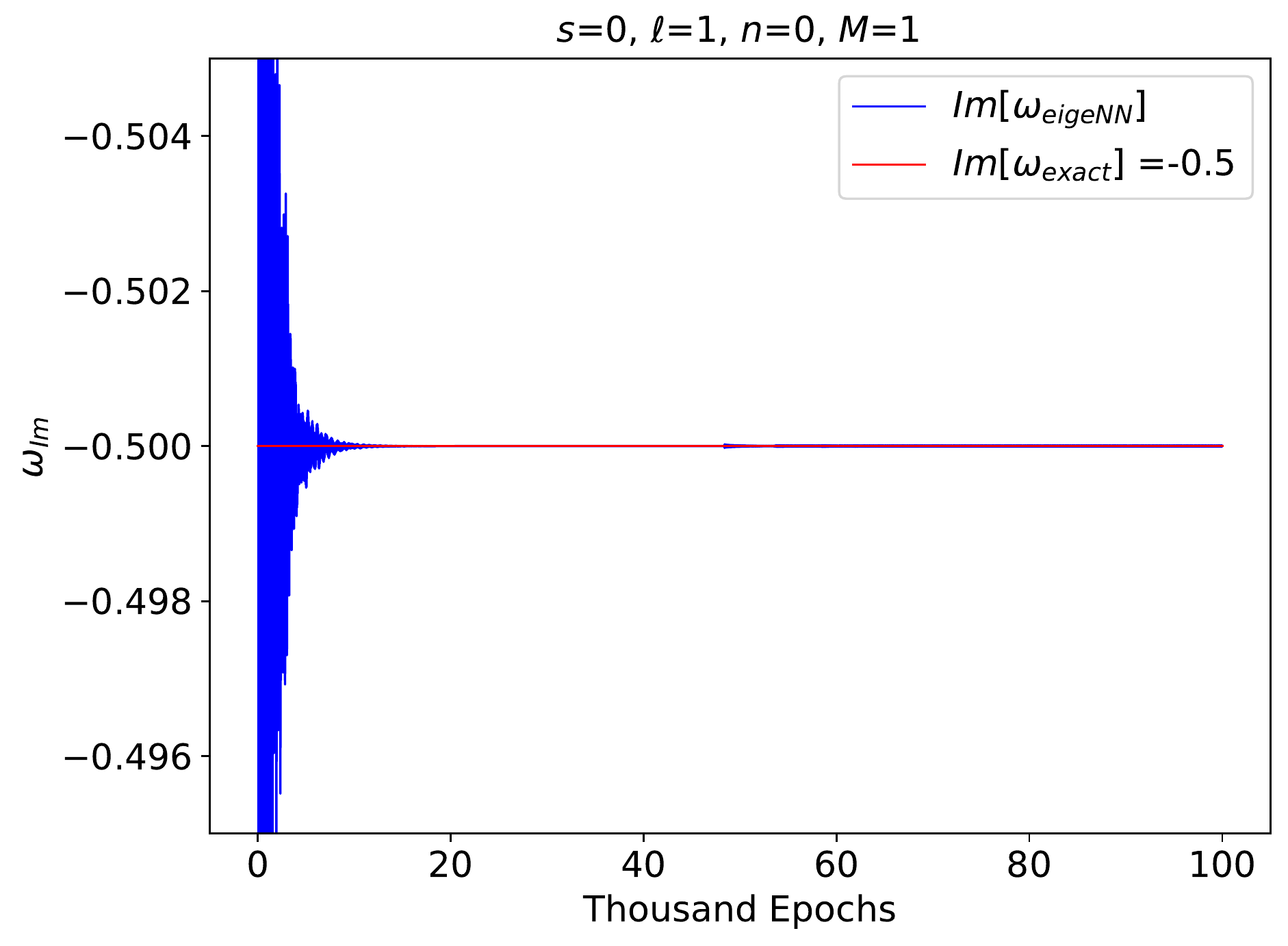}
\end{minipage}\hspace{0pc}%
\begin{minipage}{18pc}
\includegraphics[width=18pc]{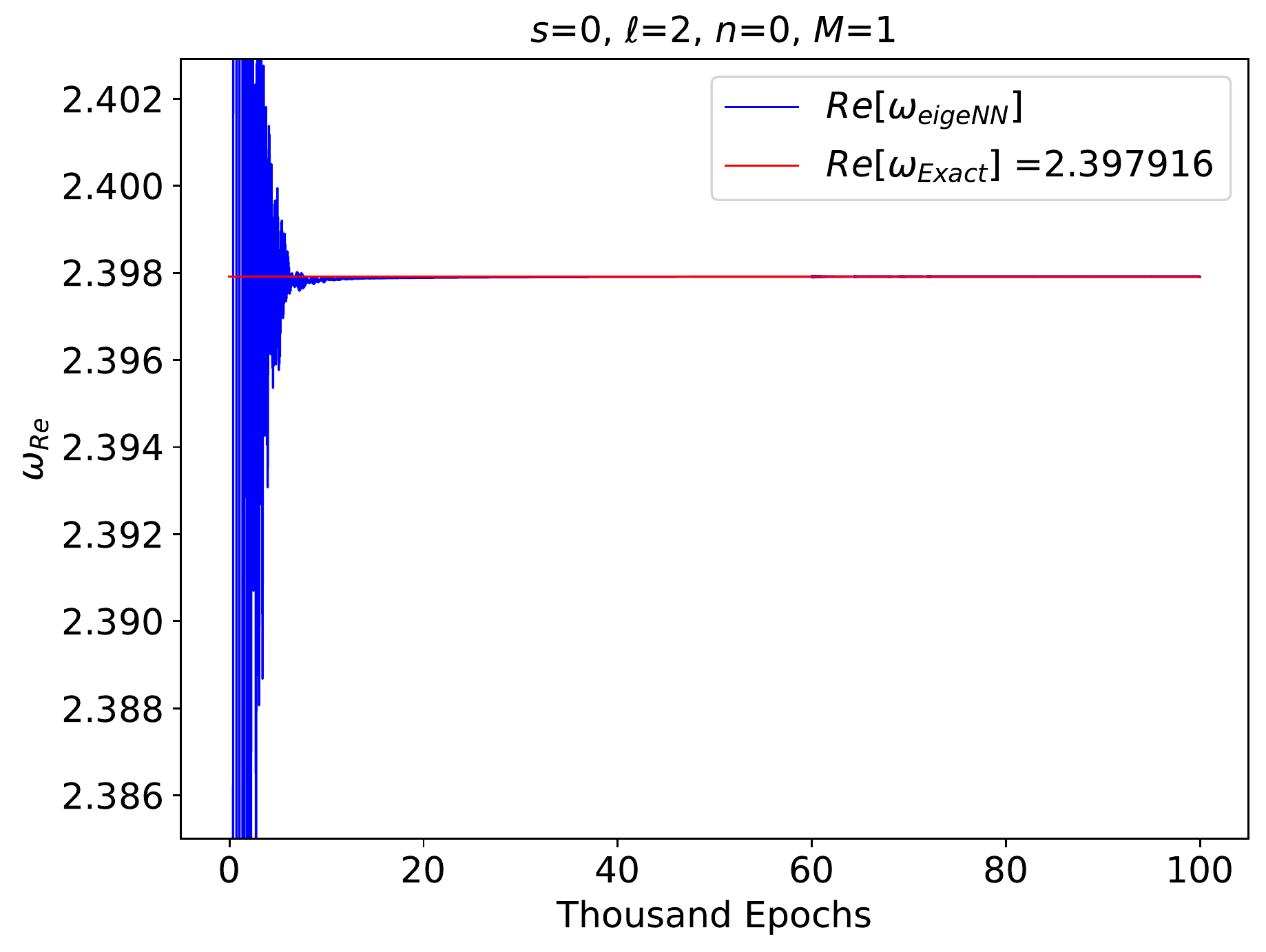}
\end{minipage}\hspace{0pc}%
\begin{minipage}{18pc}
\includegraphics[width=18pc]{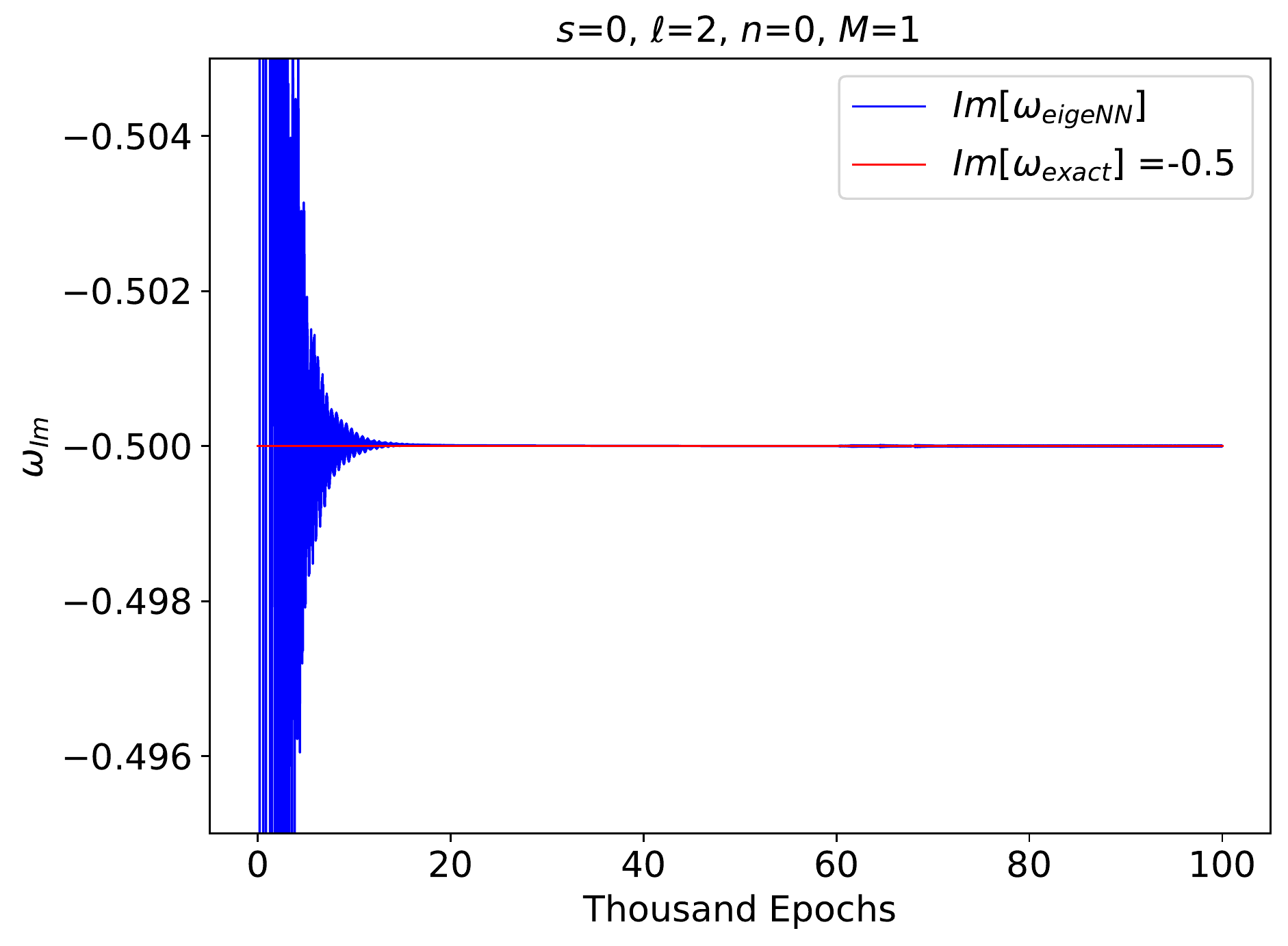}
\end{minipage}
~
\begin{minipage}{18pc}
\includegraphics[width=18pc]{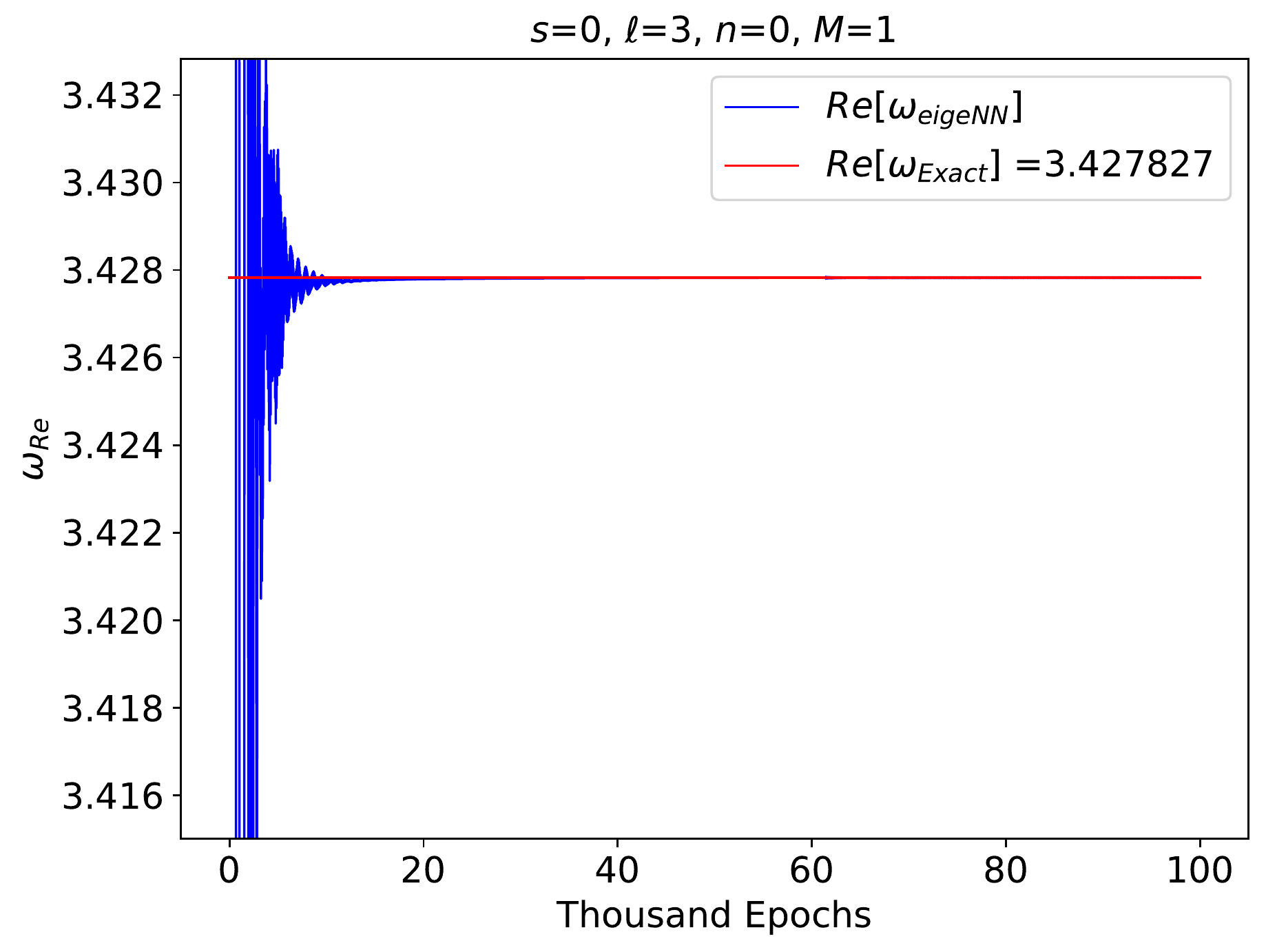}
\end{minipage}\hspace{0pc}%
\begin{minipage}{18pc}
\includegraphics[width=18pc]{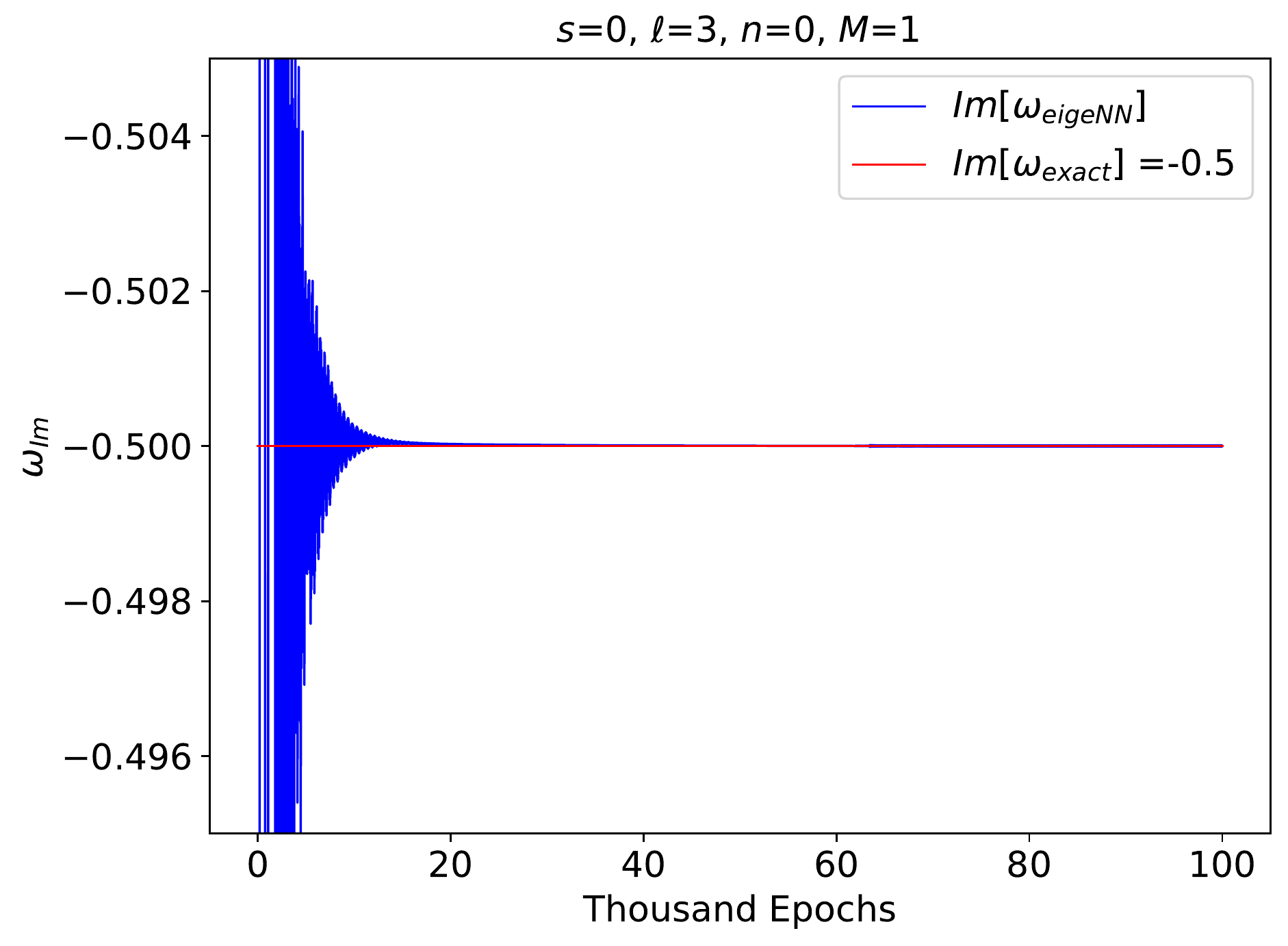}
\end{minipage}
\caption{\label{FIG: Output2} Evolution of the NN approximations of the QNM frequencies during the 100~000 epoch-long training phase. These were generated from our computations using the eigenvalue solver with no seed loss term in the loss function.}
\end{figure}

More importantly, figure \ref{FIG: Output2} shows the evolution of the real ($\omega_{Re}$) and imaginary ($\omega_{Im}$) parts of the NN's approximations of the QNMs as they train for 100~000 epochs. These plots were obtained from our computations without the seed loss term in the loss function. 

\begin{table}[t!]
\caption{\label{Table 1} The eigenvalue solver (eigeNN) approximations of the fundamental mode ($n = 0$,\\ $\ell = 1, ..., 3$) QNMs for massless scalar field perturbations of near extremal SdS and RNdS BHs.}
\begin{center}
\begin{tabular}{cccccccccccr}
\hline
$n$ & $\ell$ &&  $\omega_{\textsubscript{$Exact$}}$ && $\omega_{\textsubscript{$eigeNN$}}$ && $\omega_{\textsubscript{$eigeNN$}}$ \\
 &  && \cite{FerrariMashhoon1984} && \text(no seed) && \text(with seed) \\[0.5em]
\hline\\
\multirow{2}{0.5em}{0} & \multirow{2}{0.25em}{1} && \multirow{2}{7em}{$1.322876 - 0.5i$} && $1.322886 - 0.500004i$ && $1.322894 - 0.500011i$ \\[0.25em]
                       &   &&                                      && ($<0.001\%$)($<0.001\%$) && (0.001\%)(0.002\%) \\[0.25em]
                       & \multirow{2}{0.5em}{2} && \multirow{2}{7em}{$2.397916 - 0.5i$} && $2.397917 - 0.500001i$ && $2.397916 - 0.500000i$ \\[0.25em]
                       &   &&                                      && ($<0.001\%$)($<0.001\%$) && ($<0.001\%$)($<0.001\%$) \\[0.25em] 
                       & \multirow{2}{0.5em}{3} && \multirow{2}{7em}{$3.427827 - 0.5i$} && $3.427828 - 0.500001i$ && $3.427828 - 0.500000i$ \\[0.25em]
                       &   &&                                      && ($<0.001\%$)($<0.001\%$) && ($<0.001\%$)($<0.001\%$)  \\[0.25em]
\hline
\end{tabular}
\end{center}
\end{table}

Figure \ref{FIG: Output2} shows that the convergence of the NN approximation ($\omega_{eigeNN}$) towards the expected QNMs ($\omega_{exact}$) given by equation (\ref{5.4}), occurs swiftly after training begins. Table \ref{Table 1} and figure \ref{FIG: Output2} indicate that the NN learns QNMs with similar levels of accuracy for different multipole numbers, regardless of the presence of a seed loss term in the loss function.

\begin{table}[t!]
\caption{\label{Table 2} The eigenvalue solver ($eigeNN$) approximations of the fundamental mode ($n = 0$) QNMs given up to 4 decimal places for massless scalar field perturbations ($s = 0$) of asymptotically flat Schwarzschild BHs.} 
\begin{center}
\begin{tabular}{cccccccccccr}
\hline
$n$ & $\ell$ &&  $\omega_{\textsubscript{$Leaver$}}$  && $\omega_{\textsubscript{$eigeNN$}}$ && $\omega_{\textsubscript{$WKB$}}$ \cite{Konoplya2003} \\
 &  && \cite{Iyer1987} && (90~000 epochs) && (6th order) \\[0.5em]
\hline\\
\multirow{2}{0.5em}{0} & \multirow{2}{0.5em}{0} && \multirow{2}{7em}{$0.1105 - 0.1049i$} && $0.1106 - 0.1049i$ && $0.1105 - 0.1008i$ \\[0.5em]
                       &   &&                                      && (0.10\%)(-0.01\%) && (-0.03\%)(-3.89\%) \\[0.5em]
                       & \multirow{2}{0.5em}{1} && \multirow{2}{7em}{$0.2929 - 0.0977i$} && $0.2929 - 0.0977i$ && $0.2929 - 0.0978i$ \\[0.5em]
                       &   &&                                      && (0.02\%)(-0.02\%) && ($$<0.01\%$$)(0.06\%) \\[0.5em] 
                       & \multirow{2}{0.5em}{2} && \multirow{2}{7em}{$0.4836 - 0.0968i$} && $0.4836 - 0.0968i$ && $0.4836 - 0.0968i$ \\[0.5em]
                       &   &&                                      && (0.01\%)(-0.04\%) && ($<0.01\%$)(-0.04\%)  \\[0.5em] 
\hline
\end{tabular}
\end{center}
\end{table}

\subsection{QNMs of asymptotically flat Schwarzschild black holes}
\label{SEC: QNMs3}

For this scenario, we have considered perturbations of asymptotically flat Schwarzschild black holes by massless scalar, Dirac, electromagnetic and gravitational fields given by equations (\ref{2.4}) and (\ref{2.5}) in the tortoise co-ordinate. Tables \ref{Table 2} - \ref{Table 5} compare our NN approximations of the QNMs with those given in the literature for the CFM and WKB approaches for solving the perturbation equations.

With regards to the set-up of our eigenvalue solvers, the same FNN configuration was used for all our computations. That is, we set up 2 hidden layers, 50 nodes per layer, and sine as the nonlinear activation function. Moreover, we employed the Adam optimiser, set up 90~000 training epochs and used a learning rate of $8\times 10^{-3}$. Our training data consisted of 100 points randomly selected from the domain $\xi \in [0, 1]$. Note also that the percentage deviation values given inside the parentheses in the tables are:
\begin{eqnarray}
\text{percentage deviation} =  \frac{|Re/Im[\omega_{eigeNN/ WKB}]| - |Re/Im[\omega_{Leaver}]|}{|Re/Im[\omega_{Leaver}]|} \times 100.
\end{eqnarray}

\begin{figure}[H]
\begin{minipage}{18pc}
\includegraphics[width=18pc]{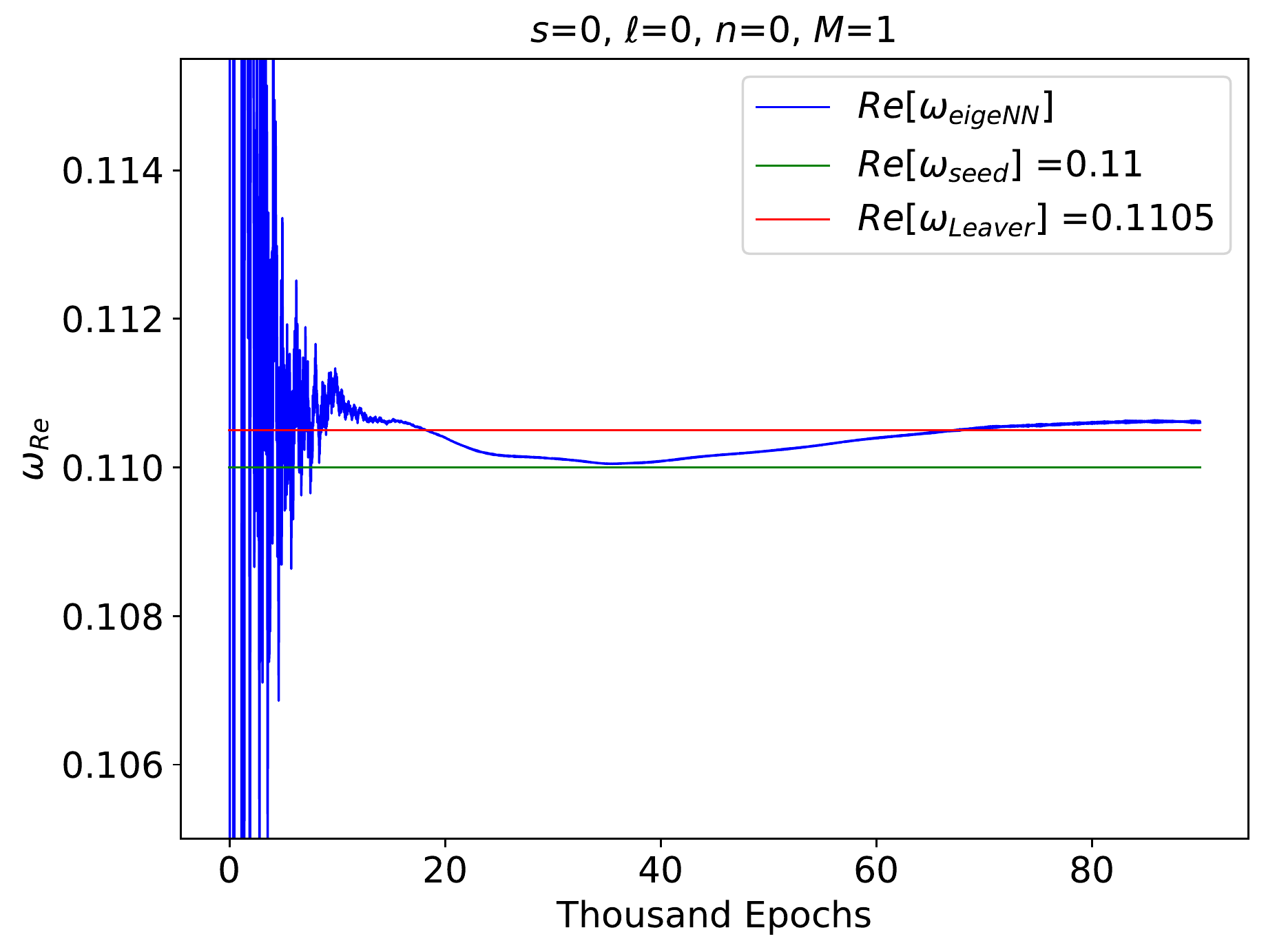}
\end{minipage}\hspace{0pc}%
\begin{minipage}{18pc}
\includegraphics[width=18pc]{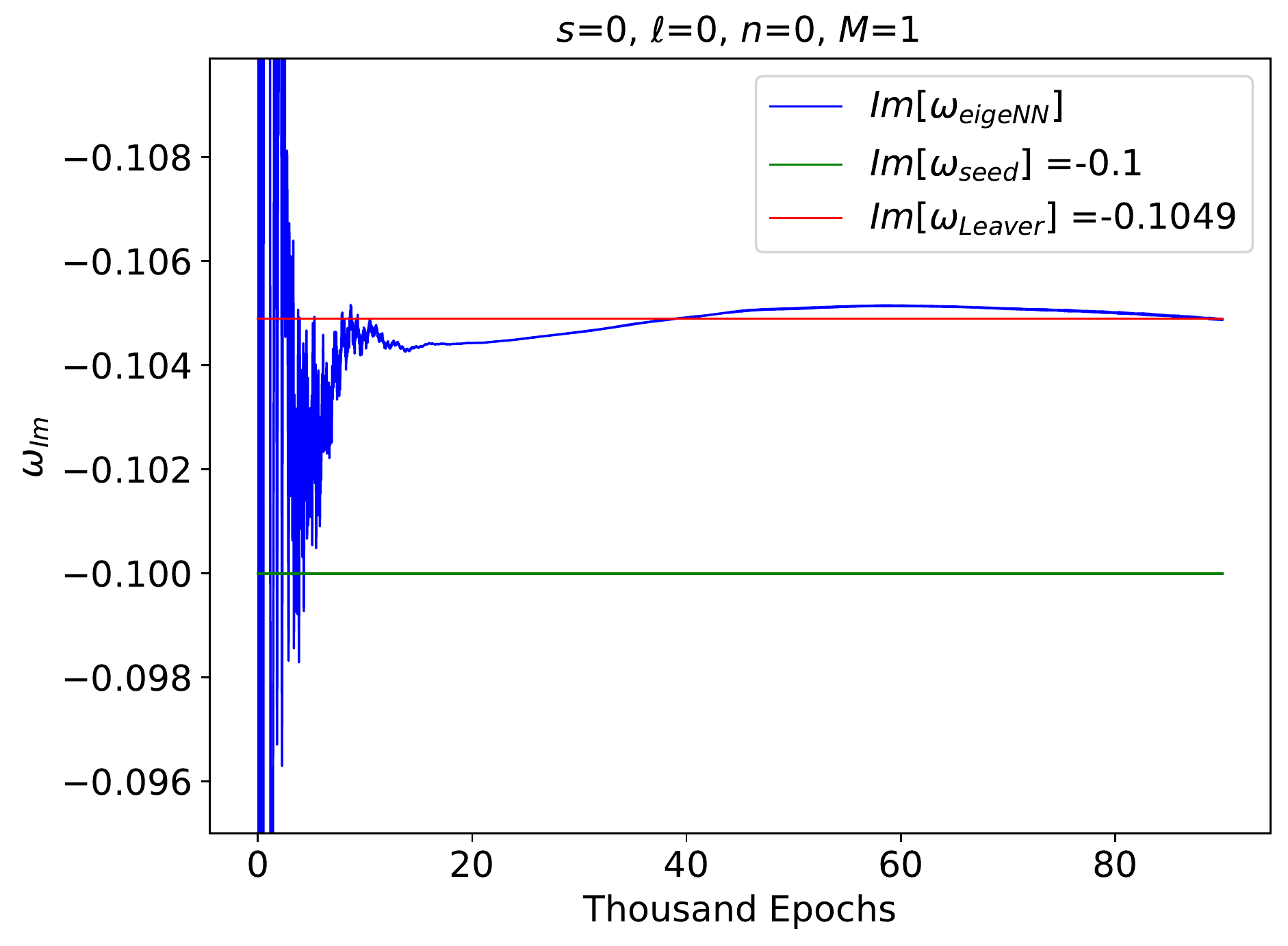}
\end{minipage}
~
\begin{minipage}{18pc}
\includegraphics[width=18pc]{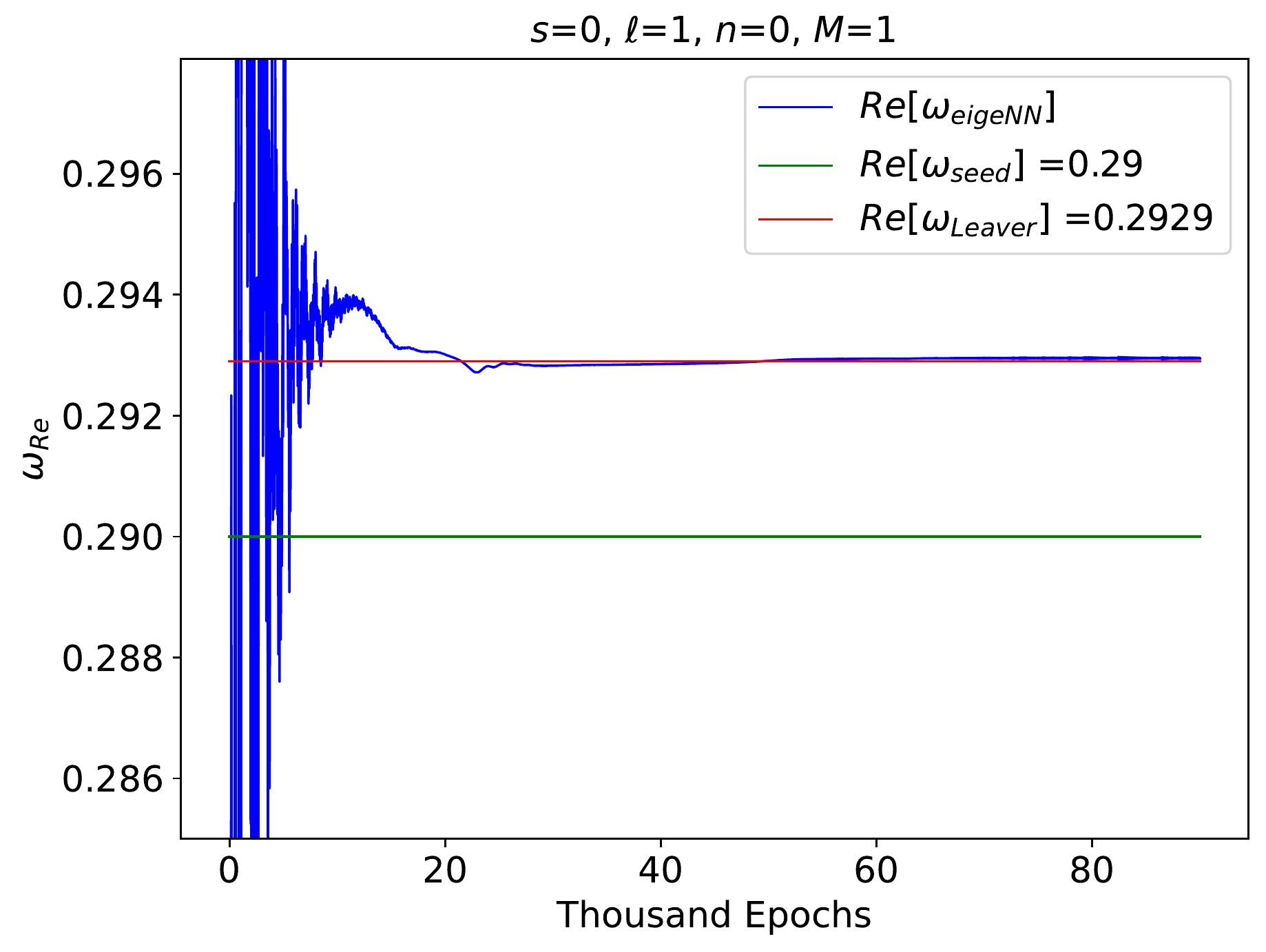}
\end{minipage}\hspace{0pc}%
\begin{minipage}{18pc}
\includegraphics[width=18pc]{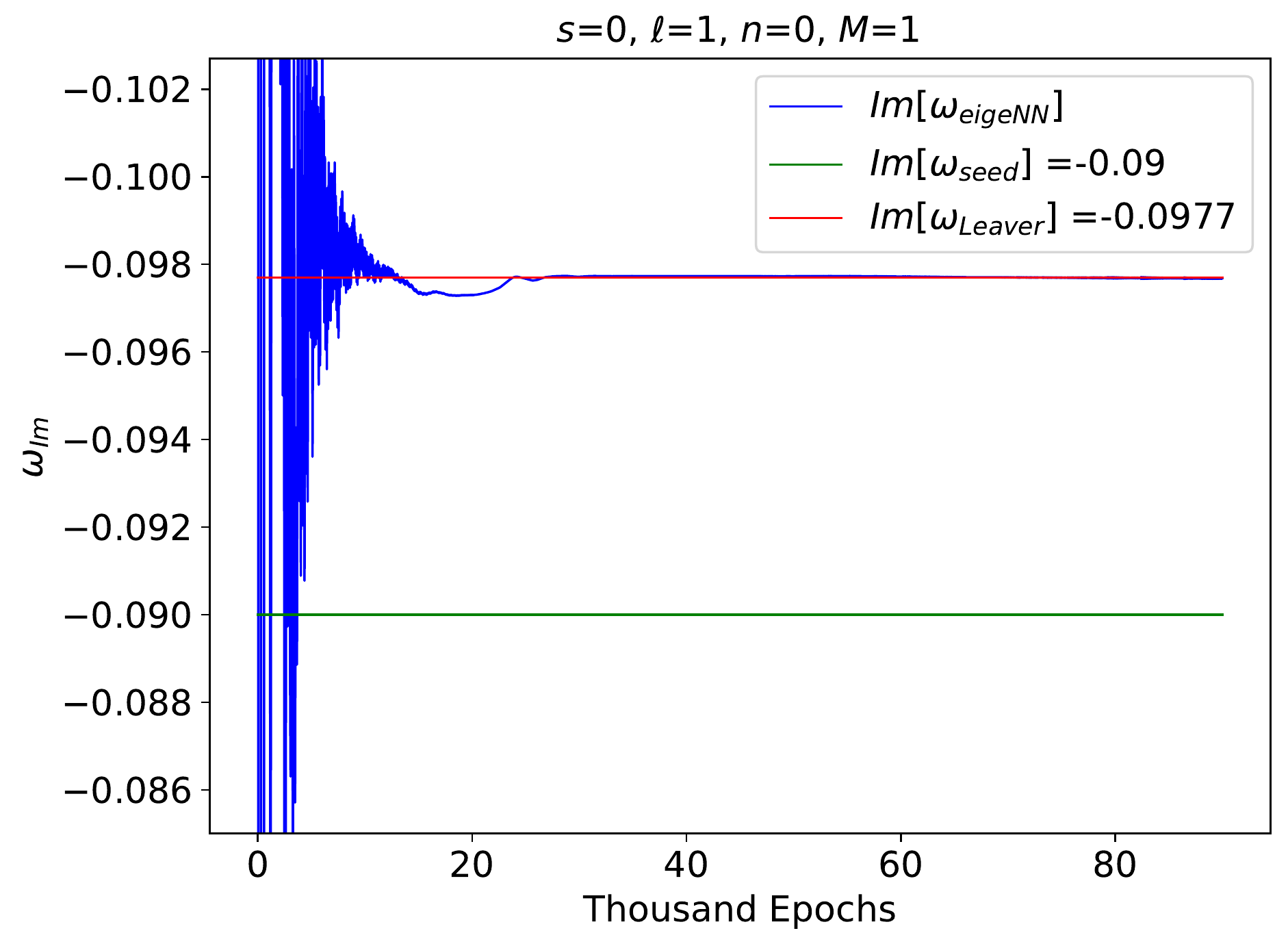}
\end{minipage}
~
\begin{minipage}{18pc}
\includegraphics[width=18pc]{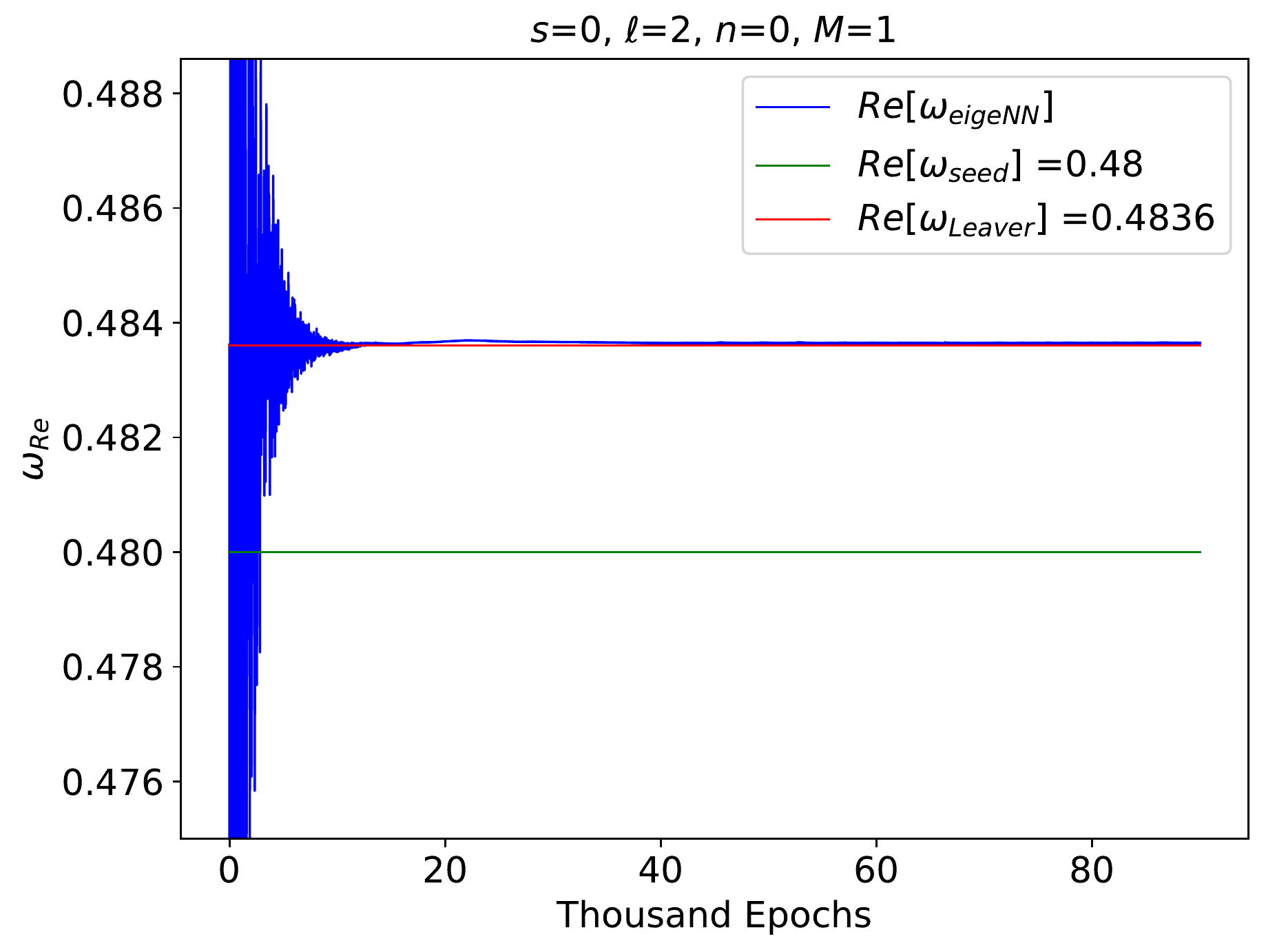}
\end{minipage}\hspace{0pc}%
\begin{minipage}{18pc}
\includegraphics[width=18pc]{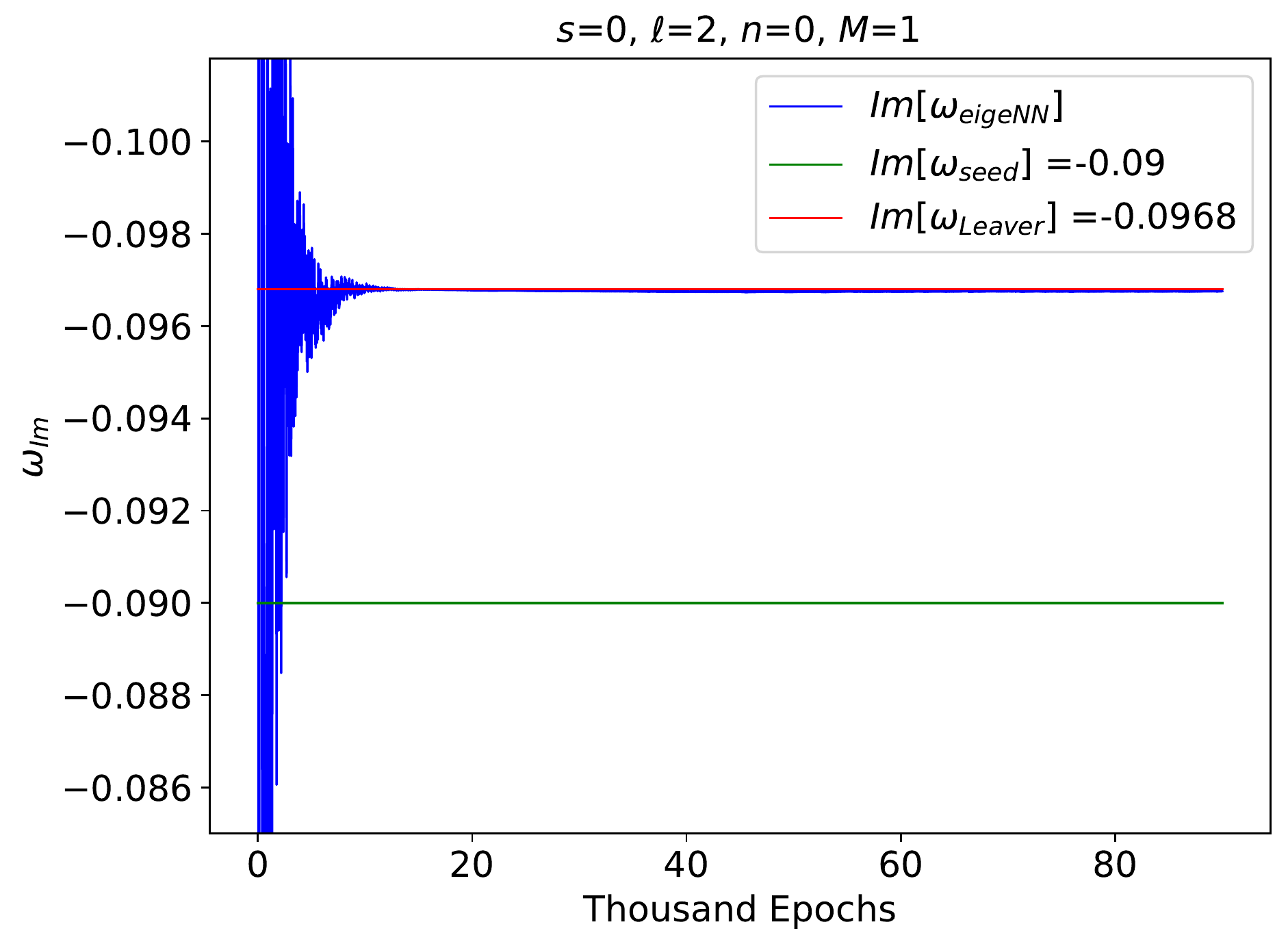}
\end{minipage}
~
\caption{\label{FIG: Output3} Evolution of the NN approximation of QNMs for an asymptotically flat Schwarzschild BH. The 90~000 epoch-long training phase takes an average time of 12 minutes. By contrast, it takes less than 1 minute to generate the QNMs (listed in table \ref{Table 2}) using the 6th order WKB method.}
\end{figure}

\begin{figure}[H]
\begin{minipage}{18pc}
\includegraphics[width=18pc]{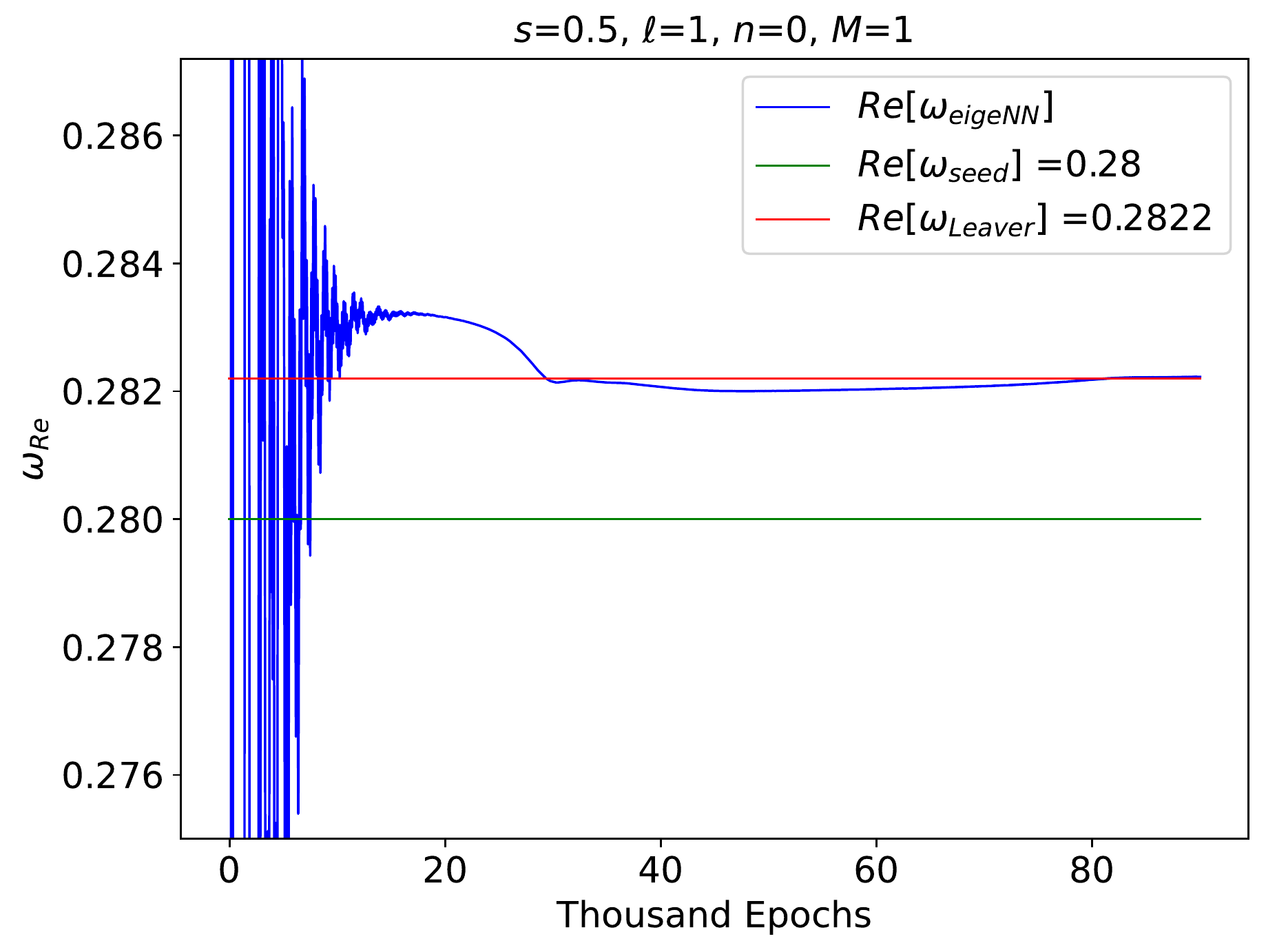}
\end{minipage}\hspace{0pc}%
\begin{minipage}{18pc}
\includegraphics[width=18pc]{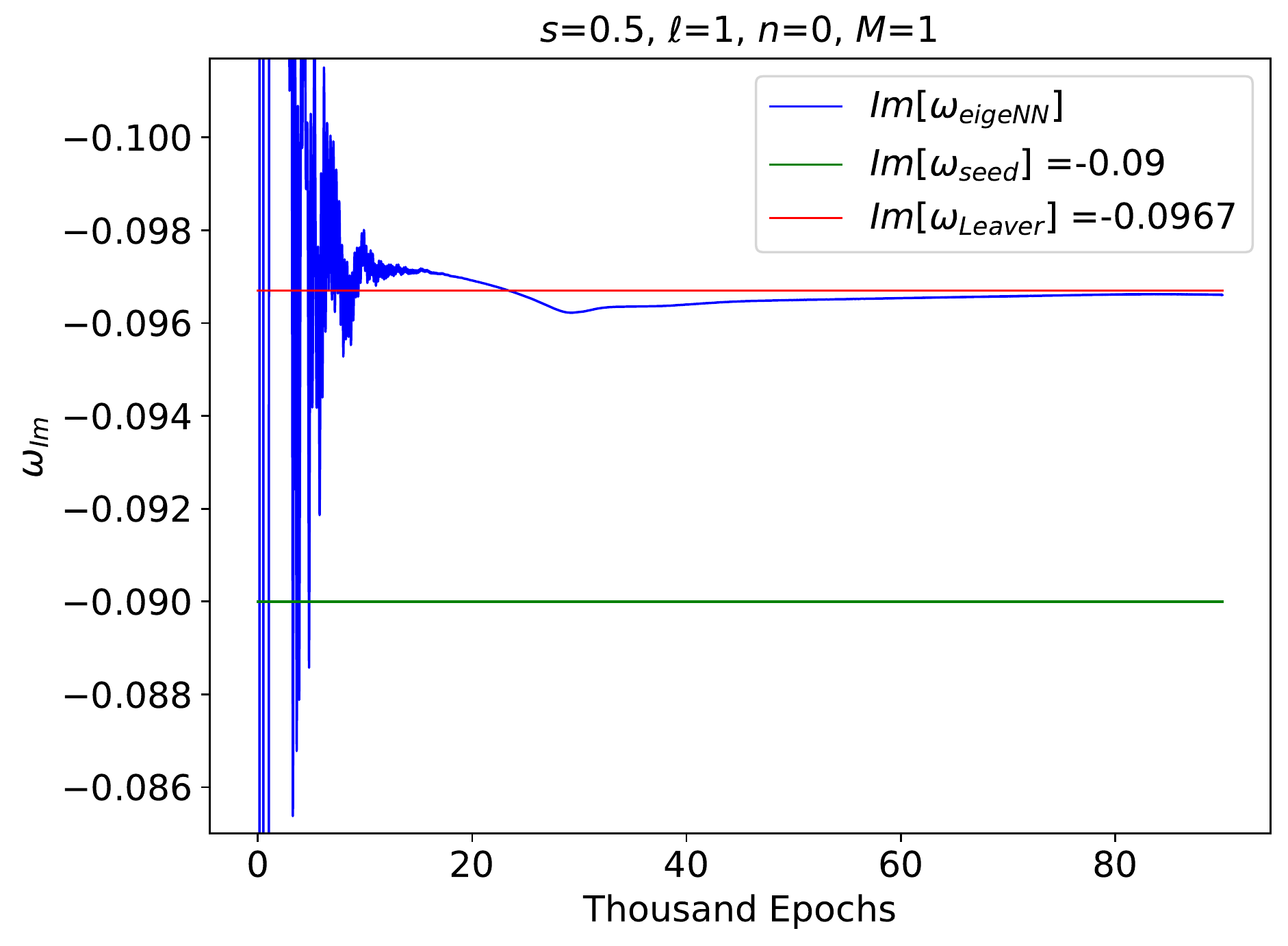}
\end{minipage}
~
\begin{minipage}{18pc}
\includegraphics[width=18pc]{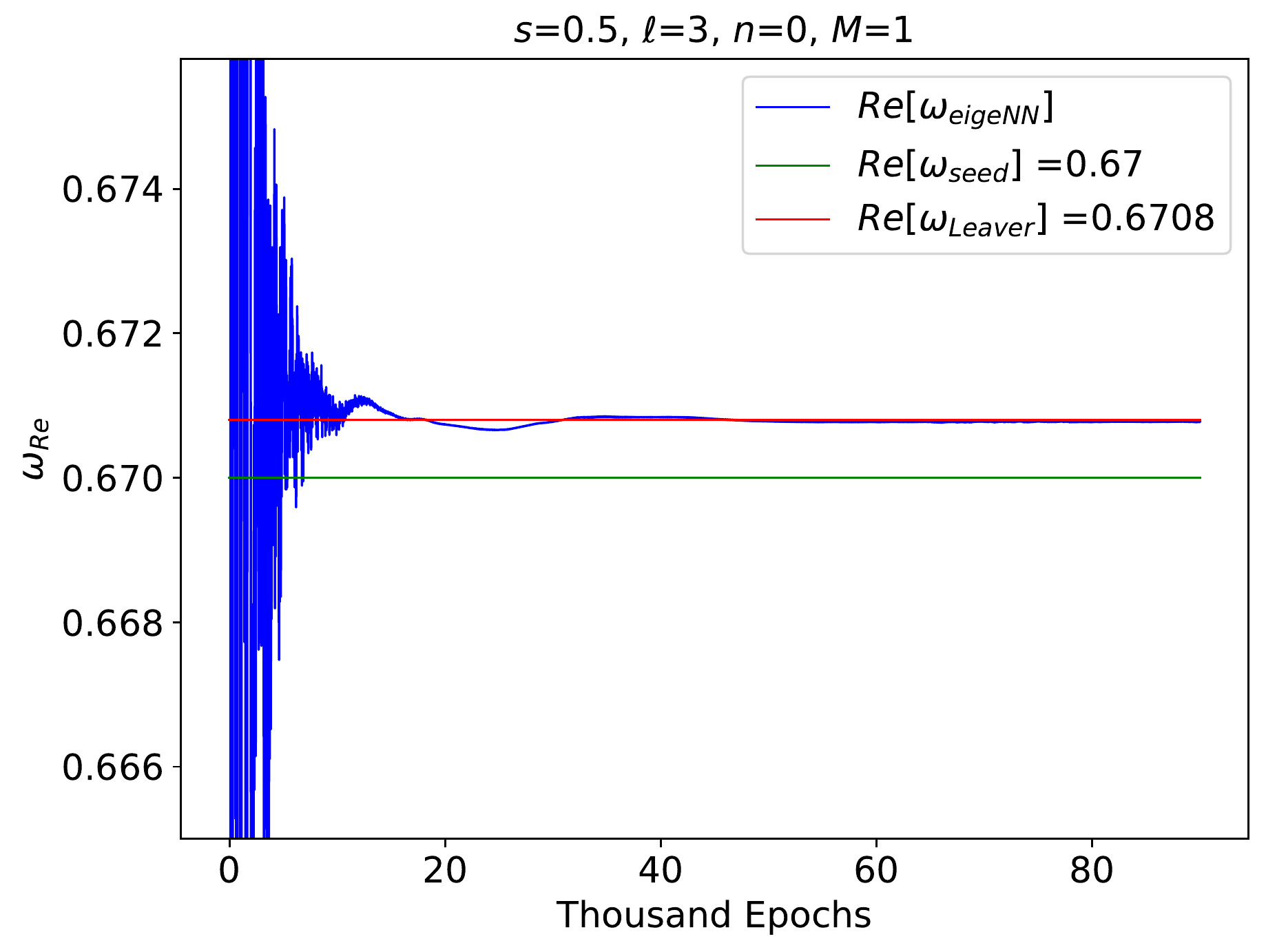}
\end{minipage}
\begin{minipage}{18pc}
\includegraphics[width=18pc]{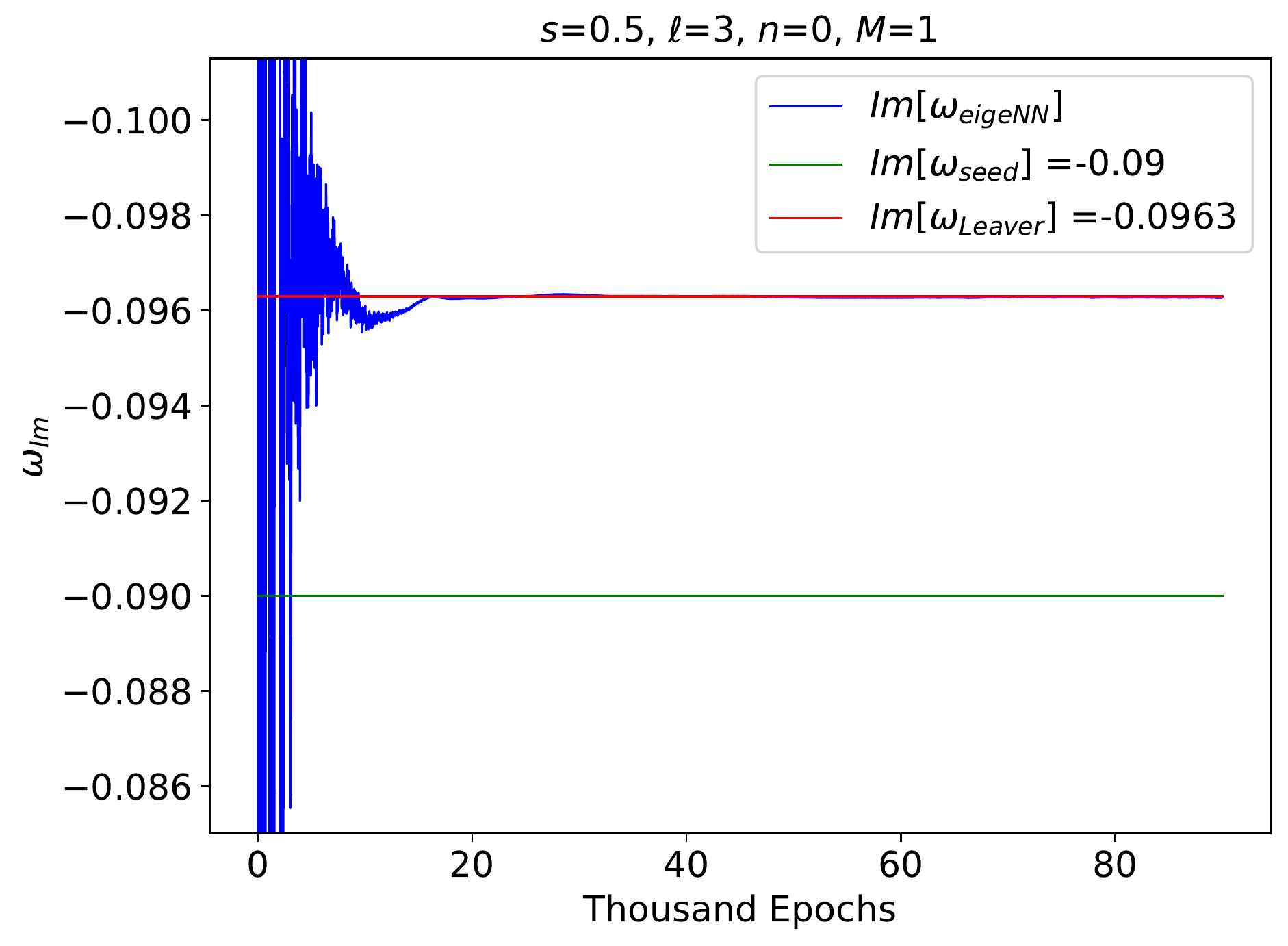}
\end{minipage}
\caption{\label{FIG: Output4}  Evolution of the NN approximation of QNMs over a 90~000 epoch-long training phase. }
\end{figure}

\begin{figure}[H]
\begin{minipage}{18pc}
\includegraphics[width=18pc]{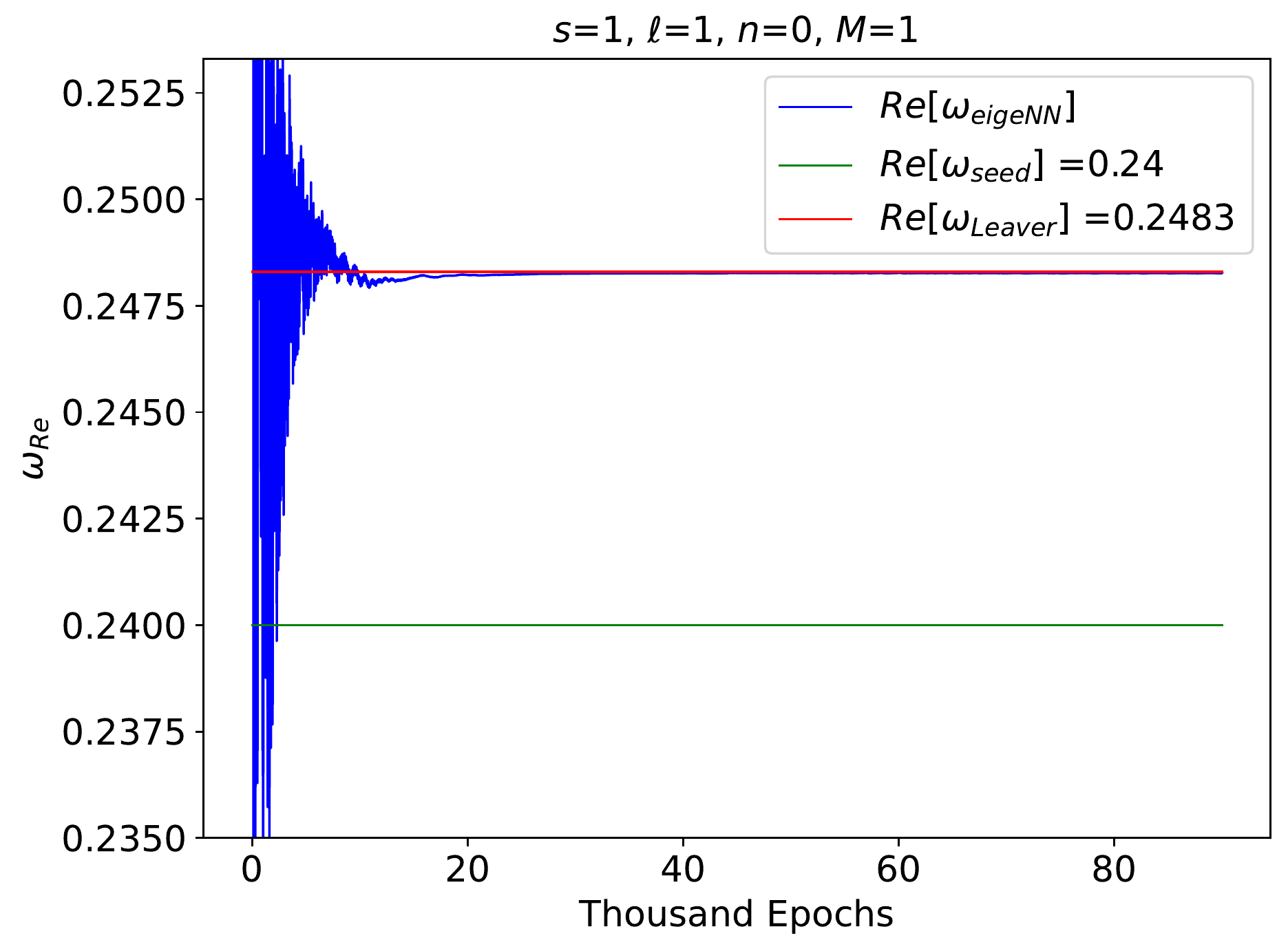}
\end{minipage}\hspace{0pc}%
\begin{minipage}{18pc}
\includegraphics[width=18pc]{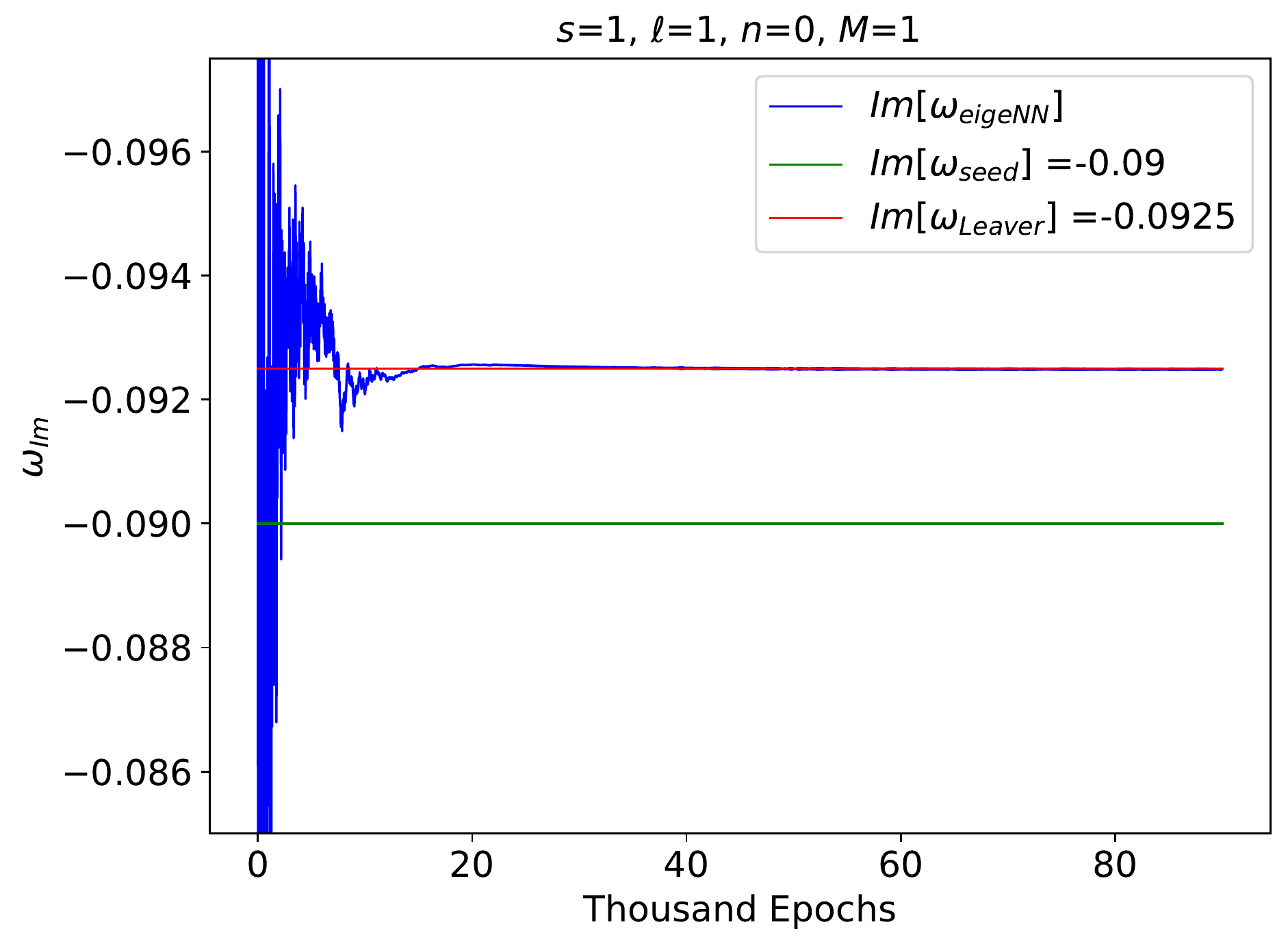}
\end{minipage}
\begin{minipage}{18pc}
\includegraphics[width=18pc]{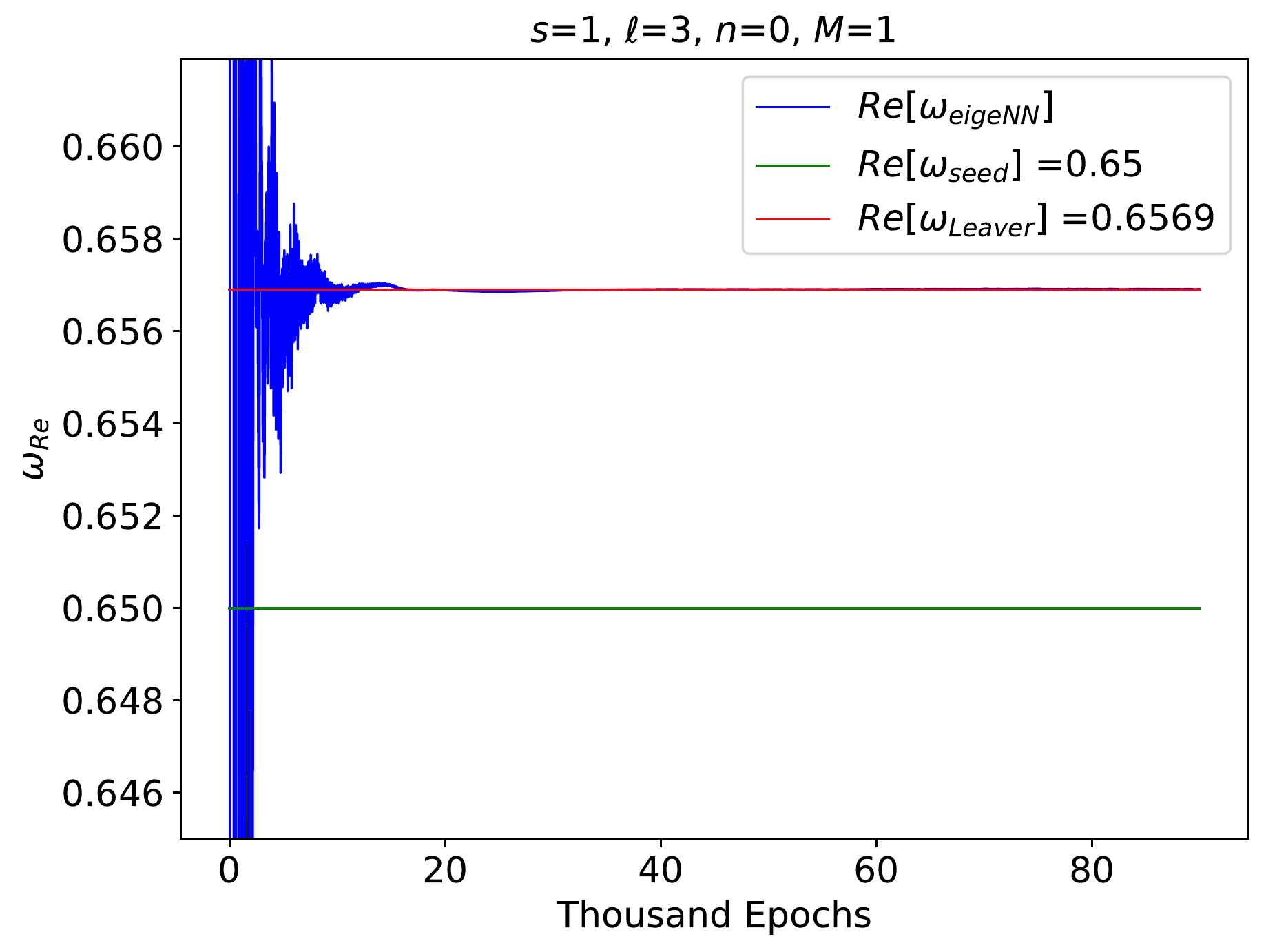}
\end{minipage}
\begin{minipage}{18pc}
\includegraphics[width=18pc]{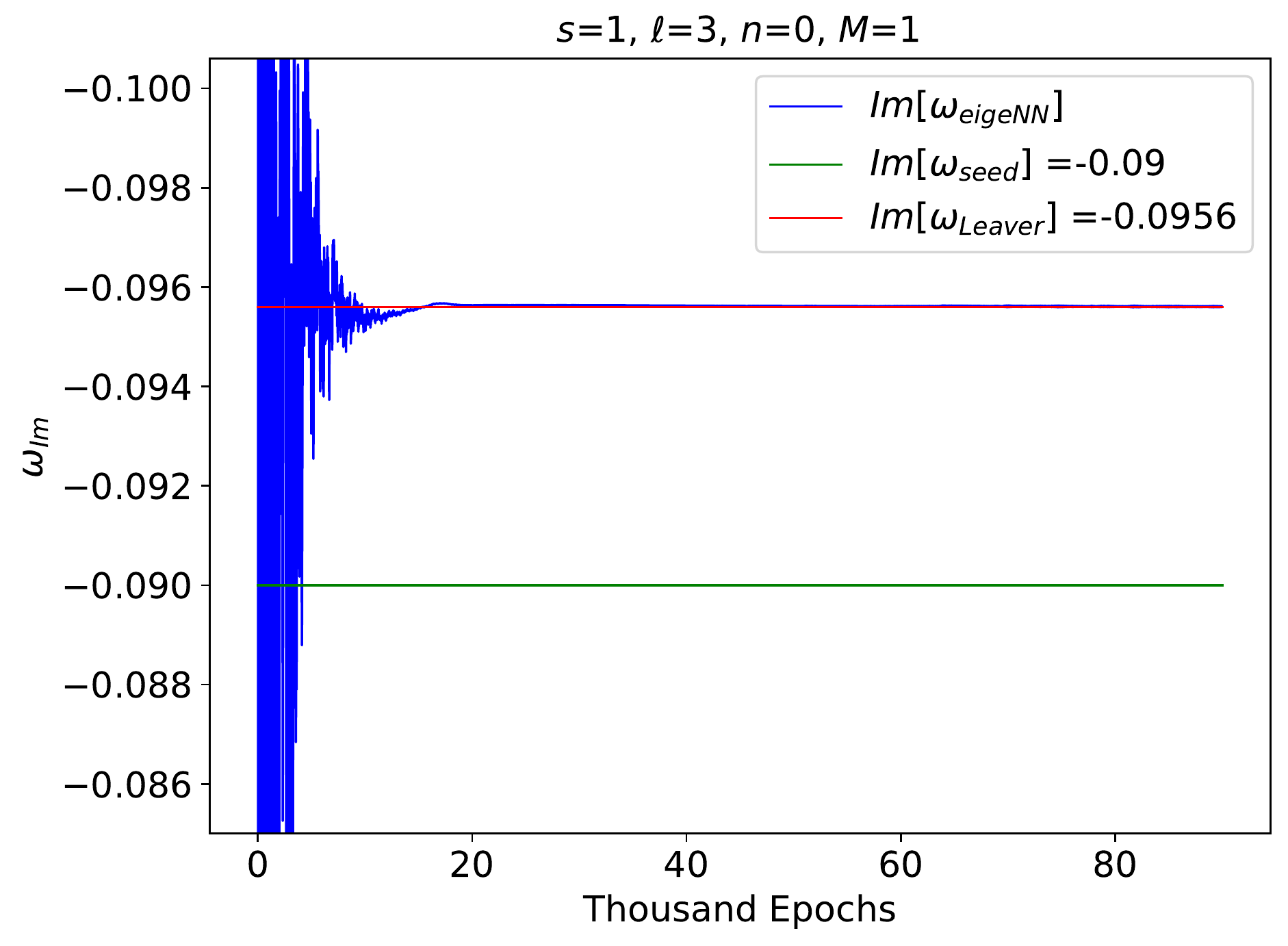}
\end{minipage}
\caption{\label{FIG: Output5}  Evolution of the NN approximation of QNMs during a 90~000 epoch-long training phase.}
\end{figure}

\begin{figure}[H]
\begin{minipage}{18pc}
\includegraphics[width=18pc]{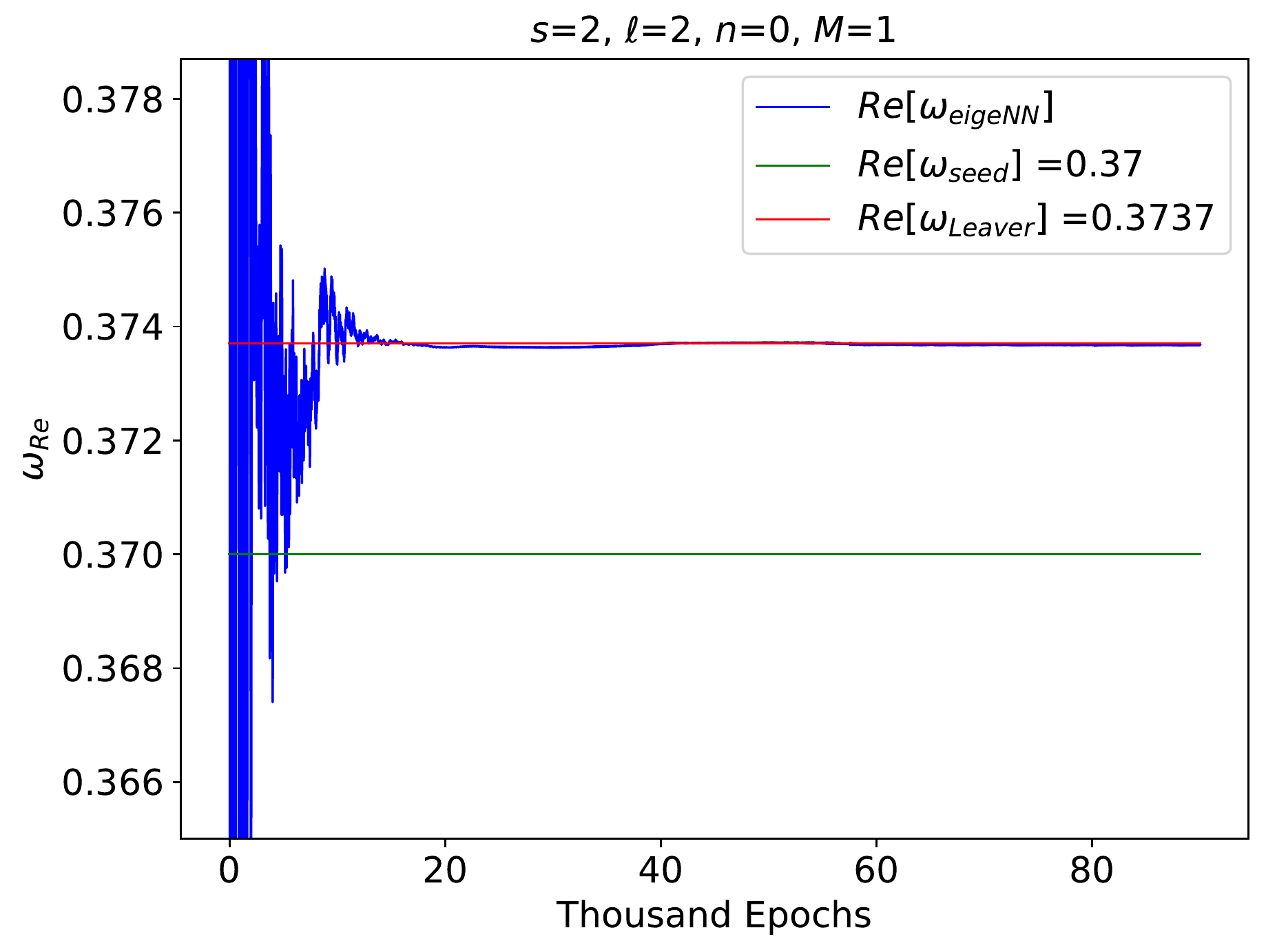}
\end{minipage}\hspace{0pc}%
\begin{minipage}{18pc}
\includegraphics[width=18pc]{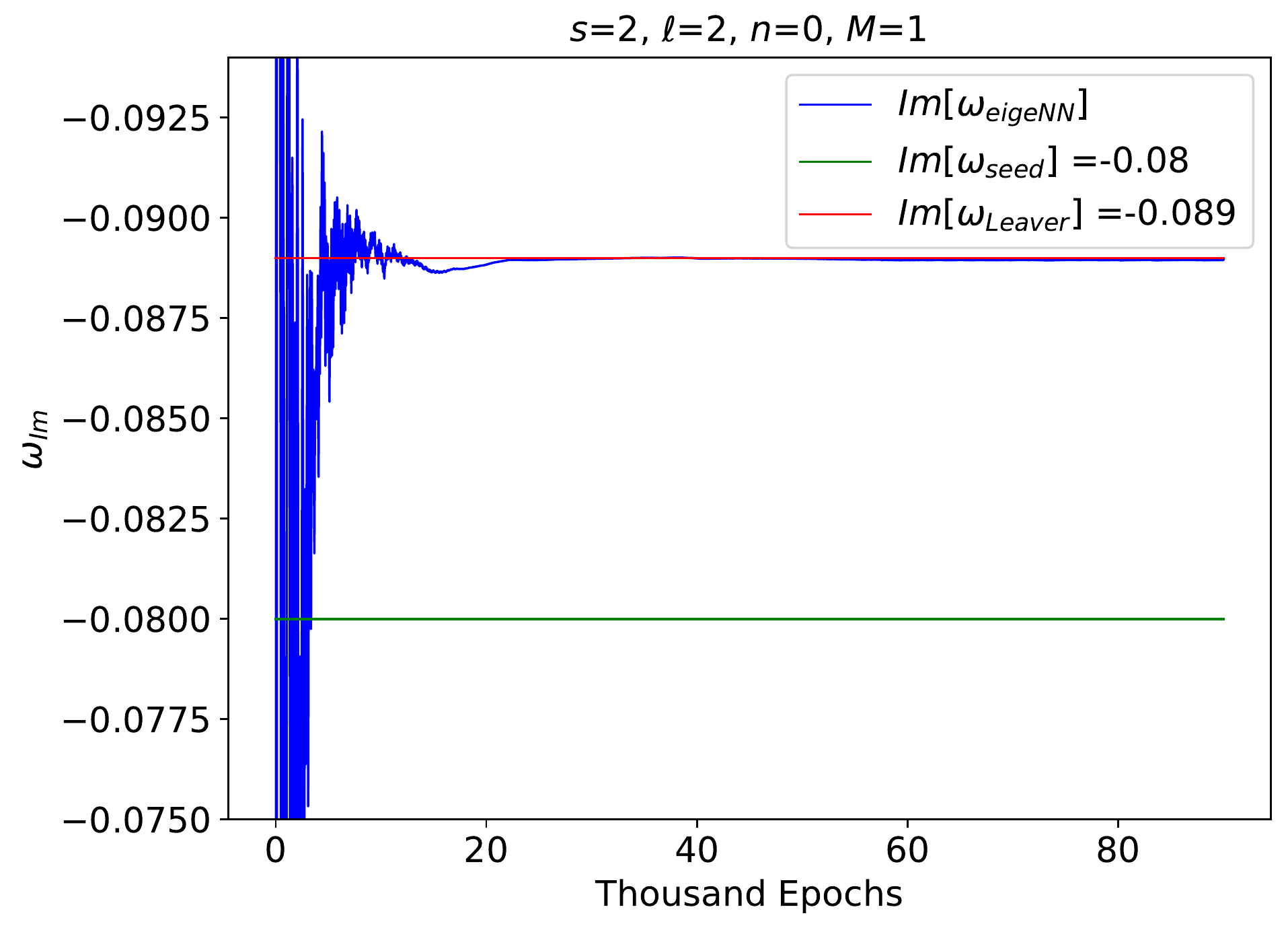}
\end{minipage}
\begin{minipage}{18pc}
\includegraphics[width=18pc]{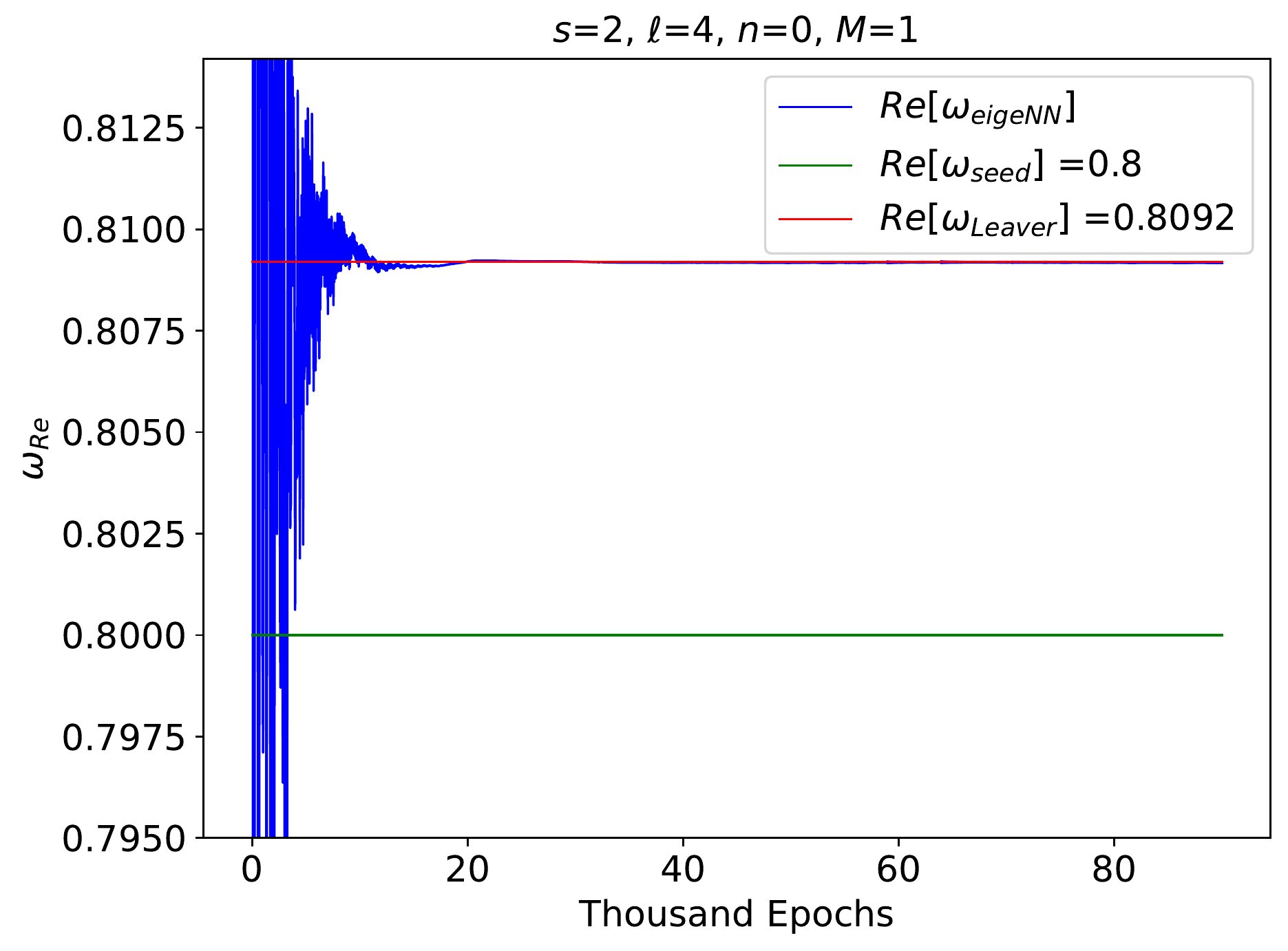}
\end{minipage}\hspace{0pc}%
\begin{minipage}{18pc}
\includegraphics[width=18pc]{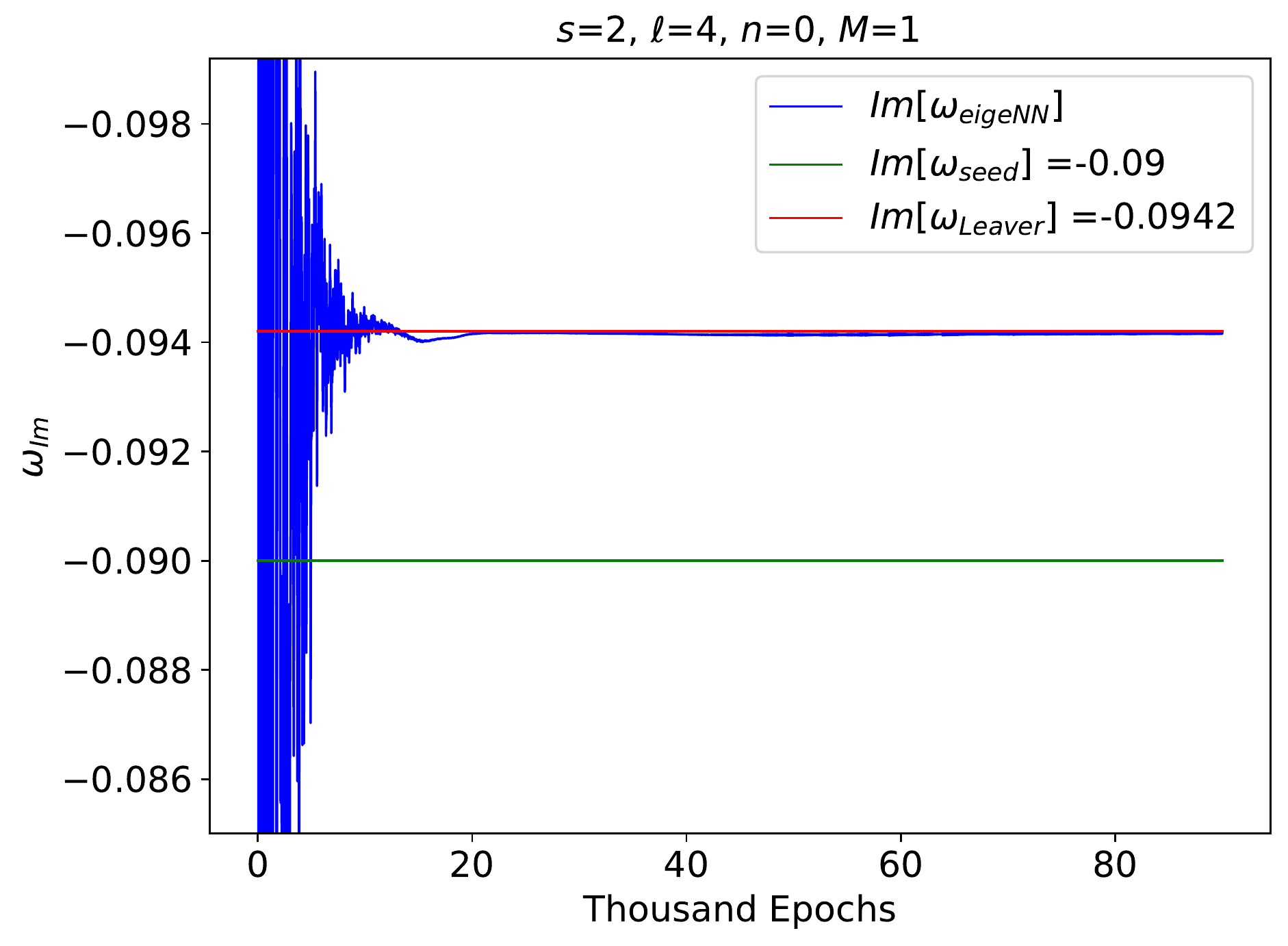}
\end{minipage}
\caption{\label{FIG: Output6}  Evolution of the NN approximation of QNMs during a 90~000 epoch-long training phase.}
\end{figure}

In the plots of figure \ref{FIG: Output3} - \ref{FIG: Output6}, the green line represents the seed values of $\omega$  that were embedded in the loss function of our eigenvalue solvers. Note that the NN converges towards the expected QNMs, rather than the seed values, which are given up to 2 decimal places. The QNM values given in tables \ref{Table 1} - \ref{Table 5} are in geometrical units. 

\begin{table}[t!]
\caption{\label{Table 3} 
The eigeNN approximations of the fundamental mode ($n = 0$) QNM frequencies for Dirac field perturbations ($s = 1/2$).} 
\begin{center}
\begin{tabular}{cccccccccccr}
\hline
$n$ & $\ell$ &&  $\omega_{\textsubscript{$Leaver$}}$  && $\omega_{\textsubscript{$eigeNN$}}$ && $\omega_{\textsubscript{$WKB$}}$ \cite{Konoplya2003} \\
 &  && \cite{Iyer1987} && (90~000 epochs) && (6th order) \\[0.25em]
\hline\\
\multirow{2}{0.5em}{0} & \multirow{2}{0.5em}{1} && \multirow{2}{7em}{$0.2822 - 0.0967i$} && $0.2822 - 0.0966i$ && $0.2822 - 0.0967i$ \\[0.25em]
                       &   &&                                      && (0.01\%)(-0.09\%) && ($<-0.01$\%)(-0.02\%) \\[0.25em]
                       & \multirow{2}{0.5em}{2} && \multirow{2}{7em}{$0.4772 - 0.0963i$} && $0.4772 - 0.0963i$ && $0.4772 - 0.0963i$ \\[0.25em]
                       &   &&                                      && (0.01\%)(0.03\%) && ($<0.01\%$)(0.05\%) \\[0.25em] 
                       & \multirow{2}{0.5em}{3} && \multirow{2}{7em}{$0.6708 - 0.0963i$} && $0.6708 -0.0963i$ && $0.6708 - 0.0963i$ \\[0.25em]
                       &   &&                                      && ($<-0.01$\%)(-0.02\%) && ($<-0.01$\%)(-0.02\%)  \\[0.5em] 
\hline
\end{tabular}
\end{center}
\end{table}

\begin{table}[H]
\caption{\label{Table 4} The $eigeNN$ approximations of the fundamental mode ($n = 0$) QNM frequencies for electromagnetic field perturbations ($s = 1$).} 
\begin{center}
\begin{tabular}{cccccccccccr}
\hline
$n$ & $\ell$ &&  $\omega_{\textsubscript{$Leaver$}}$  && $\omega_{\textsubscript{$eigeNN$}}$ && $\omega_{\textsubscript{$WKB$}}$ \cite{Konoplya2003} \\
 &  && \cite{Iyer1987} && (90~000 epochs) && (6th order) \\[0.5em]
\hline\\
\multirow{2}{0.5em}{0} & \multirow{2}{0.5em}{1} && \multirow{2}{7em}{$0.2483 - 0.0925i$} && $0.2483 - 0.0925i$ && $0.2482 - 0.0926i$ \\[0.5em]
                       &   &&                                      && (-0.01\%)(0.01\%) && (-0.04\%)(0.15\%) \\[0.25em]
                       & \multirow{2}{0.5em}{2} && \multirow{2}{7em}{$0.4576 - 0.0950i$} && $0.4576 - 0.0950i$ && $0.4576 - 0.0950i$ \\[0.25em]
                       &   &&                                      && ($<-0.01$\%)($<0.01\%$) && ($<-0.01$\%)(0.01\%) \\[0.25em] 
                       & \multirow{2}{0.5em}{3} && \multirow{2}{7em}{$0.6569 - 0.0956i$} && $0.6569 - 0.0956i$ && $0.6569 - 0.0956i$ \\[0.25em]
                       &   &&                                      && ($<0.01\%$)(0.01\%) && ($<-0.01$\%)(0.02\%)  \\[0.25em] 
\hline
\end{tabular}
\end{center}
\end{table}

\begin{table}[H]
\caption{\label{Table 5} The $eigeNN$ approximations of the fundamental mode ($n = 0$) QNM frequencies for gravitational field perturbations ($s = 2$).} 
\begin{center}
\begin{tabular}{cccccccccccr}
\hline
$n$ & $\ell$ &&  $\omega_{\textsubscript{$Leaver$}}$  && $\omega_{\textsubscript{$eigeNN$}}$ && $\omega_{\textsubscript{$WKB$}}$ \cite{Konoplya2003} \\
 &  && \cite{Iyer1987} && (90~000 epochs) && (6th order) \\[0.5em]
\hline\\
\multirow{2}{0.5em}{0} & \multirow{2}{0.5em}{2} && \multirow{2}{7em}{$0.3737 - 0.0896i$} && $0.3737 - 0.0890i$ && $0.3736 - 0.0889i$ \\[0.25em]
                       &   &&                                      && (0.01\%)(-0.05\%) && (-0.02\%)(-0.12\%) \\[0.25em]
                       & \multirow{2}{0.5em}{3} && \multirow{2}{7em}{$0.5994 - 0.0927i$} && $0.5994 - 0.0927i$ && $0.5994 - 0.0927i$ \\[0.25em]
                       &   &&                                      && (0.01\%)($<0.01\%$) && ($<0.01\%$)($<0.01\%$) \\[0.25em] 
                       & \multirow{2}{0.5em}{4} && \multirow{2}{7em}{$0.8092 - 0.0942i$} && $0.8092 - 0.0942i$ && $0.8092 - 0.0942i$ \\[0.25em]
                       &   &&                                      && ($<0.01\%$)(0.04\%) && ($<-0.01$\%)(-0.03\%)  \\[0.25em] 
\hline
\end{tabular}
\end{center}
\end{table}

In figures \ref{FIG: Output3} - \ref{FIG: Output6}, it is clear that the eigenvalue solvers are able to learn the expected values of $\omega_{Re}$ and $\omega_{Im}$ for various perturbation scenarios of an asymptotically flat Schwarzschild black hole. For the fundamental mode, $n=0$ and various choices of $s$ and $\ell$, the physical constraints provided in the loss function are sufficient to steer the NN toward the exact, non-trivial solutions of our perturbation equations. This is remarkable, considering the conceptual simplicity of the NN optimisation algorithm.

As is evidenced by the tables \ref{Table 2} - \ref{Table 5}, the  QNMs computed by our eigenvalues solvers are as accurate as the values computed with the CFM and the 6th-order WKB method. Since we have used the QNMs from Leaver \cite{Leaver1985, Iyer1987}, which are given up to 4 decimal places, as our closest approximation to an exact solution, we only confirm accuracy up to that level. Needless to say, an ideal measure of accuracy of our approximations would be exact QNMs obtained via analytical methods for solving our differential equations. Note that the QNMs obtained using the AIM (which are not listed in the tables but can be found in Ref.~\cite{Choetal2012}) were shown to be as accurate as Leaver's method.  

Regarding the time taken by eigenvalue solvers to complete executing computations, there is a significant difference between the duration for training our NNs compared to that of running the other approximation techniques. While it takes around 10 minutes to run each of the 90~000 epoch-long training sessions to solve our perturbation equations, the computation takes much less time. For example, in the case of the WKB and AIM techniques, many QNM values can be computed in less than a minute. Note that this comparison is tentative and reflects just the outcome of the present work, which is a baseline for potential future improvements. For example, the computational speeds of PINNs could be enhanced by tapping into the parallelisable nature of NNs. So far we have focussed on fine-tuning the accuracy of NN approximations as that is particularly important for black hole QNMs.

\par In terms of other performance measures, the scalability of PINNs, with regard to the ability to handle a large number of input dimensions, is one major advantage of PINNs over other numerical methods. As pointed out in Refs.~\cite{Luetal2021, Grohs2018}, NNs overcome the curse of dimensionality and, therefore, have the capacity to approximate high-dimensional functions quite efficiently. This attribute justifies the potential future extension of PINNs to compute the QNMs associated with high-dimensional perturbation scenarios, where traditional mesh-based approximation techniques could suffer.

Our results show signs of the expected limitations, listed in Ref.~\cite{Karniadakisetal2021}, of solving PDEs with NNs that have been observed in various applications of physics-informed machine learning. One is the fact that complicated loss functions (with many terms in the governing equations) lead to highly non-convex optimisation problems. As a result, the training process may not be sufficiently stable and convergence to the global minimum may not be guaranteed \cite{Karniadakisetal2021}.

We can see this by contrasting the results obtained from our computations involving the relatively simple differential equation for near extremal non-rotating black holes versus the relatively more complex perturbation equations of asymptotically flat Schwarzschild black holes. For the former, figure \ref{FIG: Output2} shows that our NN quickly converges towards the expected solution regardless of the multipole number, even in the absence of a seed loss term to further constrain the eigenvalue solvers. However, for the latter, the NN has more difficulty converging in lower multipole number cases as seen in figure \ref{FIG: Output3} where the NN converges faster for $\ell =2$ when compared to $\ell =0$. In fact, without the seed loss term to solve asymptotically flat Schwarzschild black holes, the NN fails to converge when we have $\ell = 0, 1, 2$ but does for $\ell > 3$.


In our attempts to solve even more challenging problems, such as the perturbation equations of asymptotically flat Reissner-Nordstr\"{o}m and asymptotically (anti)-de Sitter Schwarzschild black holes, the instability appears to be more pronounced as the NNs fail to converge on the expected QNMs for these cases. To alleviate this constraint and broaden the scope of PDEs to be solved we will need to add to our eigenvalue solvers some stronger constraints or features that address instability.

\section{Discussions and Conclusion}\label{SEC: Discussion} 

In summary, we have explored the possibility of implementing PINNs as a new technique to solve black hole perturbation equations. We considered two variations of PINNs built with the DeepXDE and Pytorch packages in Python. To give some background on the underlying physics, we began by revisiting the perturbation equations for static, spherically symmetric black holes, particularly asymptotically flat and (anti)-de Sitter Schwarzschild black holes whose perturbations are described by one-dimensional Schr\"{o}dinger-like eigenvalue problems. Our goal was to determine when and how PINNs can be best applied to solve these equations, which are generally difficult to solve analytically and compute the QNMs of black holes.

Since PINNs are extensions of deep neural networks, we outlined NNs in section \ref{SEC: DNN}, in terms of their structure and the mechanisms behind their function approximation abilities. Afterwards, PINNs were described with illustrative examples showing how physics constraints are embedded in the loss function of a NN. These constraints include the governing PDE, its associated boundary conditions and regularisation functions.

Of the two variations of PINNs considered in this work, the eigenvalue solvers were implemented to compute the QNMs of asymptotically flat Schwarzschild and near extremal non-rotating de Sitter black holes. Given that the latter scenario has known exact formulae for the QNM frequencies (given by Ref.~\cite{FerrariMashhoon1984, CardosoLemos2003, Molina2003}), we were able to reliably validate the accuracy of our NN approximations. We obtained QNM values with up to 6 digit accuracy and plots showing the evolution of the NN's approximation of the QNMs over a 100~000 epoch training phase. The plots showed that the NN's approximation quickly converged towards the expected solutions, regardless of the multipole number $\ell$ or the existence of a seed loss term in the loss function.

Regarding the more analytically intractable problems, we managed to solve the perturbation equations of asymptotically flat Schwarzschild black holes by embedding the equations themselves, the QNM boundary conditions and a seed loss term into the loss function of the eigenvalue solvers. The computed QNMs have the same level of accuracy as those obtained through Leaver's CFM \cite{Leaver1985} or Konoplya's 6th order WKB method \cite{Konoplya2003} (at least up to 4 decimal places as given in the literature \cite{Iyer1987}). However, in terms of efficiency, our eigenvalue solvers take several minutes to train, compared to the few seconds to a minute it takes to generate accurate results using other techniques such as the WKB method. We also found that the efficiency of PINNs could be optimised by setting up the NN using lower values in the range of values of the hyperparameters that we tested; that is, the number of training epochs, number of training points and number of nodes per layer. 

To date, we have been able to compute only the fundamental mode frequencies (i.e. $n = 0, \ell \geq 0$) that, as it turns out, are the least damped, longest-lived modes compared to higher overtones with $n > 0$. This is because we have added regularisation terms that simultaneously penalise the NN for learning trivial eigenfunctions and encourage it to learn the most energetic QNMs, which happen to occur when $n=0$ for any given $\ell$. Potential future work would seek a modification of the eigenvalue scanning mechanism similar to that introduced by Ref.~\cite{Jinetal2020}, which will allow for the computation of higher overtones for our complex-valued QNMs. 

\par Concerning the question of the stability of PINNs as they increase in depth, that is still very much an open problem, in general, within the literature and is in the early stages of investigation. When studying PINNs to mimic the analysis of numerical discretisation techniques, convergence and stability are related to how well the NN learns from the physical laws embodied by the governing PDEs and initial/boundary conditions \cite{Salvatoretal2022}. It is well-known that there is a bias-variance trade-off that comes with choosing the right depth of a neural network, where too deep a neural network may lead to overfitting and inefficiency. The most recent literature in laying out a theoretical framework for PINNs includes Refs.~\cite{Shinetal2020, Shinetal2020b} that provide formal findings regarding PINNs and their convergence when dealing with linear problems including second order elliptic, hyperbolic and parabolic type PDEs. Utilising these recent developments is important to understand and improve on PINNs in continued use to compute QNMs.

As discussed in section \ref{SEC: Schwarzschild}, our NNs exhibit signs of instability which we suspect to be a result of the level of complexity in the loss function, which makes for a highly non-convex optimisation process \cite{Karniadakisetal2021}. This is counter-intuitive to our initial expectation that PINNs can accurately solve any PDE (regardless of complexity) if they are formulated in a finite domain and their associated boundary conditions are properly set up. This was not the case for our attempts when applying eigenvalue solvers to the Reissner-Nordstr\"{o}m case. To overcome this instability in future work, one plausible approach would be to consider the recent work in Ref.~\cite{Gnanasambandam2022} that shows that a ``self-scalable'' activation function leads to PINNs which are
less susceptible to spurious stationary points, an obstacle in highly non-convex loss functions.

A final point to note concerning the limitations of PINNs is their relative inefficiency compared to the extant methods for computing QNMs. Further investigation needs to be done to improve the performance of the eigenvalue solvers as they currently do not surpass the efficiency of established methods such as the WKB method and the AIM. Overall, PINNs have not developed far enough to be applied broadly in the study of black hole perturbations. In conclusion to their seminal work on PINNs, Ref.~\cite{Raissietal2019} pointed out that this method should not be viewed as a replacement for classical numerical methods for solving PDEs, but rather as methods that can bring added merits such as implementation simplicity to accelerate the rate of testing new ideas. In a similar vein, the application of PINNs to QNMs brings at least a new angle to study the perturbation equations, even though they are not as efficient as canonical methods.

As is, the PINN approach may only work in computing the fundamental QNMs of not only four-dimensional Schwarzschild black holes, but also general dimensional Schwarzschild black holes (described in Ref.~\cite{KonoplyaZhidenko2011}) given the similarity of the differential equations. Despite the present challenges (which are expected for a burgeoning method) this approach to computing QNMs is worth pursuing further as it demonstrates the same level of accuracy as the leading existing methods. 

\begin{acknowledgments}
The authors would like to thank Jan Nordstr\"{o}m for the enlightening discussions, and Wesley Doorsamy for insightful suggestions provided to drafts of this paper. ASC is supported in part by the National Research Foundation of South Africa (NRF). GEH is supported by UJ. AMN is supported by the UJ GES 4IR.
\end{acknowledgments}

\bibliographystyle{utphys}
\bibliography{main}

\end{document}